\documentclass[10pt]{article}
\usepackage{hyperref}
\usepackage{adjustbox}
\usepackage{graphicx}
\usepackage{lipsum}
\usepackage{enumerate}
\usepackage[compatibility=false]{caption}
\usepackage{lipsum, babel}
\usepackage{longtable}
\usepackage{graphicx}
\usepackage{dcolumn}
\usepackage[table]{xcolor}
\usepackage{amsmath}
\usepackage{multirow}
\usepackage{amssymb}
\usepackage{amsfonts}
\usepackage{orcidlink}
\usepackage{bm}
\usepackage{float}
\hypersetup{colorlinks=true, linkcolor=blue, citecolor=blue, urlcolor=blue}
\usepackage{color}
\usepackage{caption}
\usepackage{subfigure}
\usepackage{setspace}
\usepackage[numbers,sort&compress]{natbib}
\begin{document}
\baselineskip=14pt

\begin{center}
\LARGE{Spherically Symmetric Quantum-Corrected Black Holes with String Clouds: A Multi-Observable Analysis}
\end{center}

\vspace{0.3cm}

\begin{center}
{\bf Faizuddin Ahmed\orcidlink{0000-0003-2196-9622}}\footnote{\bf faizuddinahmed15@gmail.com}\\
{\it Department of Physics, The Assam Royal Global University, Guwahati, 781035, Assam, India}\\
\vspace{0.15cm}

{\bf Ahmad Al-Badawi\orcidlink{0000-0002-3127-3453}}\footnote{\bf ahmadbadawi@ahu.edu.jo}\\
{\it Department of Physics, Al-Hussein Bin Talal University 71111, Ma’an, Jordan}\\
\vspace{0.15cm}

{\bf Orhan Dönmez\orcidlink{0000-0001-9017-2452}}\footnote{\bf orhan.donmez@aum.edu.kw} \\
{\it College of Engineering and Technology, American University of the Middle East, Egaila 54200, Kuwait}\\
\vspace{0.15cm}

{\bf \.{I}zzet Sakall{\i}\orcidlink{0000-0001-7827-9476}}\footnote{\bf izzet.sakalli@emu.edu.tr}\\
{\it Physics Department, Eastern Mediterranean University, Famagusta 99628, North Cyprus via Mersin 10, Turkey}\\
\vspace{0.15cm}

{\bf Saeed Noori Gashti\orcidlink{0000-0001-7844-2640}}\footnote{\bf saeed.noorigashti70@gmail.com}\\
{\it School of Physics, Damghan University, Damghan 3671645667, Iran}\\
\vspace{0.15cm}

{\bf Behnam Pourhassan\orcidlink{0000-0003-1338-7083}}\footnote{\bf b.pourhassan@du.ac.ir}\\
{\it School of Physics, Damghan University, Damghan 3671645667, Iran}\\
\vspace{0.15cm}

{\bf Fatih Doğan\orcidlink{0000-0001-7992-1723}}\footnote{\bf fatih.dogan@aum.edu.kw}\\
{\it College of Engineering and Technology, American University of the Middle East, Egaila 54200, Kuwait}\\
\end{center}

\vspace{0.35cm}

\begin{abstract}
We present an investigation of quantum-corrected black hole spacetimes coupled with clouds of strings, examining two distinct theoretical models that incorporate quantum gravitational effects through different implementations of correction terms. Our study explores the geodesic structure, focusing on photon sphere properties, black hole shadows, and innermost stable circular orbits of test particles around these exotic geometries. The analysis reveals fundamental modifications to particle trajectories that distinguish quantum-corrected solutions from their classical counterparts, with observable implications for high-energy astrophysics. We investigate quasi-periodic oscillations arising from test particle motion, deriving frequency relationships that could serve as observational probes of quantum gravity effects in accreting black hole systems. Through rigorous gravitational lensing analysis using the Gauss-Bonnet theorem, we calculate weak-field deflection angles and identify distinctive signatures that enable discrimination between the two quantum correction models. The gravitational lensing study reveals opposite dependencies on quantum parameters between the models, providing unambiguous observational discriminants. Additionally, we analyze the thermodynamic properties including temperature, entropy, and heat capacity modifications, exploring topological characteristics and phase transition behavior in these quantum-corrected systems. The investigation reveals that the two models exhibit contrasting behaviors across multiple observational channels, from gravitational lensing deflection angles to quasi-periodic oscillation frequencies, providing remarkable discriminants for testing quantum gravity theories.
\end{abstract}

\newpage

\tableofcontents

\small

\section{Introduction}
\label{isec1}

The intersection of quantum gravity and Black Hole (BH) physics represents one of the most profound frontiers in modern theoretical physics, where the fundamental principles of quantum mechanics and General Relativity (GR) converge in the extreme gravitational environments surrounding these cosmic entities~\cite{1,2,3}. As our understanding of BH physics has evolved from purely classical descriptions to incorporate quantum corrections, the need for theoretical frameworks that can bridge the gap between classical GR and quantum gravity has become increasingly apparent. Quantum-corrected BH solutions offer a promising avenue for exploring how quantum gravitational effects might manifest in astrophysically relevant scenarios, potentially providing observable signatures that could distinguish between competing theories of quantum gravity~\cite{4,5,6}.

The Letelier spacetime, originally proposed as a solution describing BHs surrounded by a Cloud of Strings (CoS), has emerged as a particularly interesting theoretical laboratory for investigating modified gravity effects~\cite{7,8}. This framework naturally incorporates topological defects that can arise in various cosmological and string-theoretical scenarios, providing a rich phenomenological playground where the interplay between classical modifications and quantum corrections can be studied~\cite{9,10}. The CoS parameter $\alpha$ quantifies the density of the string cloud environment, effectively modifying the gravitational field strength and leading to observable deviations from the standard Schwarzschild geometry. Recent developments in loop quantum gravity and other quantum gravitational approaches have provided new theoretical tools for constructing quantum-corrected BH solutions that preserve the fundamental principles of general covariance while incorporating quantum effects in a controlled manner~\cite{11,12,13}.

Geodesic analysis in quantum-corrected spacetimes reveals fundamental modifications to particle trajectories that have direct observational consequences~\cite{14,15}. The study of null geodesics provides insights into photon motion, photon sphere properties, and BH shadow characteristics, all of which are directly observable through modern high-resolution imaging techniques such as those employed by the Event Horizon Telescope (EHT)~\cite{16,17}. Timelike geodesics govern the motion of massive particles and determine crucial features such as the Innermost Stable Circular Orbit (ISCO), which plays a central role in accretion disk physics and the generation of quasi-periodic oscillations (QPOs) observed in X-ray binaries and Active Galactic Nuclei (AGN)~\cite{18,19,20}. The modifications to geodesic structure induced by quantum corrections can lead to measurable shifts in QPO frequencies, providing a potential observational pathway for detecting quantum gravitational effects in astrophysical systems.

Gravitational lensing represents one of the most powerful probes of spacetime curvature, offering direct access to the gravitational field structure in regimes where other observational techniques may be insufficient~\cite{21,22}. The weak deflection of light by BHs has been extensively studied in classical GR, yielding predictions that have been confirmed through numerous astrophysical observations including strong lensing events, Einstein rings, and multiple imaging of distant quasars. In quantum-corrected geometries, the deflection angle acquires additional contributions that encode information about both quantum corrections and topological defects, potentially providing distinctive signatures observable through Very Long Baseline Interferometry (VLBI) and future space-based interferometric missions~\cite{23,24}. The application of the Gauss-Bonnet Theorem (GBTh) to calculate deflection angles offers significant computational advantages and provides clear geometric interpretations of how metric modifications influence light propagation~\cite{25,26}.

Field perturbations around BHs provide crucial insights into their stability properties and dynamical response to external disturbances~\cite{27,28}. The study of scalar and Electromagnetic (EM) field perturbations in quantum-corrected BH backgrounds reveals how quantum effects modify the characteristic frequencies and stability behavior, which carry information about the underlying spacetime geometry~\cite{29,30}. These perturbation frequencies are potentially observable through gravitational wave detectors, offering another avenue for testing quantum gravity theories in the strong-field regime. The stability analysis of perturbations also provides important theoretical constraints on the parameter ranges where quantum-corrected solutions remain physically viable~\cite{31,32}.

Thermodynamic properties of BHs have played a central role in our understanding of the deep connections between gravity, quantum mechanics, and statistical mechanics since the pioneering work of Bekenstein and Hawking~\cite{33,34}. The introduction of quantum corrections and topological defects significantly modifies the thermodynamic relationships, affecting fundamental quantities such as temperature, entropy, and heat capacity. These modifications can lead to changes in thermodynamic stability and phase transition behavior, including alterations to the Hawking-Page (HP) transition between small and large BH phases~\cite{35,36}. The emerging field of thermodynamic topology provides new tools for characterizing phase transitions and critical phenomena in modified BH systems, offering insights into the global structure of thermodynamic phase space~\cite{37,38}.

The motivation for our investigation stems from the recognition that quantum gravitational effects, while expected to be most pronounced at the Planck scale, may manifest in observable ways in the extreme environments surrounding astrophysical BHs. Recent advances in observational capabilities, particularly the revolutionary imaging achievements of the EHT and the ongoing gravitational wave observations by LIGO/Virgo collaborations, have opened new windows for testing fundamental physics in strong gravitational fields~\cite{39,40}. The systematic study of quantum-corrected BH solutions coupled with CoS provides a unified framework for exploring how these effects might be detected and distinguished from classical predictions across multiple observational channels.

Our investigation aims to provide a particular analysis of quantum-corrected Letelier BH spacetimes through multiple complementary approaches. We examine the geodesic structure to understand how quantum corrections modify particle trajectories, photon sphere properties, and ISCO characteristics. We investigate the resulting implications for QPO frequencies observable in accretion disk systems, providing testable predictions for current and future X-ray missions. Through rigorous gravitational lensing analysis using the GBTh, we derive precise deflection angle formulas that reveal distinctive signatures capable of discriminating between different quantum correction models. Our field perturbation studies explore the stability properties of scalar and EM fields, relevant for gravitational wave astronomy and fundamental stability considerations. Finally, we examine the modified thermodynamic properties and topological characteristics, contributing to our understanding of quantum-corrected BH thermodynamics and phase transition behavior.

This paper is organized as follows: Section~\ref{isec2} introduces the quantum-corrected BH solutions coupled with CoS, establishing the theoretical framework for both Model-I and Model-II geometries. Section~\ref{isec3} presents the photon sphere analysis and BH shadow properties. Section~\ref{isec4} examines the ISCO characteristics and their implications for accretion disk physics. Section~\ref{isec5} investigates QPO frequencies arising from particle motion near quantum-corrected BHs. Section~\ref{isec6} analyzes the topological properties of photon orbits using modern differential geometry techniques. Section~\ref{isec7} provides a gravitational lensing analysis using the GBTh approach. Finally, Sec.~\ref{isec8} summarizes our main findings and discusses their possible observational implications for testing quantum gravity theories in astrophysical contexts.

\section{Quantum-Corrected BH Coupled with CoS} \label{isec2}

Recently, an interesting class of quantum-corrected BH models was proposed in~\cite{1} using the Hamiltonian constraint approach developed in loop quantum gravity~\cite{2,3,4,5,6}. These models are constructed to preserve general covariance, a fundamental feature of GR, and represent a significant step toward incorporating quantum gravitational effects into classical BH solutions. Within this framework, two alternative BH solutions emerge, depending on the specific choice of the quantum correction parameter $\zeta$. The perturbation properties and shadow characteristics of these BHs have been analyzed in~\cite{7}. More intricate optical phenomena were studied in~\cite{8}, and the application of advanced perturbation methods to these models was recently reported in~\cite{9}. 

In this work, we extend these quantum-corrected BH solutions by coupling them with a CoS, following the Letelier spacetime framework~\cite{10}. The CoS represents a topological defect that can arise in various cosmological and string-theoretical scenarios~\cite{11,12}, and its presence modifies the effective gravitational field around the BH. This coupling provides a unique opportunity to study the combined effects of quantum corrections and topological defects on BH physics, with potentially observable consequences for gravitational lensing, QPOs, and thermodynamic properties. The interplay between the quantum correction parameter $\zeta$ and the CoS parameter $\alpha$ gives rise to rich phenomenology that distinguishes these solutions from both classical Schwarzschild and pure Letelier BHs.

The metric of a spherically symmetric BH coupled with CoS is given by the following line element \cite{1}:
\begin{equation}
ds^2 = -f(r)\,dt^2 + \frac{1}{g(r)}\,dr^2 + r^2\,d\Omega^2,
\label{metric-0}
\end{equation}
where $f(r)$ and $g(r)$ are the metric functions that encode both the quantum corrections and the CoS effects. Here, $d\Omega^2 = d\theta^2 + \sin^2\theta\,d\phi^2$ represents the standard metric on a unit 2-sphere. The metric functions differ between the two models we consider, leading to distinct physical predictions for observables.

For the first type of quantum-corrected BH (Model-I), the metric functions are given by~\cite{1}:
\begin{align}
f(r) &= \left(1-\alpha - \frac{2M}{r} \right)\,\left[1+\frac{\zeta^2}{r^2}\,\left(1-\alpha - \frac{2M}{r} \right)\right], \label{metric-1} \\
g(r) &= f(r), \label{metric-2}
\end{align}
where $\zeta$ is the quantum correction parameter quantifying the strength of quantum gravitational effects, and $M$ is the ADM mass of the BH. In Model-I, the quantum corrections appear symmetrically in both the temporal and radial components of the metric, leading to the condition $f(r) = g(r)$. This symmetry has important implications for geodesic motion and the structure of null surfaces, as we shall demonstrate in subsequent sections.

For the second type of BH (Model-II), the metric functions take a distinct asymmetric form:
\begin{align}
f(r) &= 1-\alpha - \frac{2M}{r}, \label{metric-3} \\
g(r) &= f(r)\,\left(1+\frac{\zeta^2}{r^2}\,f(r)\right). \label{metric-4}
\end{align}
In Model-II, the temporal metric component $f(r)$ retains the classical Letelier form, while the radial component $g(r)$ incorporates quantum corrections through the additional term proportional to $\zeta^2$. This asymmetric modification distinguishes Model-II from Model-I and leads to different physical predictions for observable quantities such as photon sphere radii, ISCOs, and gravitational lensing deflection angles. The choice between these two models reflects different possible implementations of quantum corrections in the effective metric, and observational data may eventually favor one over the other.

In the classical limit $\zeta \rightarrow 0$, both metrics reduce to the Letelier BH solution~\cite{10}, which is characterized by the parameter $\alpha$ representing the string cloud density. The Letelier solution further reduces to the standard Schwarzschild solution when $\alpha = 0$. The parameter $\alpha$ satisfies $0 \leq \alpha < 1$ to ensure the existence of a physical event horizon and avoid naked singularities~\cite{13}. Physically, $\alpha$ quantifies the relative contribution of the string cloud to the total energy density, with larger values indicating a denser string cloud environment surrounding the BH.

Notably, the horizon structure is governed by the function $g(r)$, which remarkably remains identical for both Model-I and Model-II despite their different functional forms for $f(r)$. Thus, the location of the event horizon $r_h$ is determined by solving $g(r_h) = 0$, yielding the same horizon radius for both models:
\begin{equation}
r_h = \frac{2M}{1-\alpha} > r_{h,\,\text{Sch}},
\label{horizon-radius}
\end{equation}
where $r_{h,\,\text{Sch}} = 2M$ is the Schwarzschild horizon radius. This expression shows that the CoS parameter $\alpha$ effectively increases the BH horizon radius, reflecting an enhancement of the gravitational field strength due to the presence of the string cloud. For instance, when $\alpha = 0.2$, the horizon radius increases by 25\% compared to the Schwarzschild value, demonstrating the significant impact of the CoS on the spacetime geometry.

Table~\ref{horizons_newmodel} presents a survey of the horizon structure for quantum-corrected BHs coupled with CoS across a wide range of parameter values. The table displays the horizon radii for various combinations of the CoS parameter $\alpha \in \{0.0, 0.1, 0.2\}$ and the quantum correction parameter $\zeta \in \{0.0, 0.1, 0.5, 1.0, 1.5, 2.0, 2.5, 3.0, 4.0, 5.0\}$, with the BH mass fixed at $M = 1$ for convenience. Several important features emerge from this analysis. For $\zeta = 0$, the spacetime corresponds to the pure Letelier solution, and when both $\alpha = 0$ and $\zeta = 0$, the classical Schwarzschild BH is recovered with a single horizon at $r_h = 2.0M$. As $\alpha$ increases while keeping $\zeta = 0$, the outer horizon moves outward to $r_h = 2.2222M$ for $\alpha = 0.1$ and $r_h = 2.5M$ for $\alpha = 0.2$, confirming the prediction of Eq.~(\ref{horizon-radius}).

For $\zeta > 0$, quantum corrections introduce an additional inner horizon, transforming the spacetime into a non-extremal BH configuration with two distinct horizons. The inner horizon radius increases with $\zeta$: for example, at $\alpha = 0.1$, the inner horizon grows from $r_{\text{in}} \approx 0.26M$ at $\zeta = 0.1$ to $r_{\text{in}} \approx 1.91M$ at $\zeta = 5.0$. This behavior indicates that stronger quantum corrections lead to a larger region between the inner and outer horizons, potentially affecting the causal structure and thermodynamic properties of the BH. The outer horizon remains nearly constant as $\zeta$ varies, being determined primarily by $\alpha$ through Eq.~(\ref{horizon-radius}), since the quantum correction terms vanish at the location where $f(r_h) = 0$ in the Letelier limit. This demonstrates that while quantum corrections significantly modify the interior structure and create an inner horizon, they leave the location of the event horizon largely unchanged, which has important implications for observational signatures.

\begin{figure}[ht!]
    \centering
        \includegraphics[width=0.3\textwidth]{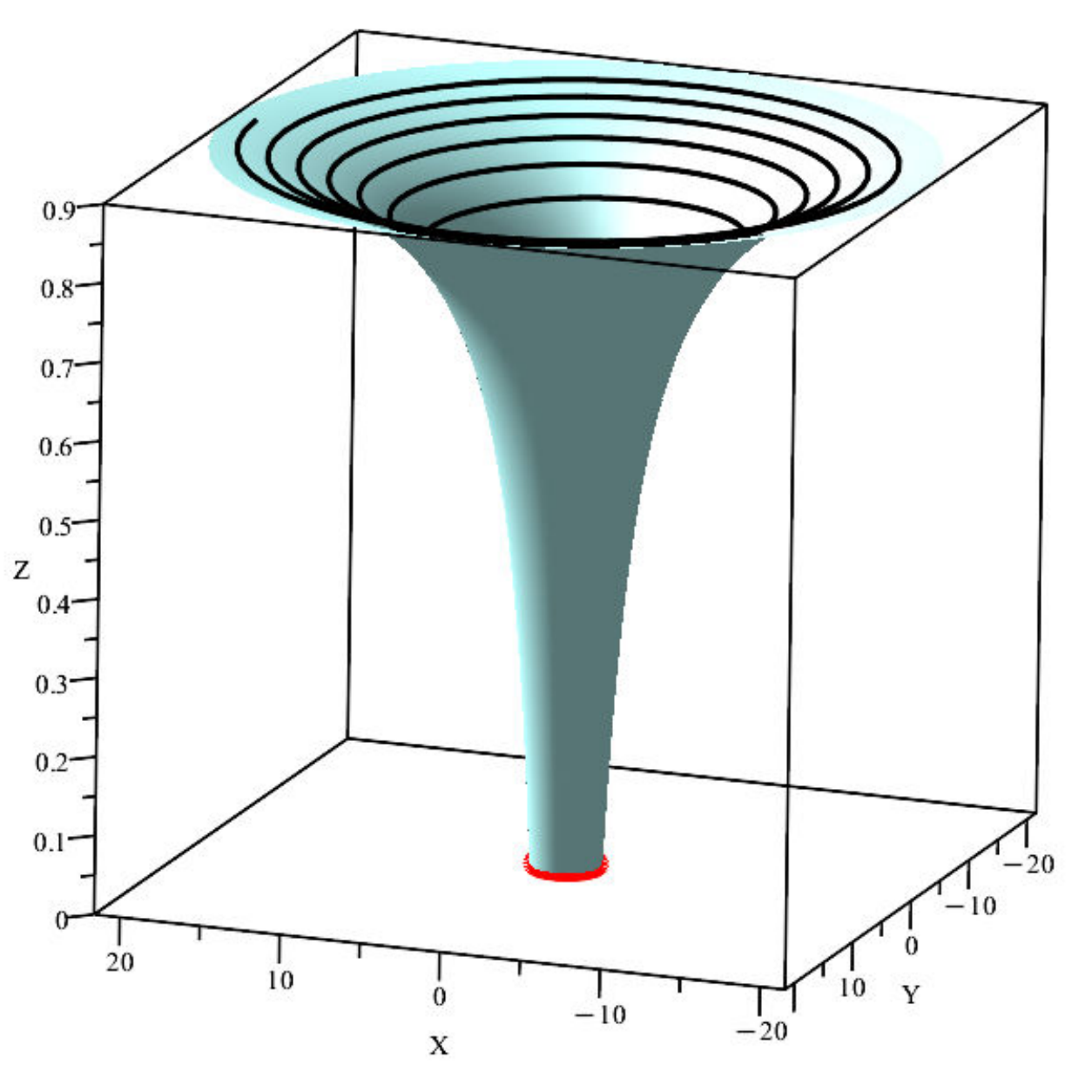}
        \includegraphics[width=0.3\textwidth]{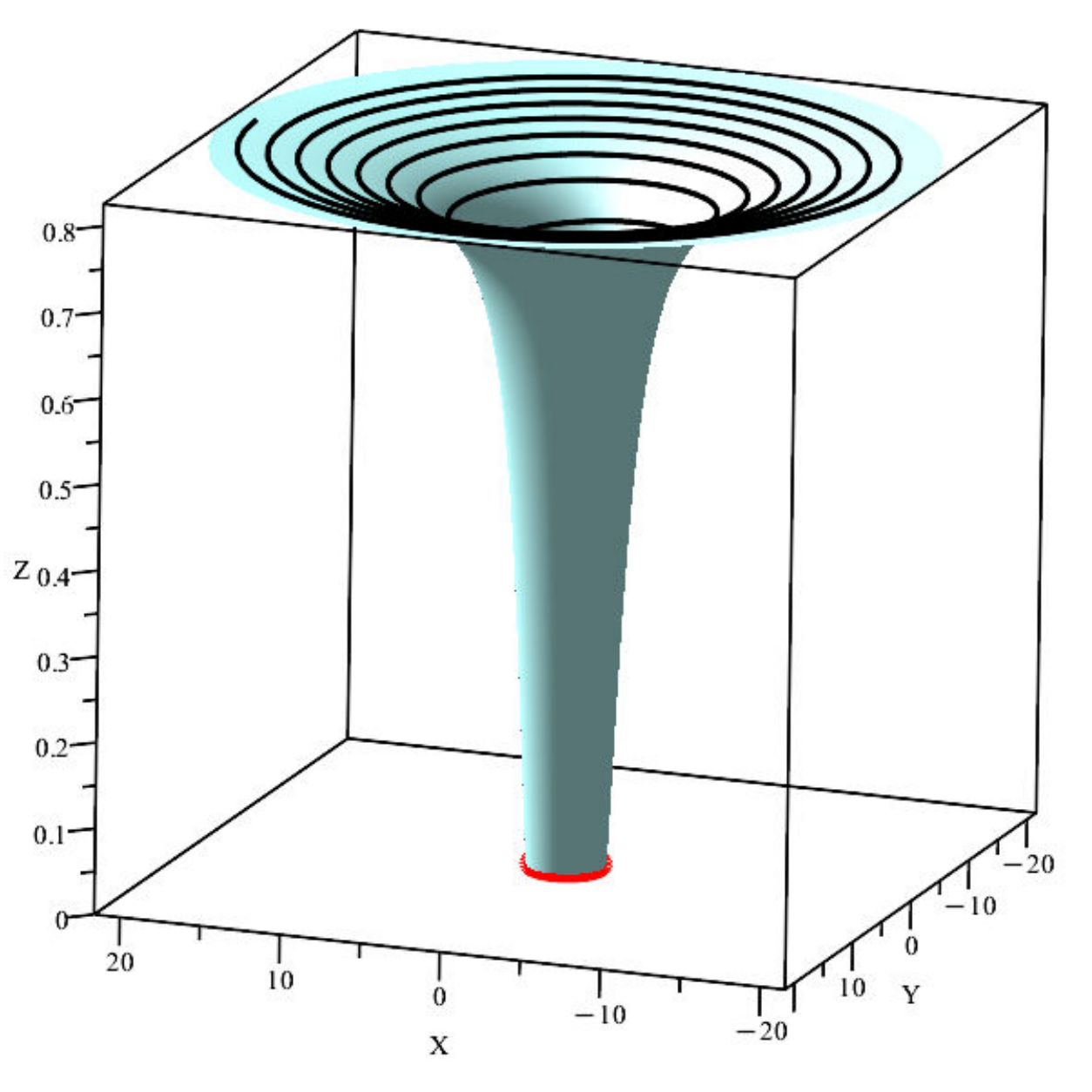}
        \includegraphics[width=0.3\textwidth]{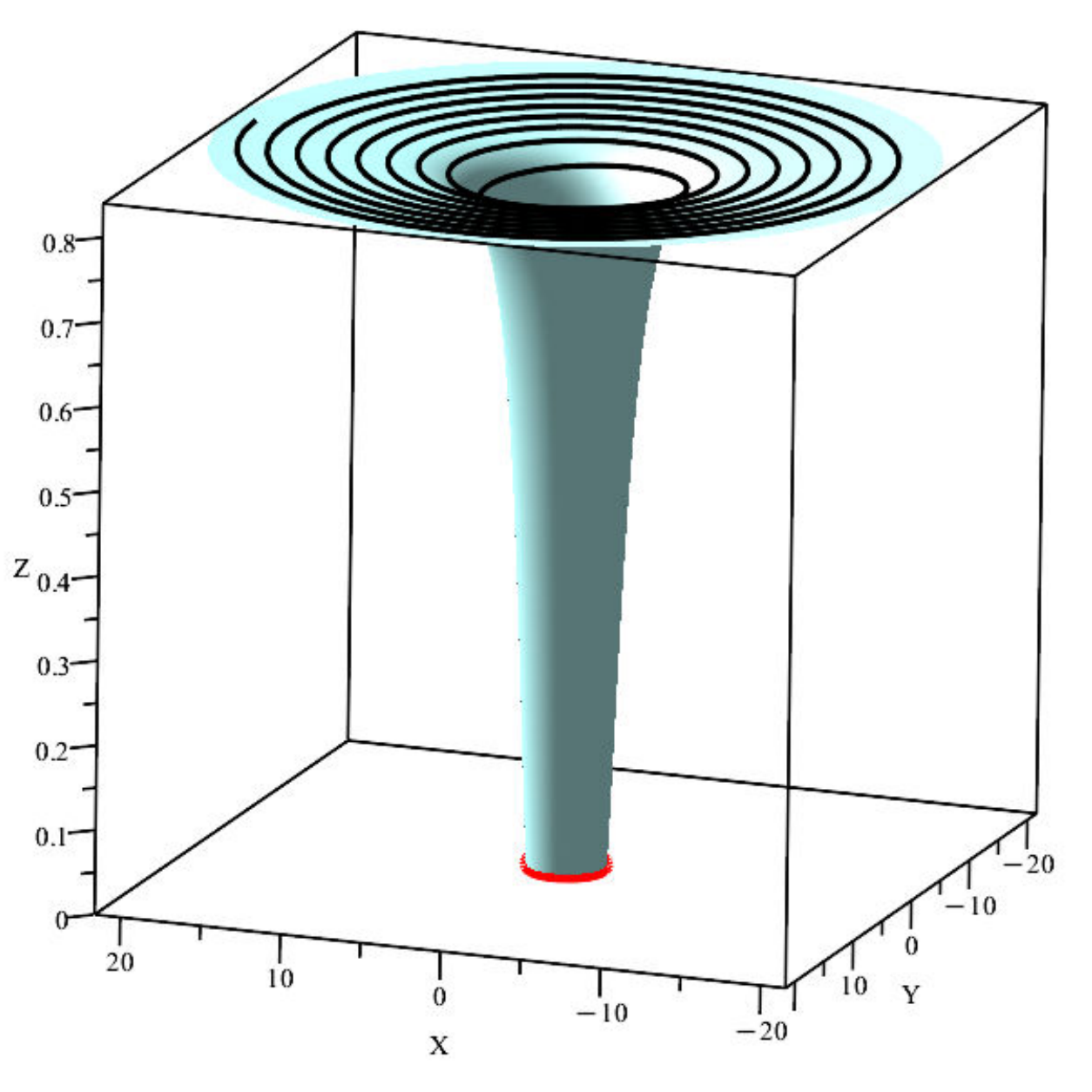}\\
        (i) $\zeta=0$, $\alpha=0$, $r_h=2.0$ (Sch. BH) \hspace{0.5cm}
        (ii) $\zeta=4.0$, $\alpha=0.1$, $r_h=2.22$ \hspace{0.5cm}
        (iii) $\zeta=5.0$, $\alpha=0.1$, $r_h=2.22$
        \hfill
        \includegraphics[width=0.3\textwidth]{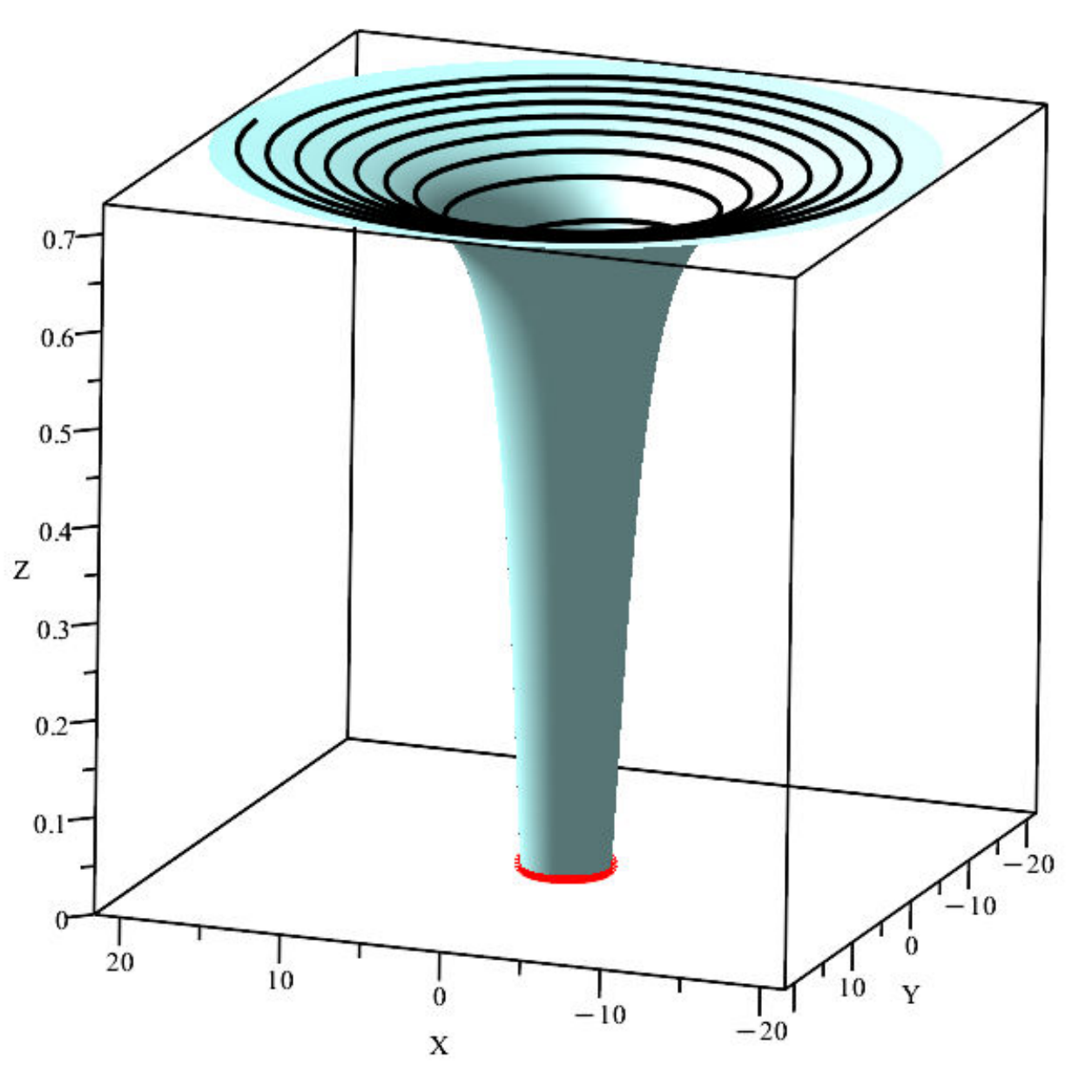}
        \includegraphics[width=0.3\textwidth]{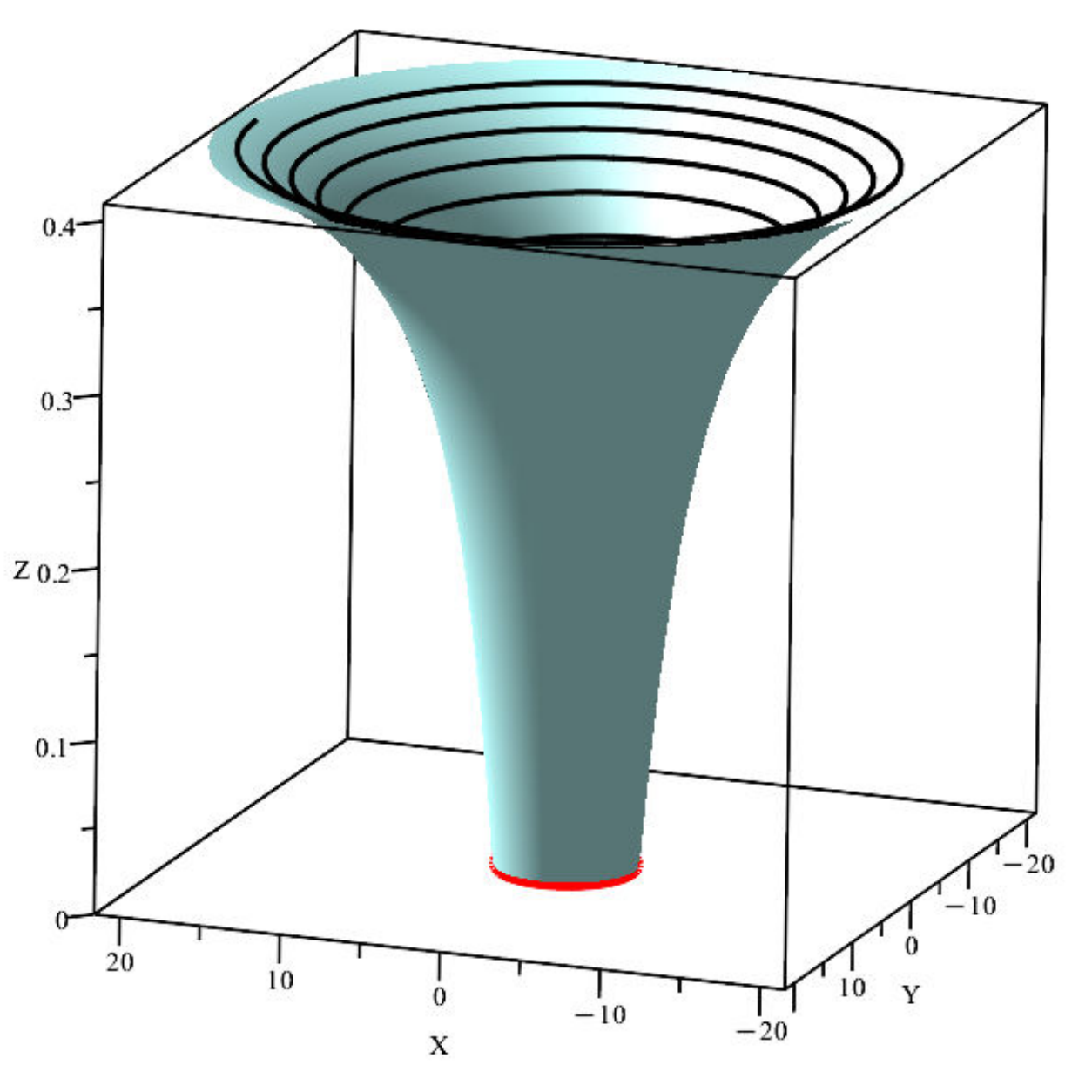}
        \includegraphics[width=0.3\textwidth]{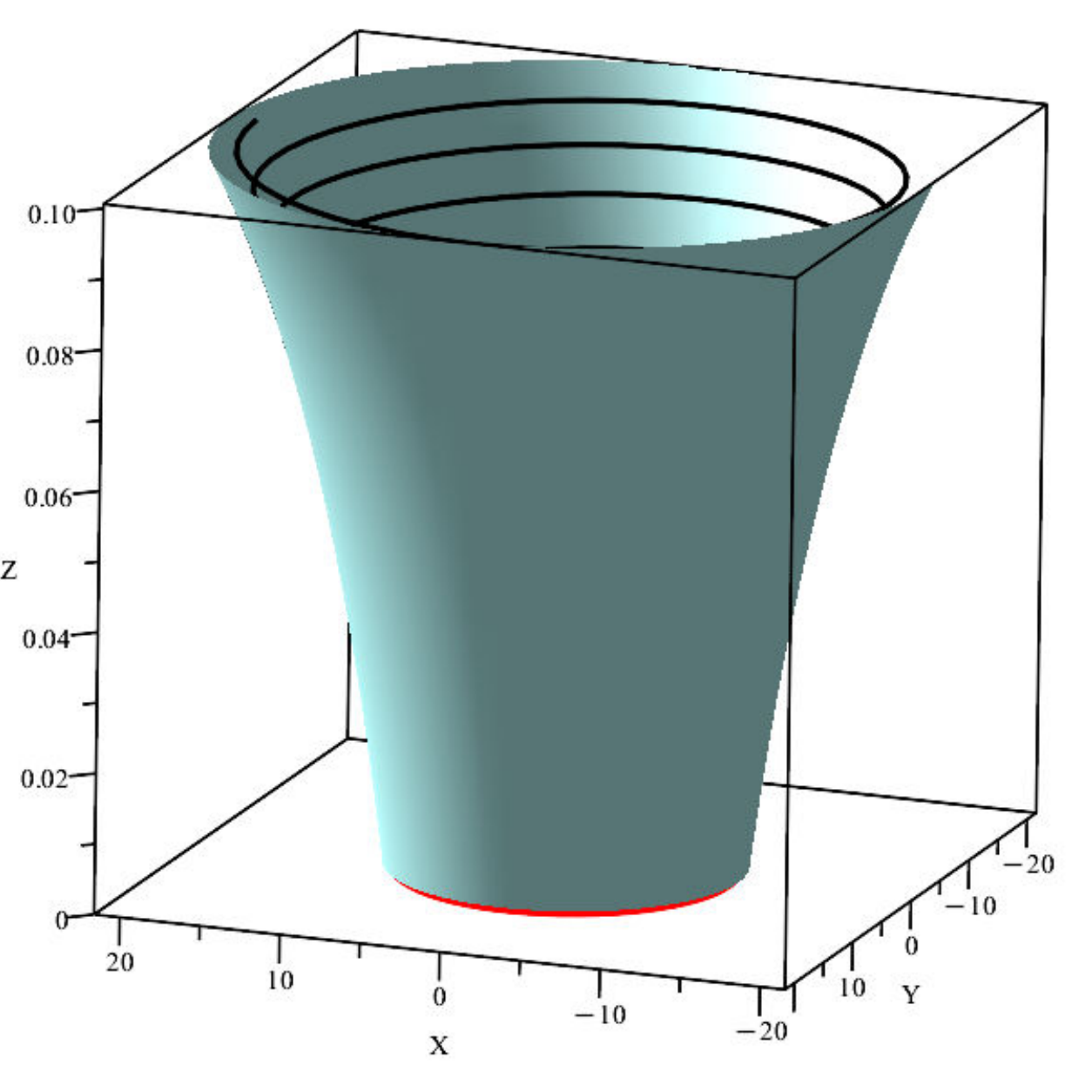}\\
        (iv) $\zeta=5.0$, $\alpha=0.2$, $r_h=2.50$ \hspace{0.5cm}
        (v)  $\zeta=5.0$, $\alpha=0.5$, $r_h=4.00$ \hspace{0.5cm}
        (vi) $\zeta=5.0$, $\alpha=0.8$, $r_h=10.00$
\caption{\footnotesize 3D diagrams of the quantum-corrected BHs (Model-I\&II) with metric function $g(r) = (1 - \alpha) - \frac{2M}{r} + \frac{\zeta^2}{r^2} \left( (1 - \alpha) - \frac{2M}{r} \right)^2$, where $M = 1$ is fixed, $\alpha \in \{0.0, 0.1, 0.2, 0.5, 0.8\}$ varies the Letelier modification, and $\zeta \in \{0.0, 4.0, 5.0\}$ is the quantum correction parameter. Each diagram features a turquoise surface representing the embedding from the event horizon $r_h$ to $r = 15$, a black falling trajectory, and a red ring at the event horizon. The transition from an extremal or single-root BH at $\zeta = 0$ to non-extremal BHs with two horizons for $\zeta > 0$ reflects the modification of the spacetime's structure. $M=1$}
\label{izzetfig-2}
\end{figure}

Figure~\ref{izzetfig-2} provides a visual representation of the spacetime geometry through embedding diagrams for six representative parameter configurations spanning the range from classical Schwarzschild to strongly quantum-corrected regimes with significant CoS effects. These 3D diagrams illustrate the spatial geometry of constant-time hypersurfaces extending from the event horizon $r_h$ to $r = 15M$. Each diagram features a turquoise surface representing the embedded geometry, a black trajectory showing representative geodesic motion, and a red ring marking the event horizon location. Panel (i) shows the classical Schwarzschild BH ($\zeta = 0$, $\alpha = 0$, $r_h = 2.0M$), which serves as the reference configuration displaying the familiar throat structure and asymptotic flatness. Panels (ii) and (iii) demonstrate the effects of introducing quantum corrections while maintaining $\alpha = 0.1$ fixed. As $\zeta$ increases from $4.0M$ to $5.0M$, the throat region becomes slightly modified, though the outer horizon location remains at $r_h = 2.22M$ as predicted by Eq.~(\ref{horizon-radius}). The subtle changes in the embedding surface reflect the quantum modifications to the metric in the interior region.

Panels (iv)–(vi) illustrate the dramatic effect of increasing the CoS parameter $\alpha$ while keeping $\zeta = 5.0M$ fixed. As $\alpha$ increases from $0.2$ to $0.8$, the event horizon moves progressively outward from $r_h = 2.5M$ to $r_h = 10.0M$, and the throat becomes significantly wider and more pronounced. This substantial geometric modification reflects the enhanced effective gravitational mass due to the string cloud, demonstrating that CoS effects dominate over quantum corrections in determining the large-scale spacetime structure and the global geometry. The transition from an extremal or single-root BH at $\zeta = 0$ to non-extremal BHs with two horizons for $\zeta > 0$ clearly reflects the modification of the spacetime's causal structure induced by quantum gravitational effects.

These quantum-corrected Letelier BH solutions provide a rich theoretical laboratory for exploring the interplay between quantum gravity and topological defects, with implications for gravitational wave astronomy, shadow observations by the EHT, and tests of fundamental physics in strong-field regimes. 

\setlength{\tabcolsep}{12pt}
\begin{longtable}{|c|c|c|c|}
\hline
\rowcolor{orange!50}
\textbf{$\alpha$} & \textbf{$\zeta$} & \textbf{Horizon(s) } & \textbf{Configuration} \\
\hline
\endfirsthead
\hline
\rowcolor{orange!50}
\textbf{$\alpha$} & \textbf{$\zeta$} & \textbf{Horizon(s) [$r_h$]} & \textbf{Configuration} \\
\hline
\endhead
0.0 & 0.0 & [2.0] & Schwarzshild BH \\
\hline
0.0 & 0.1 & [0.25917041, 2.0] & Non-extremal BH \\
\hline
0.0 & 0.5 & [0.68939835, 2.0] & Non-extremal BH \\
\hline
0.0 & 1.0 & [1.0, 2.0] & Non-extremal BH \\
\hline
0.0 & 1.5 & [1.2108937, 2.0] & Non-extremal BH \\
\hline
0.0 & 2.0 & [1.3646556, 2.0] & Non-extremal BH \\
\hline
0.0 & 2.5 & [1.4806400, 2.0] & Non-extremal BH \\
\hline
0.0 & 3.0 & [1.5700065, 2.0] & Non-extremal BH \\
\hline
0.0 & 4.0 & [1.6954152, 2.0] & Non-extremal BH \\
\hline
0.0 & 5.0 & [1.7759473, 2.0] & Non-extremal BH \\
\hline
0.1 & 0.0 & [2.2222222] & Extremal or Single Root BH \\
\hline
0.1 & 0.1 & [0.26039602, 2.2222222] & Non-extremal BH \\
\hline
0.1 & 0.5 & [0.69970491, 2.2222222] & Non-extremal BH \\
\hline
0.1 & 1.0 & [1.0251504, 2.2222222] & Non-extremal BH \\
\hline
0.1 & 1.5 & [1.2523247, 2.2222222] & Non-extremal BH \\
\hline
0.1 & 2.0 & [1.4225582, 2.2222222] & Non-extremal BH \\
\hline
0.1 & 2.5 & [1.5544638, 2.2222222] & Non-extremal BH \\
\hline
0.1 & 3.0 & [1.6587595, 2.2222222] & Non-extremal BH \\
\hline
0.1 & 4.0 & [1.8102583, 2.2222222] & Non-extremal BH \\
\hline
0.1 & 5.0 & [1.9117077, 2.2222222] & Non-extremal BH \\
\hline
0.2 & 0.0 & [2.5000000] & Extremal or Single Root BH \\
\hline
0.2 & 0.1 & [0.26162212, 2.5000000] & Non-extremal BH \\
\hline
0.2 & 0.5 & [0.71005197, 2.5000000] & Non-extremal BH \\
\hline
0.2 & 1.0 & [1.0505780, 2.5000000] & Non-extremal BH \\
\hline
0.2 & 1.5 & [1.2945977, 2.5000000] & Non-extremal BH \\
\hline
0.2 & 2.0 & [1.4822705, 2.5000000] & Non-extremal BH \\
\hline
0.2 & 2.5 & [1.6314835, 2.5000000] & Non-extremal BH \\
\hline
0.2 & 3.0 & [1.7524771, 2.5000000] & Non-extremal BH \\
\hline
0.2 & 4.0 & [1.9344559, 2.5000000] & Non-extremal BH \\
\hline
0.2 & 5.0 & [2.0617771, 2.5000000] & Non-extremal BH \\
\hline
\caption{\footnotesize Horizons of the quantum-corrected BH (Model-I and Model-II) with metric function $g(r) = (1 - \alpha) - \frac{2M}{r} + \frac{\zeta^2}{r^2} \left( (1 - \alpha) - \frac{2M}{r} \right)^2$, where $M = 1$ is fixed, $\alpha \in \{0.0, 0.1, 0.2\}$ varies the Letelier modification, and $\zeta \in \{0.0, 0.1, 0.5, 1.0, 1.5, 2.0, 2.5, 3.0, 4.0, 5.0\}$ is the quantum correction parameter. The spacetime transitions from an extremal or single-root BH at $\zeta = 0.0$ (Letelier BH, reducing to Schwarzschild at $\alpha = 0$) to non-extremal BHs with two horizons for $\zeta > 0$, with the outer horizon increasing with $\alpha$ (e.g., $2.0$, $2.2222222$, $2.5000000$) and the inner horizon growing with $\zeta$.}
\label{horizons_newmodel}
\end{longtable}

%%%%%%%%%PROF FA AA SECTION  REVISED BY IS
\section{Geodesic Structure of Quantum-Corrected BHs with CoS} 
\label{isec3}

Geodesic motion describes the trajectories of free-falling particles and light in curved spacetime around BHs, governed by the geodesic equation derived from the spacetime metric. Timelike geodesics correspond to massive particles, such as stars or gas, while null geodesics describe photon paths. Around BHs, such as Schwarzschild or Kerr solutions, geodesics exhibit unique properties including the existence of an ISCO for massive particles and a photon sphere where light can orbit in unstable circular paths. The effective potential method is commonly employed to analyze these orbits, revealing key features like orbital stability and perihelion precession caused by the strong gravitational field. Understanding geodesic motion is fundamental for interpreting phenomena such as accretion disk behavior, gravitational lensing, and the formation of BH shadows. These aspects have been extensively studied in classical works~\cite{48}. In the context of quantum-corrected BHs coupled with CoS, the geodesic structure encodes information about both quantum gravitational effects (parametrized by $\zeta$) and topological defects (parametrized by $\alpha$), offering potential observational signatures that could be detected by current and future instruments such as the EHT~\cite{49}.

\subsection{Null Geodesics and Photon Sphere Properties}

Null geodesics describe the paths of massless particles such as photons and satisfy the condition $ds^2 = 0$. The geodesic equations can be derived via the Euler-Lagrange method or Hamilton-Jacobi formalism. These yield conserved quantities like energy $\mathrm{E}$ and angular momentum $\mathrm{L}$, reducing the problem to an effective potential analysis~\cite{48}. The effective potential determines photon trajectories and reveals the photon sphere, which consists of unstable circular orbits critical for phenomena such as BH shadows~\cite{50}. Studying null geodesics is essential for understanding gravitational lensing, light deflection, and the observational signatures of BHs, providing stringent tests for GR and its alternatives~\cite{51,49}. In quantum-corrected spacetimes, deviations from the classical photon sphere radius provide direct evidence of quantum gravitational modifications to the light propagation in strong-field regimes.

We start with the Lagrangian density function in curved space given by 
\begin{equation}
    \mathcal{L}=\frac{1}{2}\,g_{\mu\nu}\,\left(\frac{dx^{\mu}}{d\lambda}\right)\,\left(\frac{dx^{\nu}}{d\lambda}\right),\label{bb1}
\end{equation}
where $\lambda$ is an affine parameter and $g_{\mu\nu}$ is the metric tensor. Using the metric tensor given in Eq.~(\ref{metric-0}), the Lagrangian density function becomes
\begin{equation}
    \mathcal{L}=\frac{1}{2}\,\left[-f(r)\,\left(\frac{dt}{d\lambda}\right)^2+\frac{1}{g(r)}\,\left(\frac{dr}{d\lambda}\right)^2+r^2\,\left(\frac{d\theta}{d\lambda}\right)^2+r^2\,\sin^2 \theta\,\left(\frac{d\phi}{d\lambda}\right)^2\right].\label{bb2}
\end{equation}

As the spacetime is static and spherically symmetric, there are two conserved quantities associated with temporal coordinate $t$ and azimuthal coordinate $\phi$. The corresponding Killing vectors are $\xi_{(t)} \equiv \partial_t$ and $\xi_{(\phi)} \equiv \partial_{\phi}$, respectively. The corresponding conserved quantities are the conserved energy $\mathrm{E}=-g_{\mu\nu}\,\xi^{\mu}_{(t)}\,u^{\nu}$ and the conserved angular momentum $\mathrm{L}=g_{\mu\nu}\,\xi^{\mu}_{(\phi)}\,u^{\nu}$, where $u^{\mu}$ is the timelike four-velocity. These are given by
\begin{equation}
    \mathrm{E}=f(r)\,\dot{t}\qquad,\qquad \mathrm{L}=r^2\,\sin^2 \theta\,\dot{\phi}.\label{null1}
\end{equation}

For null geodesics, $ds^2=0$, Eq.~(\ref{bb2}) simplifies to
\begin{equation}
    -\frac{\mathrm{E}^2}{f(r)}+\frac{1}{g(r)}\, \left(\frac{dr}{d\lambda}\right)^2+\frac{\mathrm{L}^2}{r^2\,\sin^2 \theta}+r^2\,\left(\frac{d\theta}{d\lambda}\right)^2=0.\label{null2}
\end{equation}

Employing the Carter constant $\mathcal{K}$~\cite{52} to separate variables in Eq.~(\ref{null2}), two additional equations of motion for photons can be easily computed as:
\begin{equation}
    r^2\,\left(\frac{dr}{d\lambda}\right)=\sqrt{\mathcal{R}(r)}\qquad, \qquad r^2\,\left(\frac{d\theta}{d\lambda}\right)=\sqrt{\Theta(\theta)}\label{null2a}
\end{equation}
with the functions $\mathcal{R}(r)$ and $\Theta(\theta)$ defined as:
\begin{equation}
    \mathcal{R}(r)=\frac{r^4\,\mathrm{E}^2\,g(r)}{f(r)}-(\mathcal{K}+\mathrm{L}^2)\,r^2\,g(r)\qquad, \qquad \Theta(\theta)=\mathcal{K}-\mathrm{L}^2\,\cot^2 \theta.\label{null2b}
\end{equation}
These equations~(\ref{null1}) and~(\ref{null2b}) describe the propagation of light around a quantum-corrected Letelier BH, incorporating both quantum gravitational effects through $\zeta$ and CoS effects through $\alpha$.

The equation of motion for the radial component $r$ from Eq.~(\ref{null2}) can be written as
\begin{equation}
    \left(\sqrt{\frac{f(r)}{g(r)}}\,\frac{1}{\mathrm{E}}\,\frac{dr}{d\lambda}\right)^2+V_\text{eff}(r)=0\label{null3}
\end{equation}
with the effective potential of the system given by
\begin{align}
    V_\text{eff}(r)=-\frac{f(r)}{r^4\,\mathrm{E}^2\,g(r)}\,\mathcal{R}(r)=-1+\frac{(\eta+\beta^2)}{r^2}\,f(r),\label{null3b}
\end{align}
where we have used the relation~(\ref{null2b}) and defined the following impact parameters~\cite{48,53,54}
\begin{equation}
    \beta=\mathrm{L}/\mathrm{E}\qquad, \qquad \eta=\mathcal{K}/\mathrm{E}^2.\label{null3a}
\end{equation}
Equation~(\ref{null3}) represents the one-dimensional equation of motion of a photon in curved spacetime having effective potential $V_\text{eff}(r)$ given in Eq.~(\ref{null3b}).

For circular null orbits of radius $r=r_{\rm ph}$ (photon sphere), the following conditions must be satisfied~\cite{55,56}:
\begin{equation}
    V_\text{eff}(r)=0\qquad, \qquad \partial_r V_\text{eff}(r)=0.\label{condition}
\end{equation}

We observe that the equations governing the shadow of the BH depend only on the metric function $f(r)$ since the effective potential depends only on $f(r)$ and is independent of the function $g(r)$. This means that the quantum corrections introduced by Model-II are not reflected in the BH shadow, making it identical to the Letelier case, where the photon sphere radius $r_{\rm ph}$ and the shadow radius $R_s$ are obtained as:
\begin{equation}
    r_{\rm ph} = 3M (1-\alpha)^{-1}\qquad,\qquad R_s = 3\sqrt{3}M (1-\alpha)^{-3/2}.\label{photon-shadow}
\end{equation}

In the absence of the CoS ($\alpha \to 0$), these results reduce to those findings reported in~\cite{57}. Therefore, we now focus on the BH spacetime corresponding to Model-I of static BHs with quantum corrections. The above conditions~(\ref{condition}) imply the following relations:
\begin{align}
    &\eta+\beta^2-\frac{r^2_{\rm ph}}{f(r_{\rm ph})}=0,\label{null5a}\\
    &\frac{d}{dr}\,\left(\frac{f(r)}{r^2}\right)\Big{|}_{r=r_{\rm ph}}=0.\label{null5b}
\end{align}

Solving the above relation~(\ref{null5b}) will give us the photon sphere radius. Substituting the metric function $f(r)$ from Eq.~(\ref{metric-1}), we find the following quartic relation in $r$:
\begin{align}
\lambda\, r^4 - 3\,M\, r^3 + \zeta^2\, \lambda^2\, r^2 + 2\, \zeta^2\, \lambda\, M\, r -4\, \zeta^2\, M^2 = 0,\label{null6}
\end{align}
where we have defined $\lambda = 1 - \alpha$ for notational convenience.

An exact real-valued analytical solution of the above quartic equation in $r$ would give us the photon sphere radius at $r=r_{\rm ph}$. However, this exact analytical expression is quite challenging to obtain in closed form. Nevertheless, one can find numerical values of the photon sphere by selecting suitable values of $\alpha$, $\zeta$, and $M$. In Table~\ref{tab:photon-shadow}, we present numerical values of both the photon sphere radius $r_{\rm ph}$ and the shadow radius $R_s$ for various combinations of the CoS parameter $\alpha$ and the quantum correction parameter $\zeta$. The table reveals several important trends. First, for fixed $\alpha$, increasing $\zeta$ decreases both $r_{\rm ph}$ and $R_s$, indicating that quantum corrections effectively reduce the photon capture region. Second, for fixed $\zeta$, increasing $\alpha$ increases both quantities, reflecting the enhanced gravitational field strength due to the CoS. For instance, at $\alpha = 0.30$ and $\zeta = 0.5$, the photon sphere radius is $r_{\rm ph} \approx 4.24M$, which is approximately 41\% larger than the Schwarzschild value of $3M$, demonstrating the significant combined impact of quantum corrections and CoS effects on photon orbits.

\begin{table}[ht!]
\centering
\renewcommand{\arraystretch}{1.2}
\setlength{\tabcolsep}{8pt}
\begin{tabular}{|c||c|c||c|c||c|c||c|c||c|c|}
\hline
\multirow{2}{*}{$\alpha$} & \multicolumn{2}{c||}{$\zeta = 0.1$} & \multicolumn{2}{c||}{$\zeta = 0.2$} & \multicolumn{2}{c||}{$\zeta = 0.3$} & \multicolumn{2}{c||}{$\zeta = 0.4$} & \multicolumn{2}{c|}{$\zeta = 0.5$} \\
\cline{2-11}
& $r_{\rm ph}$ & $R_s$ & $r_{\rm ph}$ & $R_s$ & $r_{\rm ph}$ & $R_s$ & $r_{\rm ph}$ & $R_s$ & $r_{\rm ph}$ & $R_s$ \\
\hline\hline
0.05 & 3.154 & 5.611 & 3.143 & 5.608 & 3.125 & 5.605 & 3.098 & 5.601 & 3.064 & 5.598 \\
\hline
0.10 & 3.330 & 6.085 & 3.320 & 6.083 & 3.303 & 6.079 & 3.280 & 6.075 & 3.249 & 6.072 \\
\hline
0.15 & 3.526 & 6.630 & 3.518 & 6.628 & 3.503 & 6.624 & 3.482 & 6.620 & 3.455 & 6.617 \\
\hline
0.20 & 3.747 & 7.261 & 3.740 & 7.259 & 3.726 & 7.256 & 3.708 & 7.252 & 3.684 & 7.248 \\
\hline
0.25 & 3.998 & 7.999 & 3.991 & 7.998 & 3.979 & 7.995 & 3.963 & 7.991 & 3.942 & 7.987 \\
\hline
0.30 & 4.284 & 8.872 & 4.278 & 8.870 & 4.268 & 8.867 & 4.254 & 8.864 & 4.235 & 8.860 \\
\hline
\end{tabular}
\caption{\footnotesize Photon sphere radius $r_{\rm ph}/M$ and shadow radius $R_s/M$ for Model-I quantum-corrected BHs with CoS, for various values of the CoS parameter $\alpha$ and quantum correction parameter $\zeta$. The values demonstrate that increasing $\alpha$ enlarges both quantities due to enhanced gravitational effects, while increasing $\zeta$ reduces them through quantum corrections. The mass is set to $M = 1$.}
\label{tab:photon-shadow}
\end{table}

\begin{figure}[ht!]
    \centering
    \includegraphics[width=0.45\linewidth]{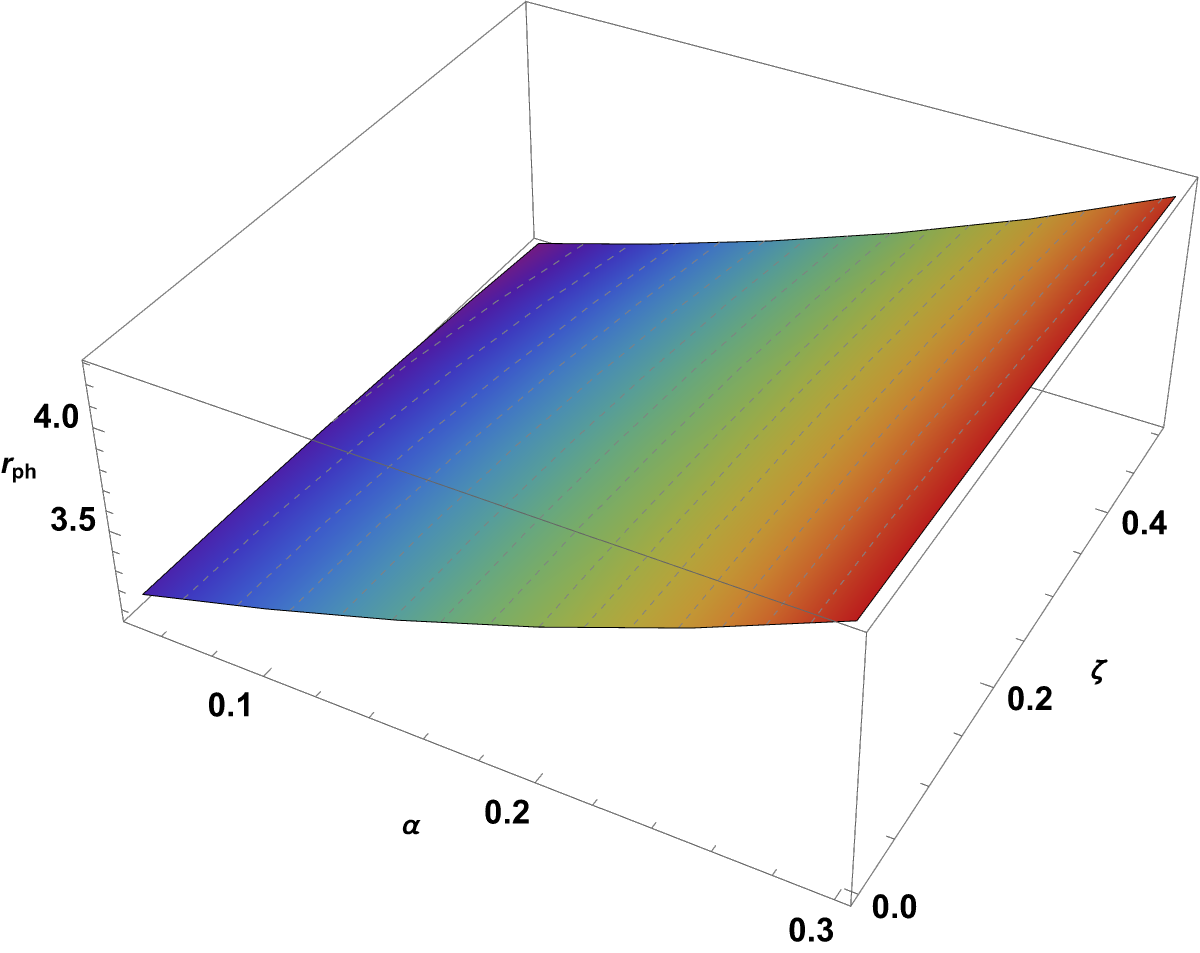}\qquad
    \includegraphics[width=0.45\linewidth]{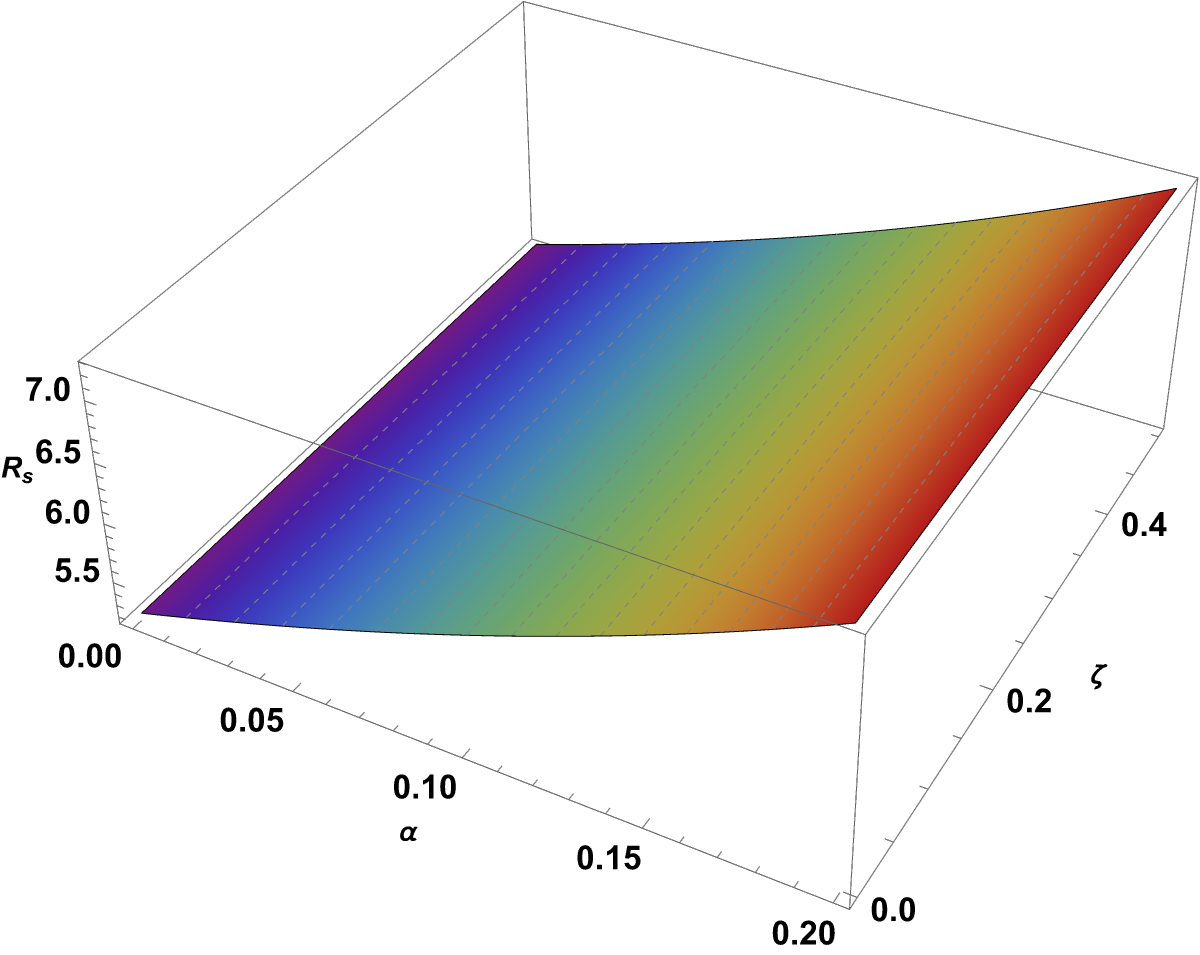}
    \caption{\footnotesize Three-dimensional plots of the photon sphere radius $r_{\rm ph}$ (left) and shadow radius $R_s$ (right) as functions of $(\alpha, \zeta)$ for Model-I quantum-corrected BHs with CoS. Both quantities increase with the CoS parameter $\alpha$ and decrease with the quantum correction parameter $\zeta$, demonstrating the competing effects of topological defects and quantum gravity on photon orbits. Mass is set to $M = 1$.}
    \label{fig:photon-shadow-3D}
\end{figure}

Figure~\ref{fig:photon-shadow-3D} provides a three-dimensional visualization of how the photon sphere radius $r_{\rm ph}$ and shadow radius $R_s$ vary as functions of both $\alpha$ and $\zeta$. The left panel shows $r_{\rm ph}(\alpha, \zeta)$, revealing a smooth surface that increases monotonically with $\alpha$ (moving along the horizontal axis) and decreases gradually with $\zeta$ (moving along the depth axis). The curvature of this surface indicates that the CoS parameter $\alpha$ has a dominant effect compared to the quantum correction parameter $\zeta$ in determining the photon sphere location. The right panel displays $R_s(\alpha, \zeta)$, exhibiting similar qualitative behavior but with larger absolute values, consistent with the relation $R_s > r_{\rm ph}$. These plots provide a parameter-space overview of how quantum corrections and CoS effects jointly influence the photon capture region, which is directly relevant for interpreting BH shadow observations by the EHT.

\begin{figure}[ht!]
    \centering
    \includegraphics[width=0.32\linewidth]{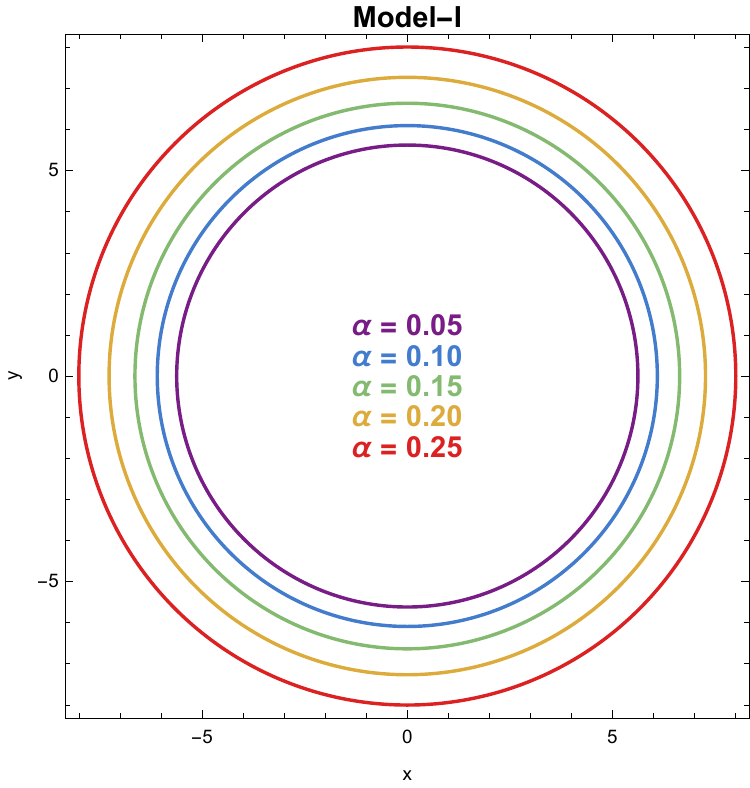}
    \includegraphics[width=0.32\linewidth]{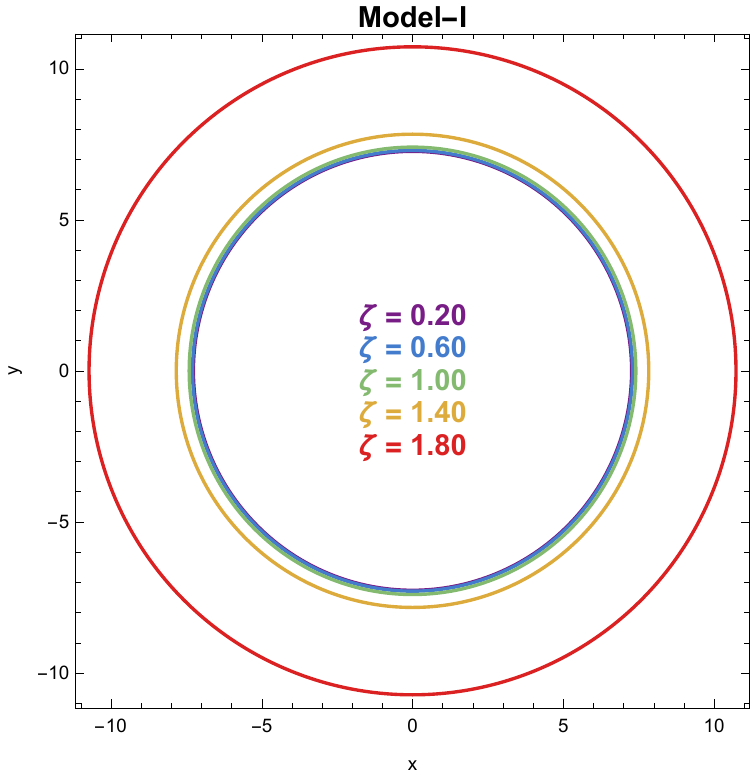}
    \includegraphics[width=0.32\linewidth]{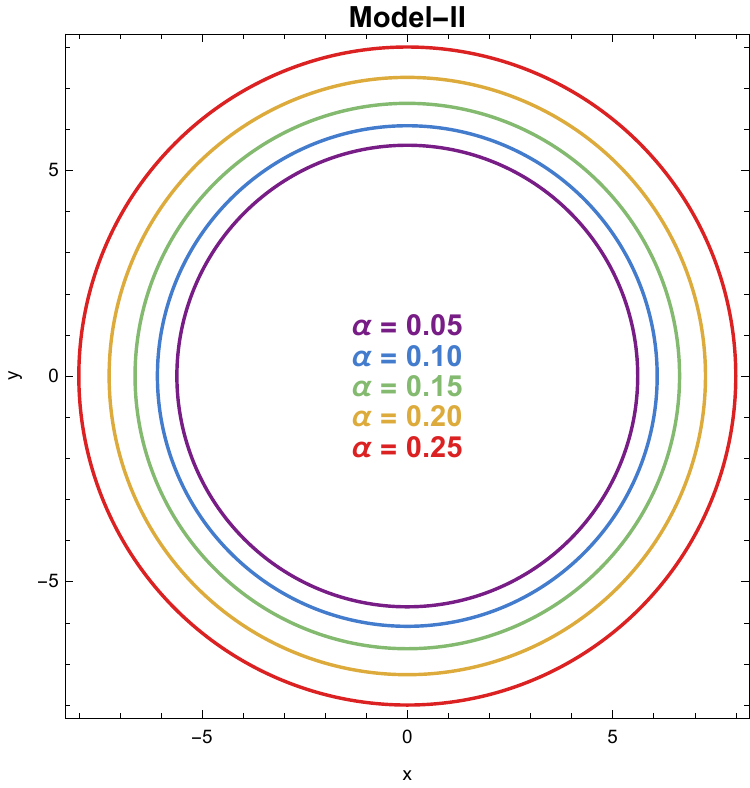}\\
    (i) Model-I: $\zeta = 0.3$ \hspace{1.5cm} (ii) Model-I: $\alpha = 0.2$ \hspace{1.5cm} (iii) Model-II: $\zeta = 0.3$
    \caption{\footnotesize BH shadow profiles for quantum-corrected BHs with CoS. Left (i): Model-I shadows for varying $\alpha$ with fixed $\zeta = 0.3$. Middle (ii): Model-I shadows for varying $\zeta$ with fixed $\alpha = 0.2$. Right (iii): Model-II shadows for varying $\alpha$ with fixed $\zeta = 0.3$. The shadow size increases with $\alpha$ in both models due to enhanced CoS effects, while in Model-I it decreases with $\zeta$ due to quantum corrections. Model-II shadows are independent of $\zeta$ as quantum corrections appear only in $g(r)$. Mass is set to $M = 1$.}
    \label{fig:BH-shadow}
\end{figure}

Figure~\ref{fig:BH-shadow} illustrates the BH shadow profiles for both Model-I and Model-II across different parameter values. For Model-I (panels i and ii), we show shadow contours in the observer's sky coordinates $(x, y)$ for varying CoS parameter $\alpha$ (panel i, with fixed $\zeta = 0.3$) and varying quantum correction parameter $\zeta$ (panel ii, with fixed $\alpha = 0.2$). As $\alpha$ increases from 0.05 to 0.25, the shadow radius grows significantly, reflecting the enhanced gravitational lensing due to the CoS. Conversely, as $\zeta$ increases from 0.20 to 1.80, the shadow size decreases slightly, indicating that quantum corrections weaken the photon capture region. Panel iii shows the shadow contours for Model-II, which, as predicted by our analytical results, are identical to the Letelier case since Model-II's quantum corrections appear only in $g(r)$ and not in $f(r)$. These shadow profiles provide testable predictions that could be compared with high-resolution observations from the EHT and next-generation VLBI arrays to constrain the parameters $\alpha$ and $\zeta$ and distinguish between different quantum gravity models.

\subsection{Topological Classification of Photon Spheres}

Beyond the geodesic analysis, a topological perspective is introduced to characterize unstable photon spheres. In this framework, photon spheres are assigned topological charges reflecting their dynamical stability. A positive topological charge typically corresponds to a stable photon sphere, whereas a negative charge is associated with an unstable configuration. This characterization provides insight into the underlying spacetime geometry and the nature of light propagation around compact objects.

To investigate the topological features of the photon sphere, we introduce a regular potential function $H(r, \theta)$, following the approach proposed in Refs.~\cite{Cunha2017, Cunha2020, Wei2020, Afshar2024}. This potential function serves as the foundation for constructing the corresponding vector field, whose zeros determine the locations and topological nature of the photon spheres. This vector potential is defined as:
\begin{equation}
H(r, \theta) = \sqrt{\frac{-g_{tt}}{g_{\phi\phi}}} = \frac{\sqrt{f(r)}}{r \sin \theta} = \frac{1}{r \sin \theta} \sqrt{\lambda - \frac{2M}{r} + \frac{\zeta^2}{r^2} \left(\lambda - \frac{2M}{r}\right)^2},\label{potential-H}
\end{equation}
where $g_{tt}$ and $g_{\phi\phi}$ are metric tensor components.

\begin{figure}[ht!]
    \centering
    \includegraphics[width=0.55\linewidth]{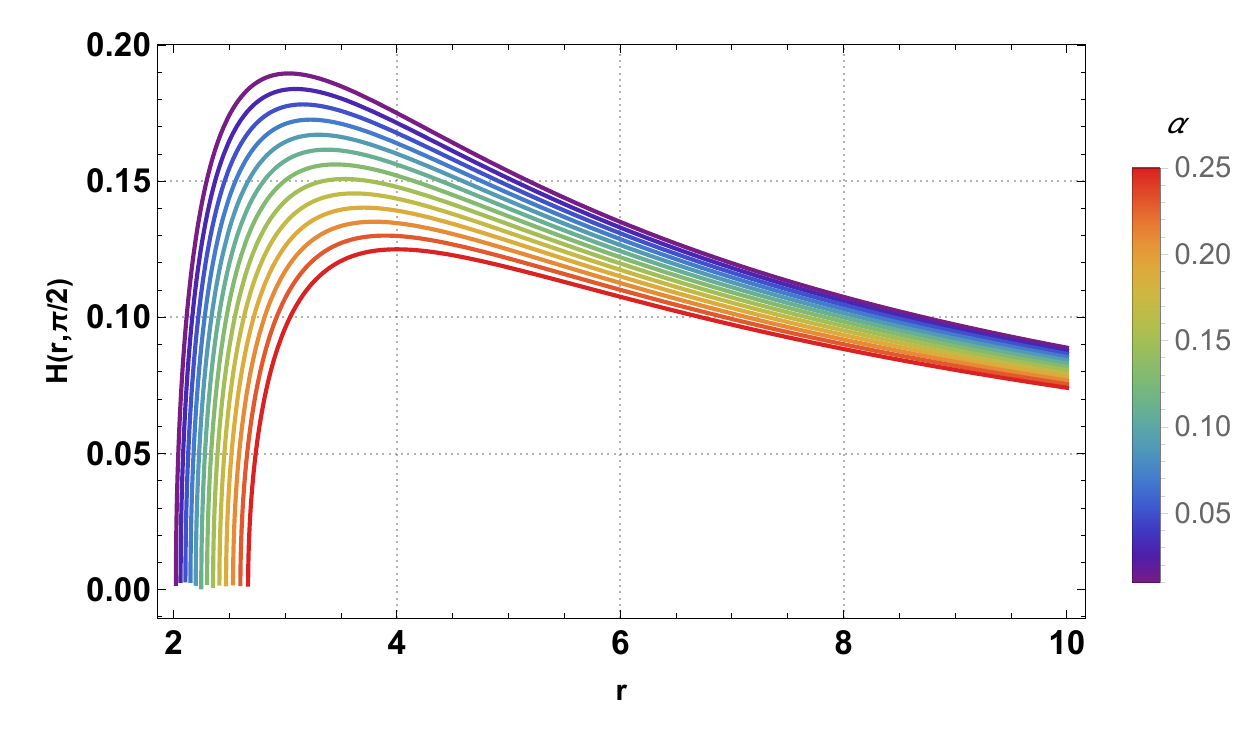}
    \caption{\footnotesize Potential function $H(r, \pi/2)$ for Model-I quantum-corrected BHs with CoS, for different values of the CoS parameter $\alpha$, while keeping fixed $\zeta = 0.1$ and $M = 1$. The peak of the potential shifts to larger radii as $\alpha$ increases, corresponding to the outward movement of the photon sphere due to enhanced gravitational effects from the CoS.}
    \label{fig:potential-H}
\end{figure}

Figure~\ref{fig:potential-H} shows the behavior of the potential function $H(r, \pi/2)$ as a function of the radial coordinate $r$ for different values of the CoS parameter $\alpha$, while keeping the quantum correction parameter fixed at $\zeta = 0.1$ and mass $M = 1$. The potential exhibits a characteristic peak structure, with the peak location shifting to larger radii as $\alpha$ increases. This shift corresponds directly to the outward movement of the photon sphere with increasing CoS density. The peak becomes more pronounced for larger $\alpha$, indicating a deeper potential well that more strongly captures photons in unstable circular orbits. Near the event horizon (left side of each curve), the potential drops sharply to zero, reflecting the fact that photons crossing the horizon cannot escape. These features of $H(r, \theta)$ encode the essential information about photon capture and are intimately connected to the topological properties we analyze next.

Finding the photon sphere radius is equivalent to identifying critical points of the potential function $H(r, \theta)$ in the radial direction:
\begin{equation}
\frac{dH}{dr} = 0.\label{critical-condition}
\end{equation}

To establish the topological structure, one constructs a two-dimensional vector field $\boldsymbol{\varphi}_H$ defined over the space spanned by the horizon radius and the polar angle, as discussed in~\cite{Wei2020}. The vector components are given by:
\begin{equation}
\varphi^r_H = \sqrt{g(r)} \frac{dH}{dr}\qquad, \qquad \varphi^{\theta}_H = \frac{1}{r} \frac{dH}{d\theta}.\label{vector-field}
\end{equation}

This vector can be represented in complex or polar form as:
\begin{equation}
\boldsymbol{\varphi}_H = ||\boldsymbol{\varphi}|| e^{i \Theta_H} = \varphi^r_H + i \varphi^{\theta}_H,\label{vector-complex}
\end{equation}
with magnitude:
\begin{equation}
||\boldsymbol{\varphi}|| = \sqrt{(\varphi^r_H)^2 + (\varphi^{\theta}_H)^2}.\label{vector-magnitude}
\end{equation}

Normalizing $\boldsymbol{\varphi}$ produces a unit vector:
\begin{equation}
\mathbf{n}^a_H = \frac{\varphi^a_H}{||\boldsymbol{\varphi}||},\label{unit-vector}
\end{equation}
where the indices $a = 1, 2$ correspond to the coordinates $r_{\rm ph}$ and $\theta$, respectively, with $\varphi^1_H = \varphi^r_H$ and $\varphi^2_H = \varphi^{\theta}_H$. To deepen the topological classification of photon spheres, scalar fields $\varphi^r_H$ and $\varphi^{\theta}_H$ are introduced. These fields encapsulate the stability and nature of photon spheres from a topological viewpoint. Their explicit forms are:
\begin{equation}
\varphi^r_H = -\frac{1}{r^2 \sin \theta} \left[\lambda\, r^4 - 3\,M\, r^3 + \zeta^2\, \lambda^2\, r^2 + 2\, \zeta^2\, \lambda\, M\, r -4\, \zeta^2\, M^2\right],\label{phi-r}
\end{equation}
and
\begin{equation}
\varphi^{\theta}_H =
-\frac{\sqrt{\lambda-\frac{2\,M}{r}+\frac{\zeta^2}{r^2}\,\left(\lambda-\frac{2\,M}{r}\right)}}{r^2}\,\frac{\cot \theta}{\sin\theta}.\label{phi-theta}
\end{equation}
These quantities facilitate a deeper understanding of the interplay between geometry, dynamics, and topology in the context of photon spheres around BHs.

\begin{figure}[ht!]
    \centering
    \includegraphics[width=0.45\linewidth]{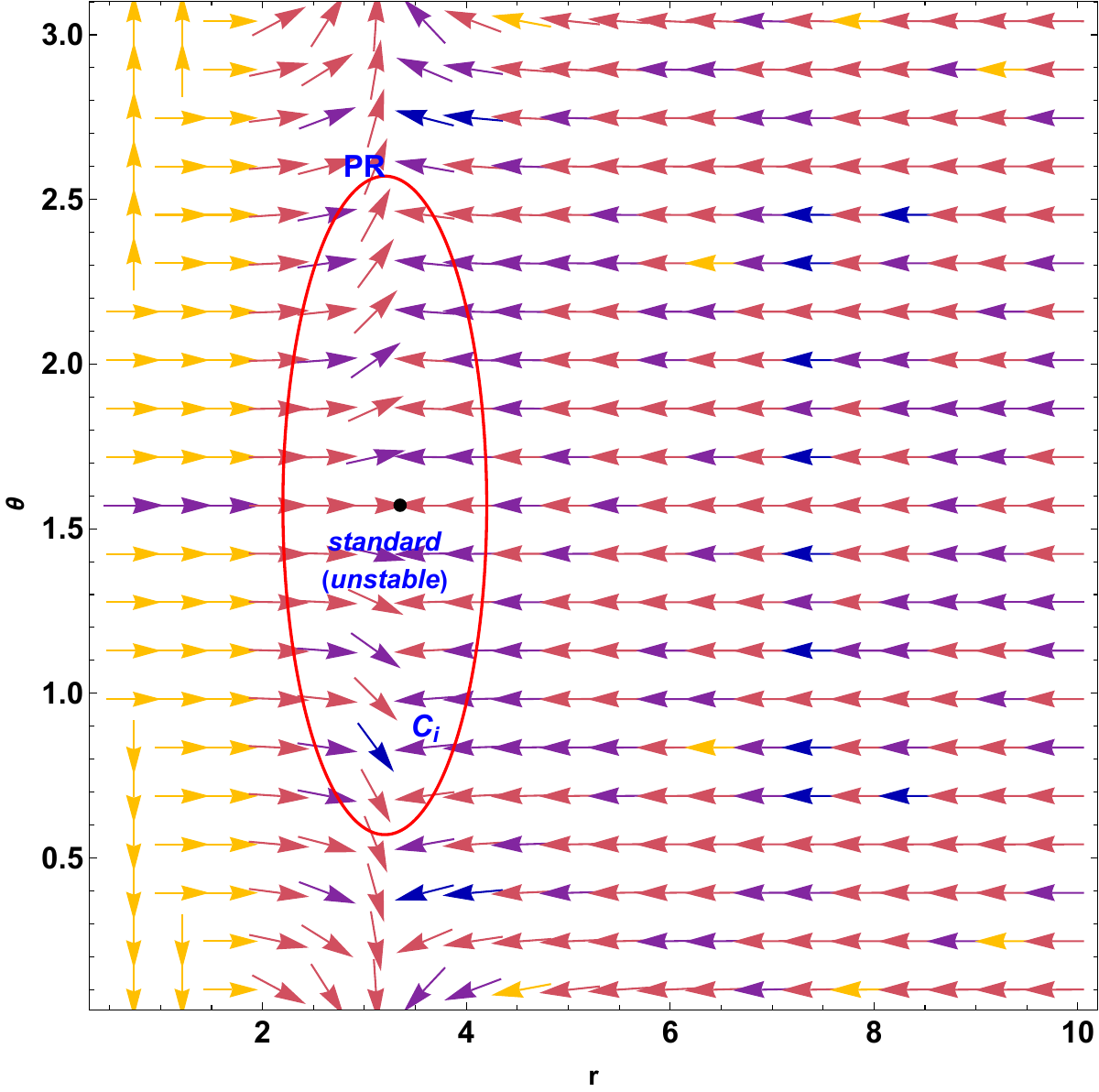}
    \includegraphics[width=0.45\linewidth]{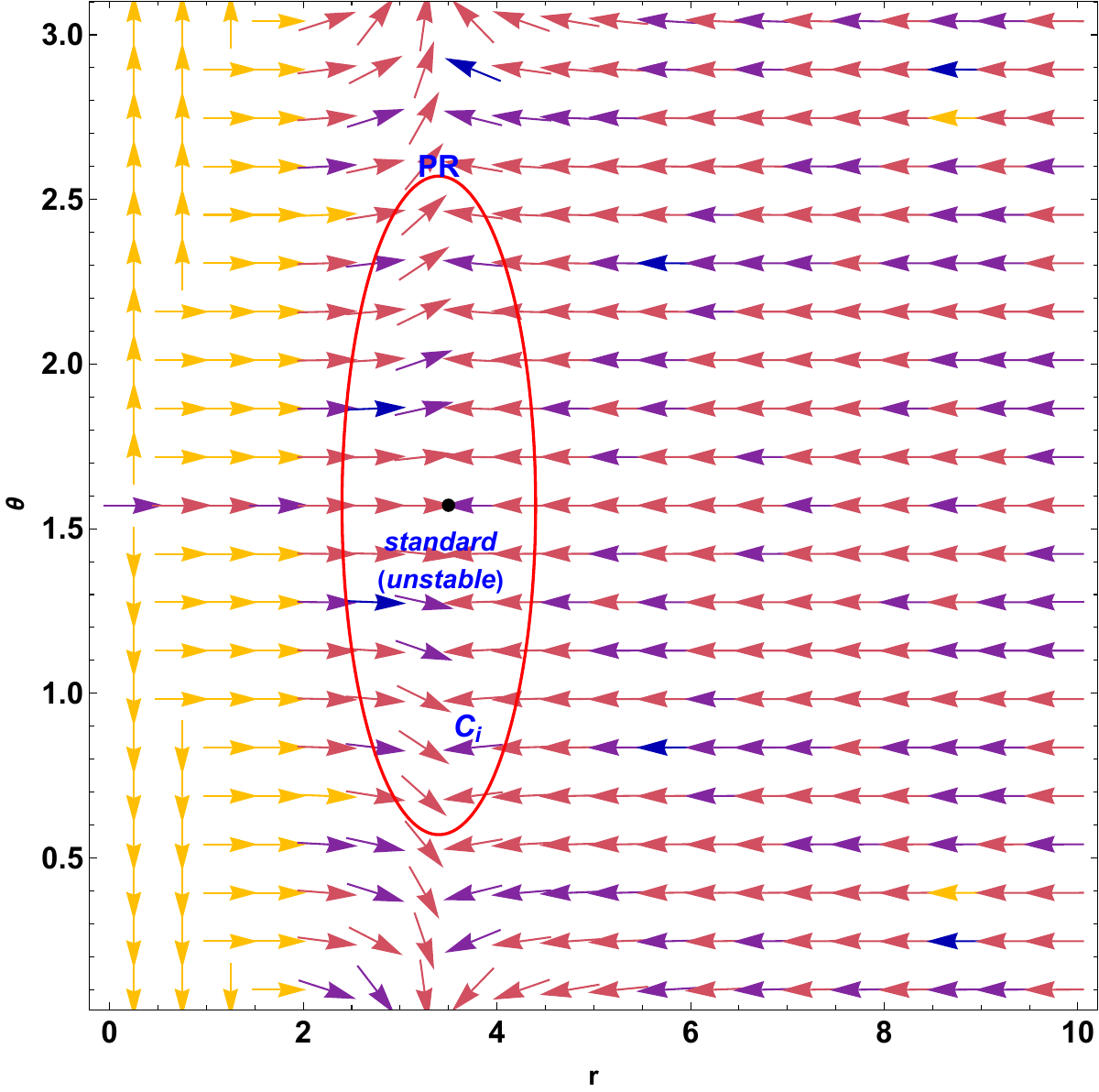}\\
    (i) $\alpha = 0.05$ \hspace{4cm} (ii) $\alpha = 0.10$
    \caption{\footnotesize The arrows represent the unit vector field $\mathbf{n}_H$ on a portion of the $r$--$\Theta$ plane for Model-I of Letelier BHs in quantum gravity with $M = 1$ and $\zeta = 0.1$. The photon ring (PR), marked with a black dot, is at $(r, \theta) = (3.35, \pi/2)$ for $\alpha = 0.05$ (left); and $(r, \theta) = (3.5, \pi/2)$ for $\alpha = 0.10$ (right). The red contour $C_i$ is a closed loop enclosing the photon ring. The topological charge of the photon ring is $Q = -1$, corresponding to a standard unstable light ring.}
    \label{fig:unit-vector}
\end{figure}

For Model-I of static BHs with CoS in quantum gravity, one can plot the unit vector field $\mathbf{n}_H$ on a portion of the $r$--$\theta$ plane, as shown in Figure~\ref{fig:unit-vector}, with parameters $M = 1$ and $\zeta = 0.1$. In this figure, it is evident that there exists a photon ring located at $(r, \theta) = (3.35, \pi/2)$ for $\alpha = 0.05$ (panel i). This corresponds to a radius slightly greater than the photon sphere radius $r_{\rm ph}$ presented in Table~\ref{tab:photon-shadow}, differing by approximately $0.20$, which arises from the discrete nature of the numerical grid used in the vector field construction. Similarly, for $\alpha = 0.10$ (panel ii), the photon ring is found at $(r, \theta) = (3.5, \pi/2)$, again at a location slightly larger than the corresponding photon sphere radius in the same table. This shift in position arises due to the influence of the CoS characterized by the parameter $\alpha$, which modifies the spacetime metric function. 

Furthermore, the winding number $W$ associated with the red contours $C_i$ characterizes the behavior of the vector field and matches that of the four-dimensional Schwarzschild BH, as discussed in~\cite{Cunha2020}. Consequently, the topological charge of the photon ring in Model-I is $Q = -1$. Based on the classification of photon rings, this configuration corresponds to a standard and unstable light ring~\cite{Cunha2020, Wei2020}. The arrows in Figure~\ref{fig:unit-vector} represent the unit vector field $\mathbf{n}_H$, which points radially outward from the photon ring (marked by the black dot), indicating its unstable nature: photons slightly perturbed from this orbit will either spiral into the BH or escape to infinity. It is worth noting that a similar analysis in the limit $\alpha = 0$ for Model-I of static BHs was carried out in~\cite{Wei2024}.

\subsection{Timelike Geodesics and ISCO Properties}

Timelike geodesics describe the motion of massive particles under the influence of gravity alone, without any non-gravitational forces acting on them. These geodesics represent the natural trajectories of particles moving through curved spacetime, especially around compact astrophysical objects such as BHs. Studying timelike geodesics provides crucial insights into phenomena like accretion disk dynamics, stellar orbits near supermassive BHs, and the structure of spacetime near event horizons. These trajectories have been extensively analyzed in various BH spacetimes, including Schwarzschild, Reissner–Nordstr\"{o}m (RN), and Kerr geometries~\cite{48}. Key aspects such as the ISCO, periastron precession, and orbital stability are fundamental to both theoretical studies and astrophysical observations. The ISCO radius marks the innermost radius at which stable circular orbits can exist; inside this radius, particles inevitably spiral into the BH. Understanding the ISCO is critical for modeling accretion disk spectra, X-ray binary systems, and QPO phenomena observed in BH candidates.

For timelike geodesics, using the normalization condition $g_{\mu\nu}\,\dot{x}^{\mu}\,\dot{x}^{\nu}=-1$, we find the equations of motion:
\begin{align}
    \left(\sqrt{\frac{f(r)}{g(r)}}\,\frac{1}{\mathrm{E}}\,\frac{dr}{d\lambda}\right)^2+U_\text{eff}(r)=0,\label{timelike-3}
\end{align}
where $U_\text{eff}(r)$ is the effective potential for massive particles given by:
\begin{equation}
    U_\text{eff}(r)=\frac{f(r)}{\mathrm{E}^2}-1+\frac{f(r)}{r^2}\,(\eta+\beta^2).\label{timelike-4}
\end{equation}

For the ISCO, we have the following conditions:
\begin{align}
    &U_{\rm eff}=0,\label{timelike-5}\\
    &\partial_r U_\text{eff}(r)=0,\label{timelike-6}\\
    &\partial_{rr} U_\text{eff}(r) \geq 0.\label{timelike-7}
\end{align}

The first condition~(\ref{timelike-5}) represents energy conservation for circular orbits, the second condition~(\ref{timelike-6}) ensures that the orbit is at an extremum of the effective potential (either maximum or minimum), and the third condition~(\ref{timelike-7}) guaranties stability by requiring that the orbit sits at a potential minimum.

Simplification of condition~(\ref{timelike-6}) with the result of~(\ref{timelike-5}) yields the specific angular momentum of test particles orbiting in circular geodesics:
\begin{equation}
    \mathcal{L}^2_0=-\mathcal{K}+\frac{r^3\,f'(r)}{2\,f(r)-r\,f'(r)}.\label{timelike-8}
\end{equation}
Substituting $\mathcal{L}^2_0$ in Eq.~(\ref{timelike-5}) and using~(\ref{timelike-4}) gives the specific energy:
\begin{equation}
    \mathcal{E}^2=\frac{2\,f^2(r)}{2\,f(r)-r\,f'(r)}.\label{timelike-9}
\end{equation}

The explicit expressions of these quantities for the two BH models are given by:
\begin{align}
&\mathcal{L}^2_0=
\begin{cases}
\displaystyle
-\mathcal{K}+r^2\,\left[\frac{\frac{M}{r} - \frac{\zeta^2 \lambda^2}{r^2} + \frac{6 \zeta^2 \lambda M}{r^3} - \frac{8 \zeta^2 M^2}{r^4}}{\lambda - \frac{3M}{r} + \frac{\zeta^2 \lambda^2}{r^2} + \frac{2 \zeta^2 \lambda M}{r^3} - \frac{4 \zeta^2 M^2}{r^4}}\right], &\, \text{Model-I} \\[10pt]
\displaystyle
-\mathcal{K}+\frac{M\,r}{\lambda-\frac{3\,M}{r}}, & \,\text{Model-II} 
\end{cases}
\label{timelike-10}\\
&\mathcal{E}^2=
\begin{cases}
\displaystyle
\frac{\left[\left(\lambda - \frac{2\,M}{r}\right)+\frac{\zeta^2}{r^2}\left(\lambda - \frac{2\,M}{r}\right)^2\right]^2}{\lambda - \frac{3M}{r} + \frac{\zeta^2 \lambda^2}{r^2} + \frac{2 \zeta^2 \lambda M}{r^3} - \frac{4 \zeta^2 M^2}{r^4}}, &\, \text{Model-I} \\[10pt]
\displaystyle
\frac{\left(\lambda - \frac{2\,M}{r}\right)^2}{\lambda-\frac{3\,M}{r}}, &\, \text{Model-II}
\end{cases}
\label{timelike-11}
\end{align}

The ISCO radius can be determined using the condition $\partial_{rr} U_\text{eff}(r)=0$, which results in the following relation for the metric function:
\begin{equation}
    3 f(r)\, \frac{f'(r)}{r} + f''(r)\, f(r) - 2\, f'(r)^2 = 0.\label{isco-condition}
\end{equation}

For Model-I, we find the following equation satisfying the ISCO radius $r=r_\text{ISCO}$:
\begin{align}
&3 \left( \lambda - \frac{2M}{r} + \frac{\zeta^2}{r^2} \left( \lambda^2 - \frac{4 \lambda M}{r} + \frac{4M^2}{r^2} \right) \right) \left( \frac{2M}{r^3} - \frac{2 \zeta^2 \lambda^2}{r^4} + \frac{12 \zeta^2 \lambda M}{r^5} - \frac{16 \zeta^2 M^2}{r^6} \right) \nonumber\\
&+ \left( -\frac{4M}{r^3} + \frac{6 \zeta^2 \lambda^2}{r^4} - \frac{48 \zeta^2 \lambda M}{r^5} + \frac{80 \zeta^2 M^2}{r^6} \right)
\left( \lambda - \frac{2M}{r} + \frac{\zeta^2}{r^2} \left( \lambda^2 - \frac{4 \lambda M}{r} + \frac{4M^2}{r^2} \right) \right) \nonumber\\
&- 2 \left( \frac{2M}{r^2} - \frac{2 \zeta^2 \lambda^2}{r^3} + \frac{12 \zeta^2 \lambda M}{r^4} - \frac{16 \zeta^2 M^2}{r^5} \right)^2 = 0.\label{timelike-12}
\end{align}

For BH Model-II, we find that the ISCO radius is given by $r_{\text{ISCO}} = 6M/\lambda$, which matches the Letelier BH case. It is important to note that in the limit $\zeta \to 0$, Eq.~(\ref{timelike-12}) also reduces to the same result, confirming that $r_{\text{ISCO}} = 6M/\lambda$ is recovered for BH Model-II. This shows that quantum corrections in Model-II, which appear only in $g(r)$, do not affect the ISCO radius, unlike in Model-I where quantum corrections modify both $f(r)$ and $g(r)$ and thereby influence the ISCO location.

Table~\ref{tab:isco-radius} presents numerical values of the ISCO radius $r_{\text{ISCO}}/M$ as a function of the CoS parameter $\alpha$ and the quantum correction parameter $\zeta$ for Model-I. Several important trends emerge from these data. First, for fixed $\zeta$, increasing $\alpha$ increases the ISCO radius, reflecting the enhanced gravitational field strength due to the CoS. For example, at $\zeta = 0.5$, the ISCO radius increases from $r_{\text{ISCO}} \approx 6.32M$ at $\alpha = 0.05$ to $r_{\text{ISCO}} \approx 8.57M$ at $\alpha = 0.30$, representing a 36\% increase. Second, for fixed $\alpha$, increasing $\zeta$ produces a very small increase in the ISCO radius. For instance, at $\alpha = 0.20$, the ISCO radius increases from $r_{\text{ISCO}} = 7.50000M$ at $\zeta = 0.1$ to $r_{\text{ISCO}} = 7.50260M$ at $\zeta = 0.9$, an increase of only 0.03\%. This indicates that quantum corrections have a much weaker effect on the ISCO radius compared to CoS effects, consistent with our earlier observations for photon spheres. The ISCO radius is mainly determined by the CoS parameter $\alpha$, with quantum corrections providing only minor perturbative modifications.

\begin{table}[ht!]
\centering
\renewcommand{\arraystretch}{1.2}
\begin{tabular}{|c||*{9}{c|}}
\hline
$\alpha \backslash \zeta$ & 0.1 & 0.2 & 0.3 & 0.4 & 0.5 & 0.6 & 0.7 & 0.8 & 0.9 \\
\hline\hline
0.05 & 6.31579 & 6.31580 & 6.31587 & 6.31603 & 6.31638 & 6.31701 & 6.31803 & 6.31959 & 6.32183 \\
\hline
0.10 & 6.66667 & 6.66668 & 6.66673 & 6.66685 & 6.66712 & 6.66760 & 6.66838 & 6.66958 & 6.67130 \\
\hline
0.15 & 7.05882 & 7.05883 & 7.05887 & 7.05896 & 7.05916 & 7.05952 & 7.06012 & 7.06102 & 7.06232 \\
\hline
0.20 & 7.50000 & 7.50001 & 7.50003 & 7.50010 & 7.50025 & 7.50052 & 7.50096 & 7.50163 & 7.50260 \\
\hline
0.25 & 8.00000 & 8.00000 & 8.00002 & 8.00007 & 8.00018 & 8.00038 & 8.00070 & 8.00118 & 8.00189 \\
\hline
0.30 & 8.57143 & 8.57143 & 8.57145 & 8.57148 & 8.57156 & 8.57170 & 8.57192 & 8.57227 & 8.57277 \\
\hline
\end{tabular}
\caption{\footnotesize ISCO radius $r_{\text{ISCO}}/M$ for Model-I quantum-corrected BHs with CoS, as a function of the CoS parameter $\alpha$ and quantum correction parameter $\zeta$. The ISCO radius increases significantly with $\alpha$ due to enhanced gravitational effects from the CoS, but exhibits only minimal variation with $\zeta$, indicating that quantum corrections have a much weaker influence on the ISCO location compared to CoS effects. Mass is set to $M = 1$.}
\label{tab:isco-radius}
\end{table}

\begin{figure}[ht!]
    \centering
    \includegraphics[width=0.55\linewidth]{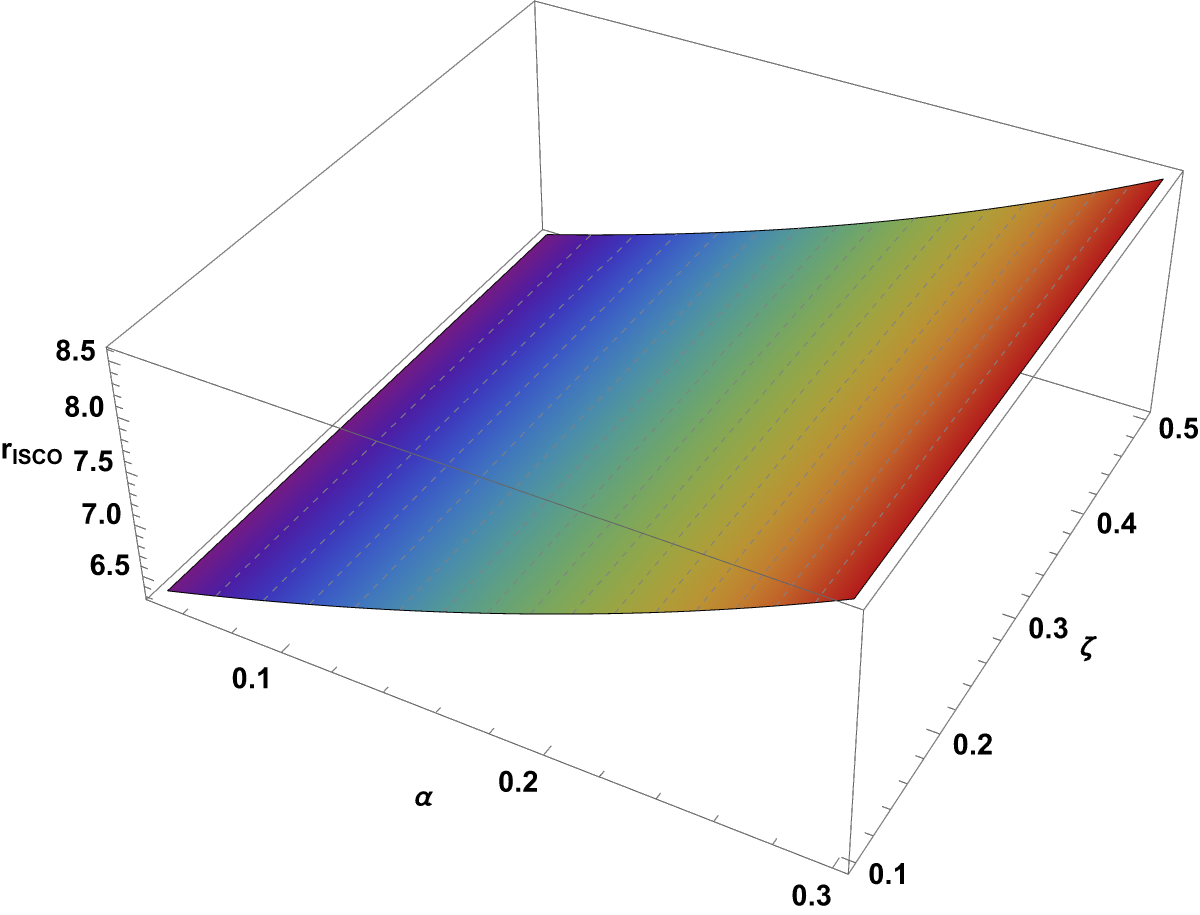}
    \caption{\footnotesize Three-dimensional plot of the ISCO radius $r_{\rm ISCO}$ as a function of $(\alpha, \zeta)$ for Model-I quantum-corrected BHs with CoS. The ISCO radius increases steeply with the CoS parameter $\alpha$ but exhibits minimal variation with the quantum correction parameter $\zeta$, demonstrating that CoS effects dominate over quantum corrections in determining the location of the ISCO. Mass is set to $M = 1$.}
    \label{fig:ISCO}
\end{figure}

Figure~\ref{fig:ISCO} provides a three-dimensional visualization of the ISCO radius $r_{\rm ISCO}$ as a function of both $\alpha$ and $\zeta$. The surface exhibits a steep gradient along the $\alpha$ direction (indicating strong dependence on the CoS parameter) and an almost flat profile along the $\zeta$ direction (indicating weak dependence on quantum corrections). The color gradient from blue (lower values) to red (higher values) clearly illustrates that the ISCO radius is primarily controlled by $\alpha$, with $\zeta$ providing only second-order corrections. This behavior contrasts with the photon sphere properties, where quantum corrections played a more significant role in Model-I. The ISCO location is crucial for determining the inner edge of accretion disks, which in turn affects the disk's radiative efficiency and spectral properties. Observations of X-ray spectra from BH binaries could potentially constrain the CoS parameter $\alpha$ by measuring the ISCO radius, although disentangling CoS effects from other astrophysical uncertainties remains a challenge.

In summary, for photon orbits and BH shadows, CoS effects dominate and increase the photon sphere radius and shadow size, while quantum corrections in Model-I provide smaller, opposite effects. The topological analysis confirms that Model-I photon spheres carry topological charge $Q = -1$, consistent with unstable light rings. For massive particle orbits, the ISCO radius is primarily determined by the CoS parameter, with quantum corrections playing a negligible role. The results might offer specific predictions for future observational tests with the EHT, VLBI arrays, and X-ray timing, potentially providing constraints on quantum gravity and CoS parameters through multi-messenger astrophysical data.

\section{Numerical Simulations of Plasma Dynamics and QPO Formation}
\label{isec4}

In addition to purely theoretical investigations, it is crucial to gain a numerical understanding of the plasma structure and the underlying physical mechanisms that emerge around quantum-corrected Letelier BHs. Such an approach allows us to examine the interplay between the quantum correction parameter $\zeta$, which originates from quantum gravity-inspired modifications of classical BH metrics, and the CoS parameter $\alpha$, which encodes the effects of a surrounding distribution of strings. Since these parameters naturally arise in different classes of BH models, their combined impact provides valuable insights into how deviations from standard GR may manifest in astrophysical environments. Importantly, identifying and quantifying these effects may play a decisive role in interpreting observational signatures, particularly QPOs detected in the vicinity of a BH, and in testing the limits of gravity with next-generation high-resolution telescopes operating in strong gravitational regimes.

For these reasons, in this part of the paper we go beyond analytical treatments and directly solve the GRH equations numerically, following the methodology outlined in \cite{Donmez2004,Donmez2006,Donmez2012,Donmez2025}. This approach enables us to capture nonlinear plasma dynamics and to demonstrate explicitly how the morphology of the accretion flow, the structure of the shock cone, and its physical properties evolve in the presence of $\zeta$ and $\alpha$. We show in detail how these parameters influence the shock profile around both Model-I and Model-II BHs, leading to distinct plasma configurations that would be inaccessible through analytical approximations alone. In addition, we explore how the formation of the shock cone and the cavity within it imprints on the frequency spectrum of the emerging QPOs. Our results highlight parameter-dependent shifts in QPO frequencies as well as their potential detectability, thereby offering a concrete link between theoretical models and observable quantities.

\subsection{Evolution of the Shock Cone Structure and Accretion Dynamics}
\label{shock_cone}

\begin{figure}[ht!]
\centering
\includegraphics[width=9.0cm,height=7.5cm]{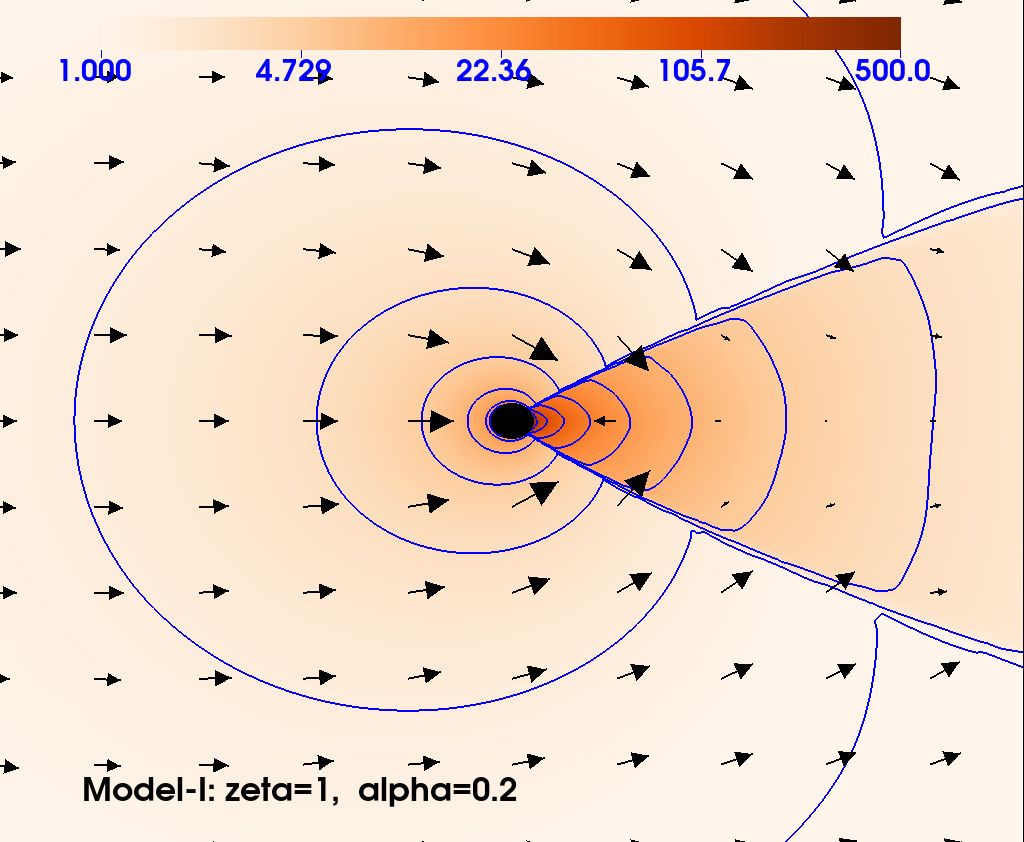}
\includegraphics[width=9.0cm,height=7.5cm]{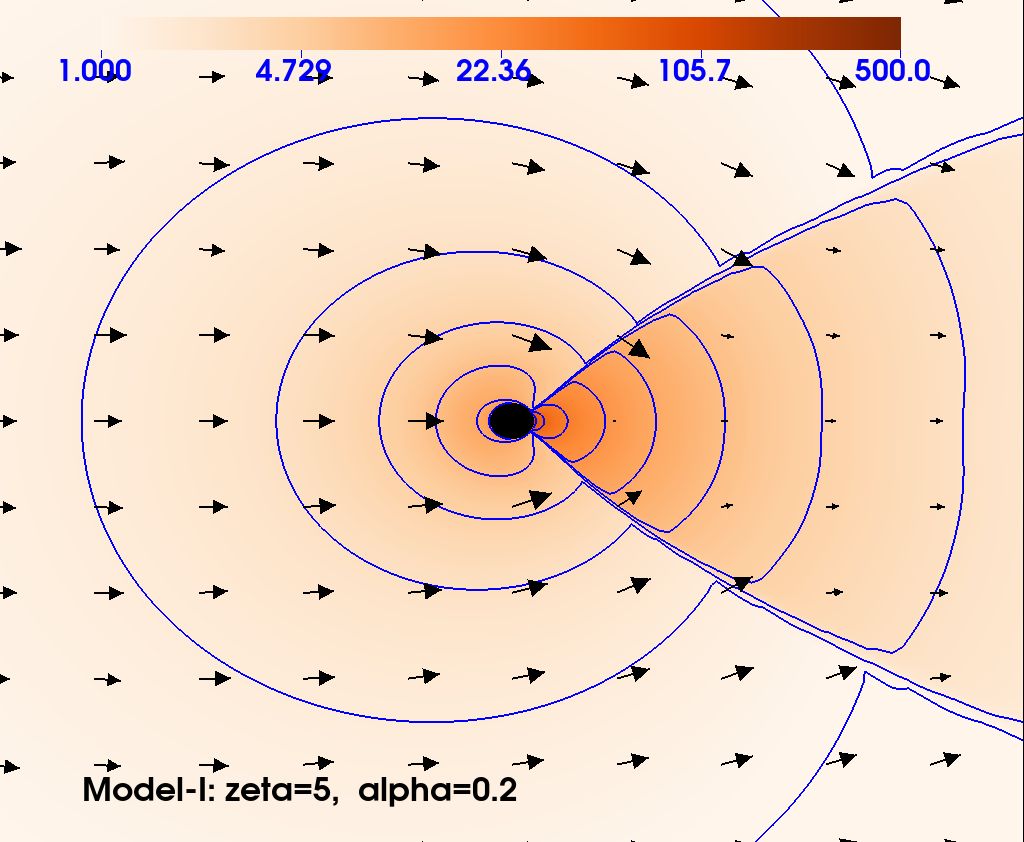} \\
\includegraphics[width=9.0cm,height=7.5cm]{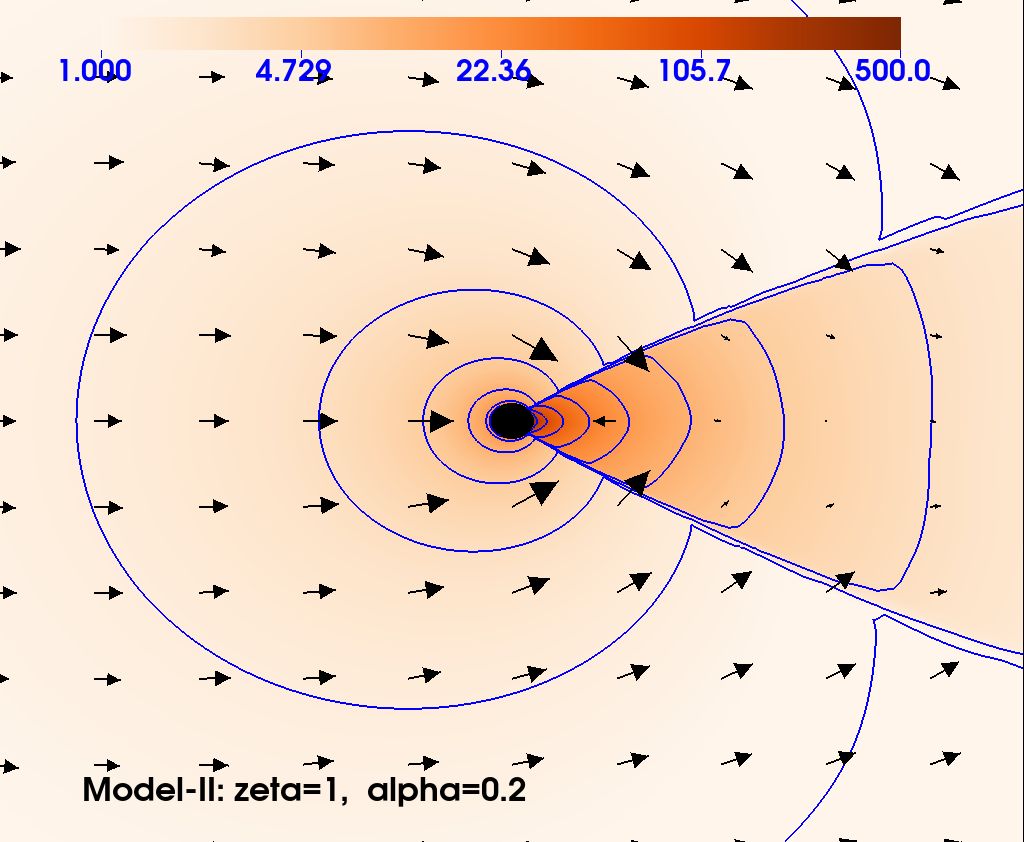}
\includegraphics[width=9.0cm,height=7.5cm]{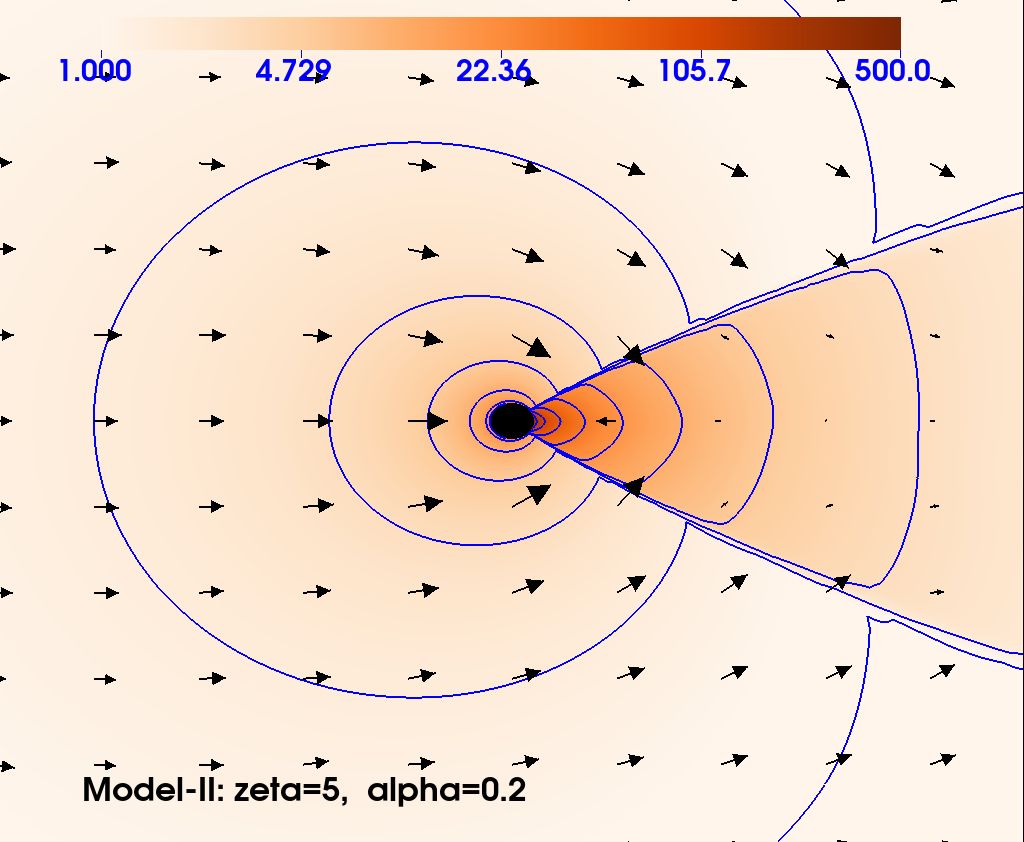} 
\caption{\footnotesize The rest-mass density of the shock cone and the plasma structure formed around two different quantum-corrected BHs are shown using color and contour plots, long after the disk has reached a stable state. The effects of the quantum correction parameter $\zeta$ and the CoS parameter $\alpha$ on the resulting dynamic structure are illustrated. In addition, vector graphics are plotted to demonstrate how matter accretes toward the BH and how it behaves inside the cone. These configurations are presented in the equatorial plane within the boundaries $x_{\min} = -70M$, $y_{\min} = -70M$, $x_{\max} = 70M$, and $y_{\max} = 70M$.}
\label{color_plazma}
\end{figure}

Revealing the physical mechanisms that arise in Model-I and Model-II and determining how the spacetime metrics influence those mechanisms, together with comparing numerical or theoretical results to observations, always enables meaningful tests of gravity. In Fig.~\ref{color_plazma}, the structure of the shock cone formed under two different quantum-corrected Letelier metrics is shown. In Model-I, the quantum term $\zeta$ appears in both $f(r)$ and $g(r)$ together with the CoS parameter $\alpha$. Consequently, $\zeta$ markedly alters the flow dynamics, as seen in the top row of Fig.~\ref{color_plazma}. Due to these changes, the cone opening angle, axial compression, and the size of the downstream cavity are significantly affected. In Model-II, $f(r)$ retains the classical Letelier form while $g(r)$ is modified by the same $\zeta^2$ term as in Model-I. Thus, although the horizons are determined by the same condition in both models, the cone morphology behaves differently away from the horizon, as visible in the bottom row of Fig.~\ref{color_plazma}, because the variations with $\zeta$ and $\alpha$ differ between the two models.

\begin{figure}[ht!]
\centering
\includegraphics[width=9.0cm,height=7.5cm]{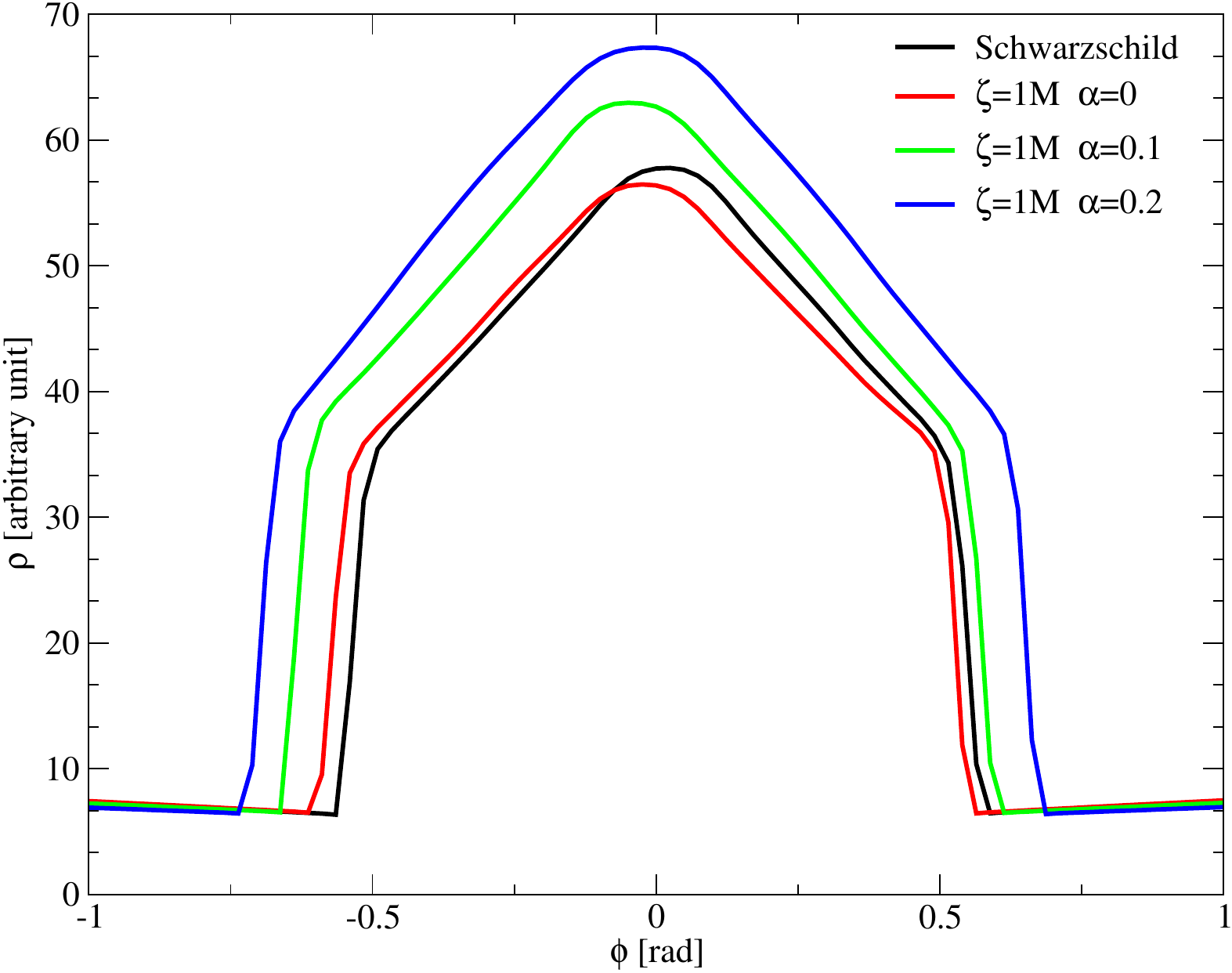}\;\; 
\includegraphics[width=9.0cm,height=7.5cm]{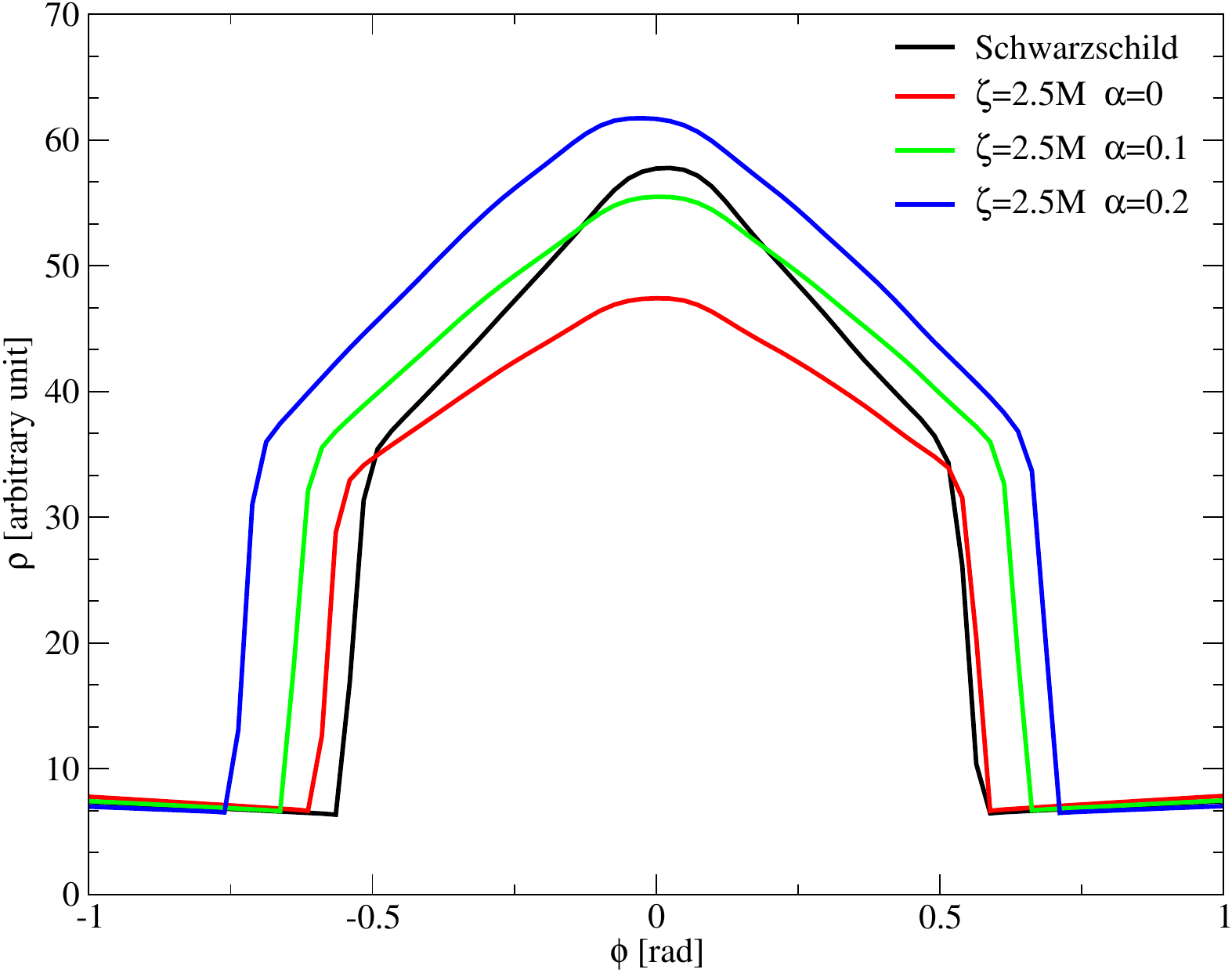} \\
\vspace{0.2cm}
\includegraphics[width=9.0cm,height=7.5cm]{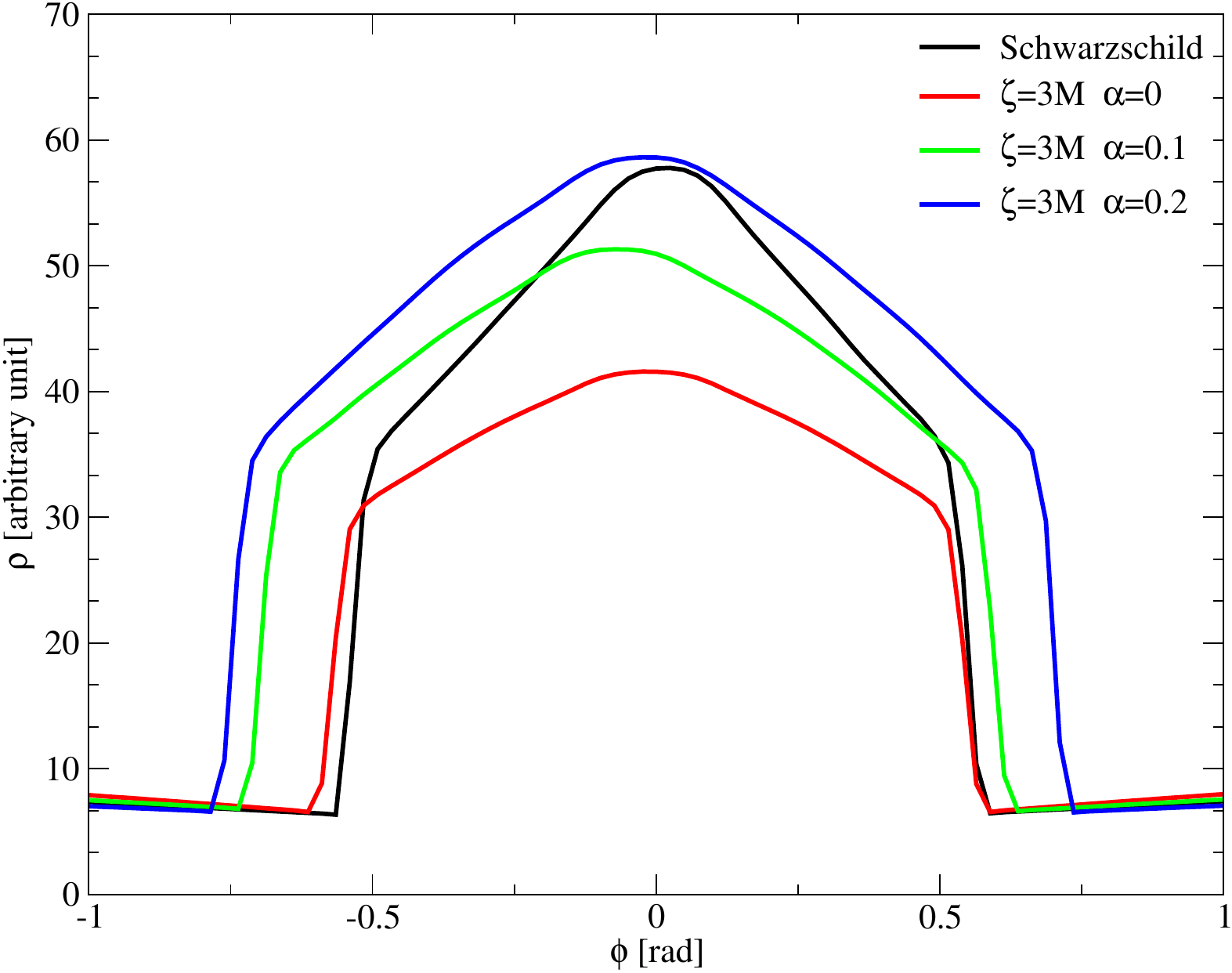}\;\; 
\includegraphics[width=9.0cm,height=7.5cm]{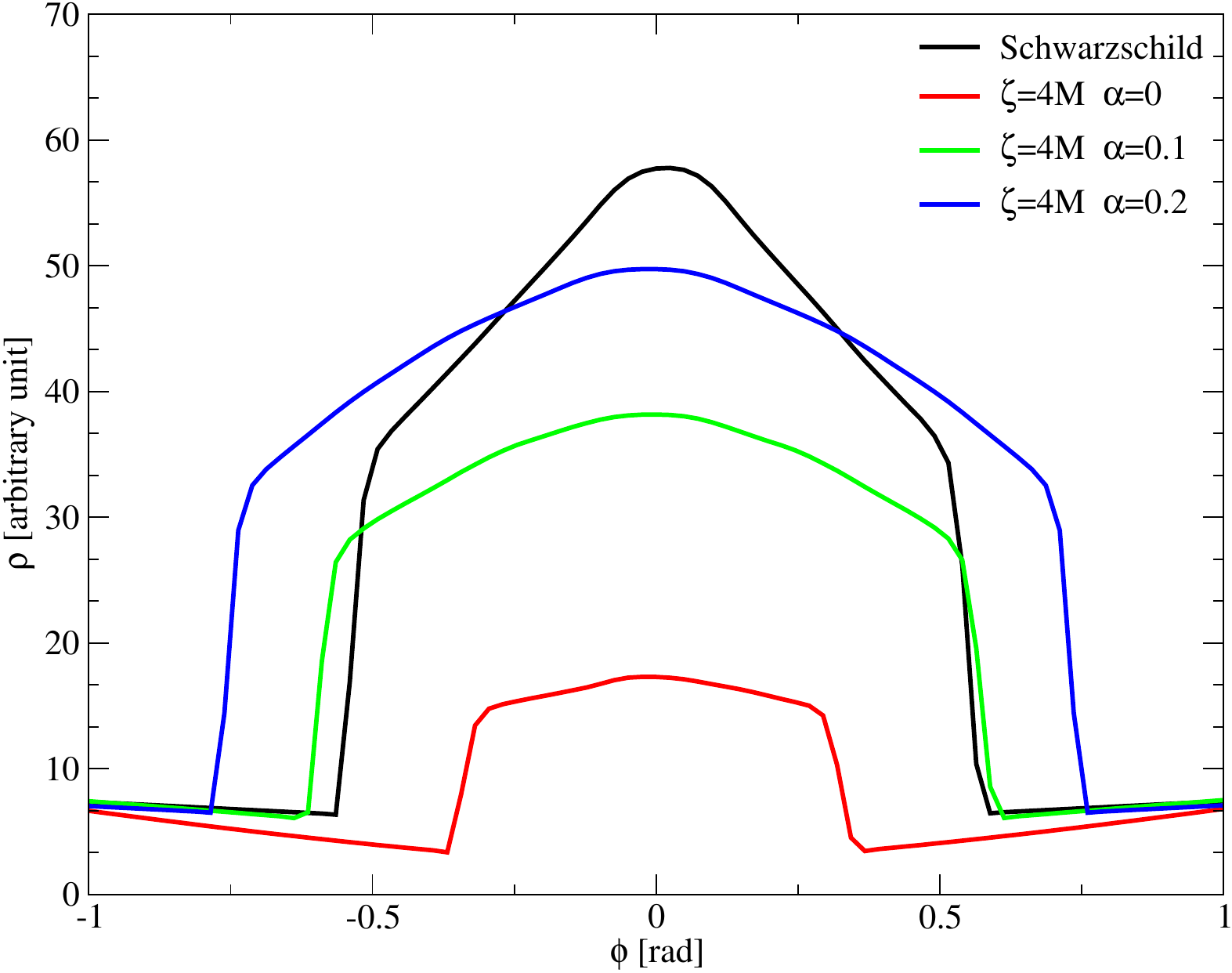} 
\caption{\footnotesize The variation of the rest-mass density in the azimuthal direction is shown for Model-I. It illustrates the structure of the shock cone formed around the BH and the density of matter trapped inside the cone. From the top-left graph to the bottom-right graph, the changes in rest-mass density are presented for increasing values of $\zeta$. In each case, Schwarzschild and different values of $\alpha$ are plotted at fixed $\zeta$, thereby revealing the effect of $\zeta$ on the cone structure and on the density of matter accreted within it.}
\label{Shock_coneBH1}
\end{figure}

\begin{figure}[ht!]
\centering
\includegraphics[width=9.0cm,height=7.5cm]{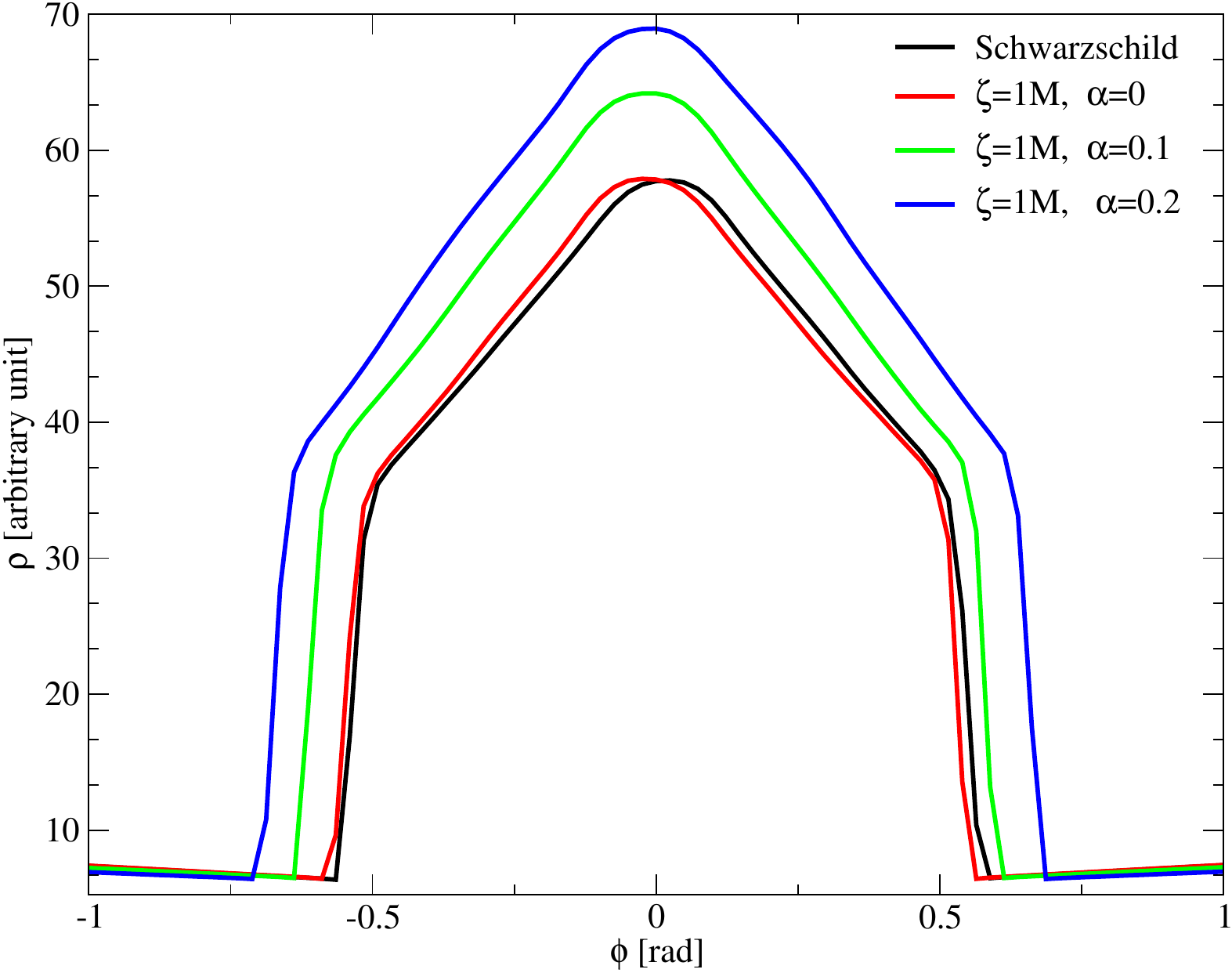}\;\; 
\includegraphics[width=9.0cm,height=7.5cm]{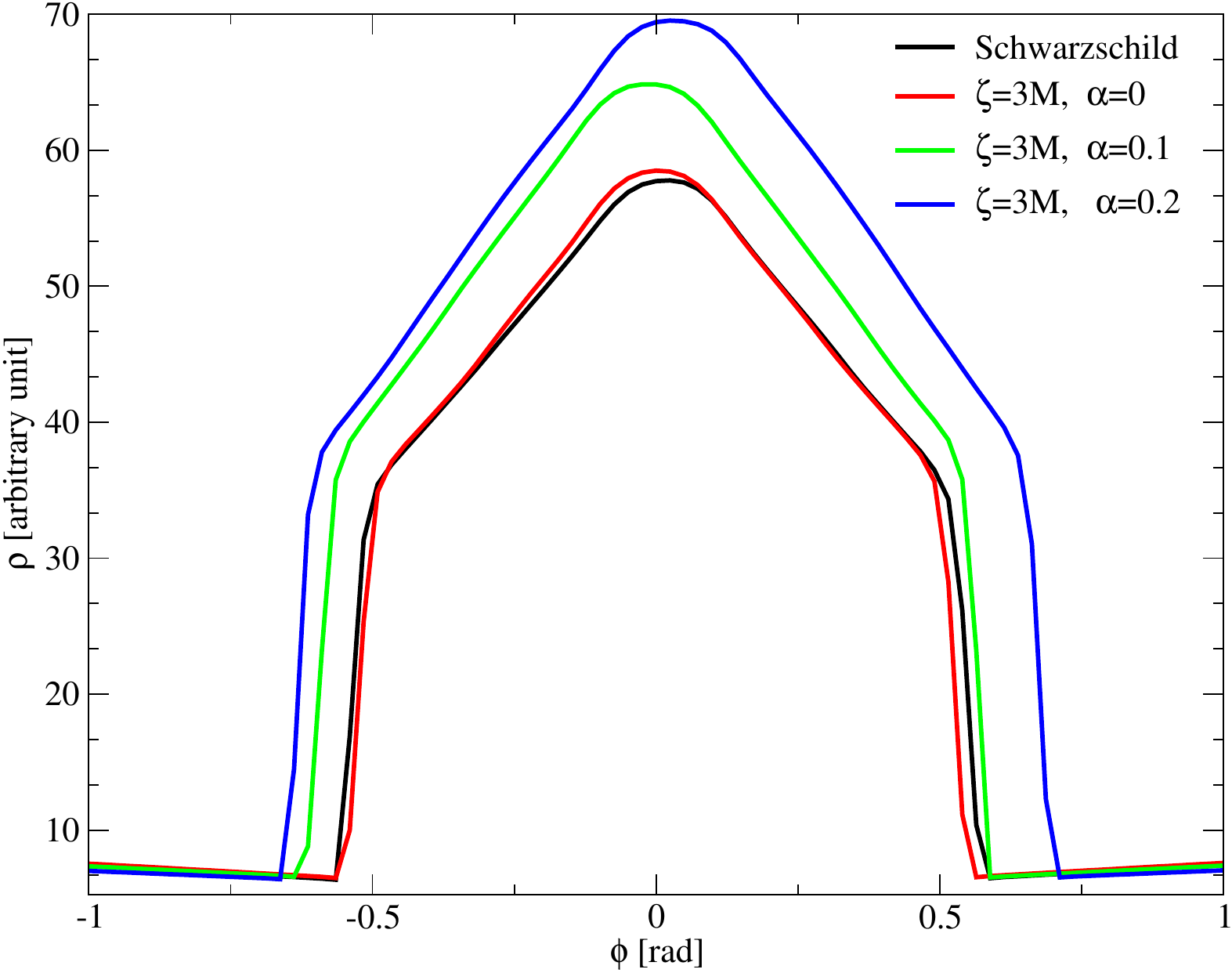} \\
\vspace{0.2cm}
\includegraphics[width=9.0cm,height=7.5cm]{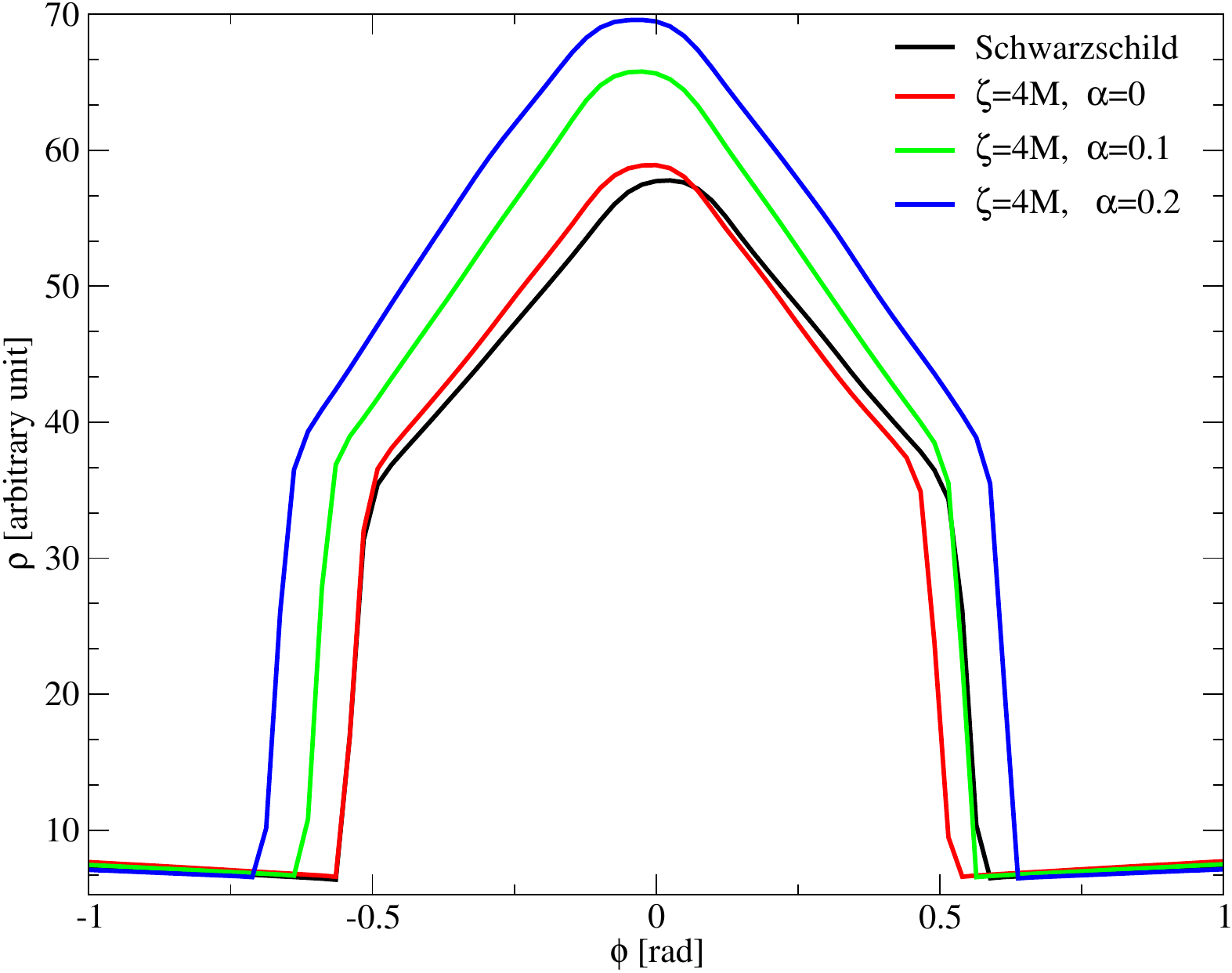}\;\; 
\includegraphics[width=9.0cm,height=7.5cm]{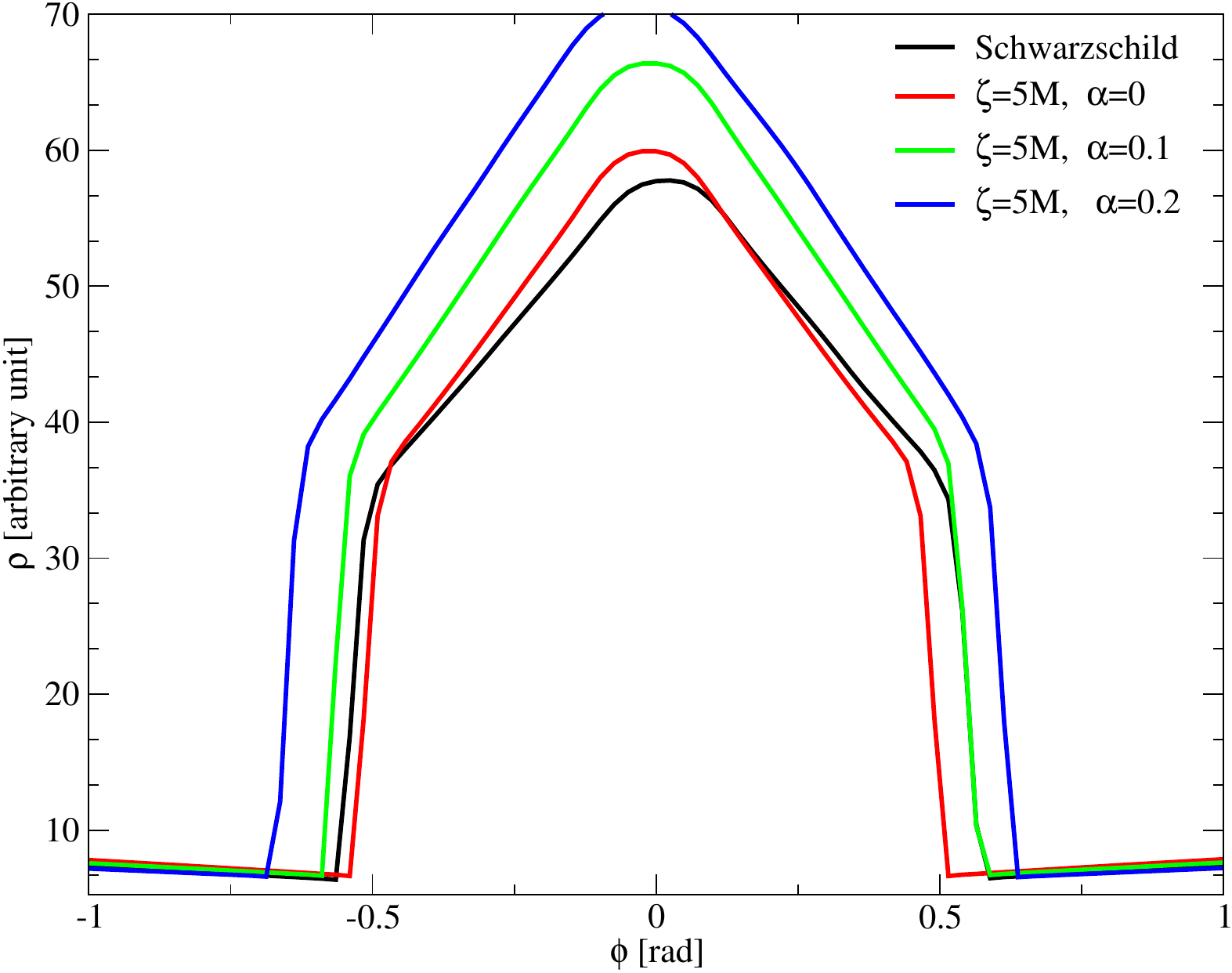} 
\caption{\footnotesize Same as Fig.~\ref{Shock_coneBH1}, but for Model-II.}
\label{Shock_coneBH2}
\end{figure}

\begin{figure}[ht!]
\centering
\includegraphics[width=18.0cm,height=14cm]{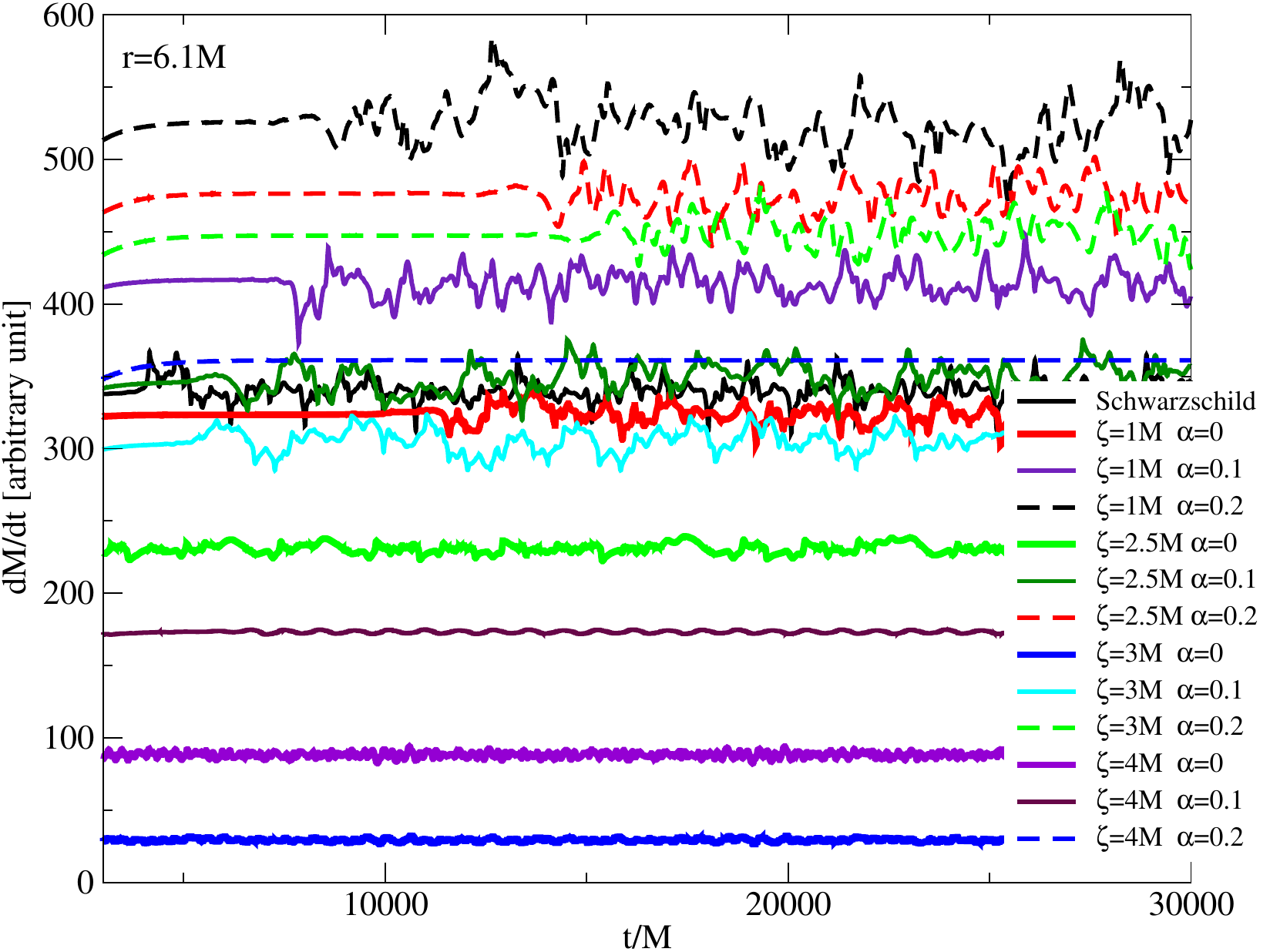}
\caption{\footnotesize The time evolution of the mass accretion rate around the ISCO is shown for different cases in Model-I. In Schwarzschild and Letelier spacetime, depending on the quantum correction parameters $\zeta$ and $\alpha$, the plasma structure formed around the BH and thus the oscillatory behavior of the shock cone are illustrated. It is observed that the instability is either enhanced or suppressed depending on the values of $\zeta$ and $\alpha$.}
\label{mass_acccBH1}
\end{figure}

\begin{figure}[ht!]
\centering
\includegraphics[width=18.0cm,height=14cm]{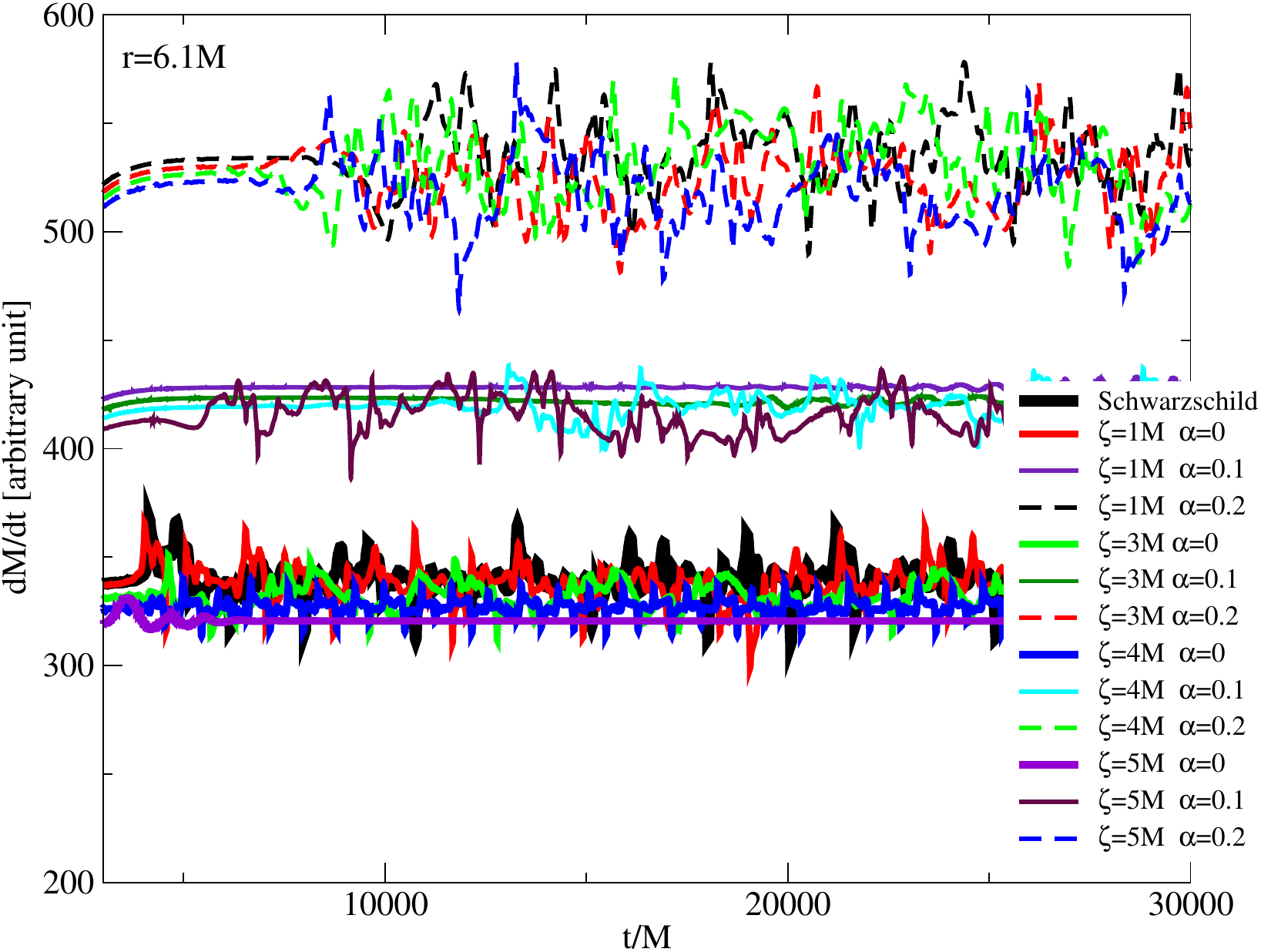}
\caption{\footnotesize Same as Fig.~\ref{mass_acccBH1}, but for Model-II.}
\label{mass_acccBH2}
\end{figure}

However, in both cases, the presence and size of the cavity depend on $\zeta$ and $\alpha$ as discussed in detail in Figs.~\ref{Shock_coneBH1} and~\ref{Shock_coneBH2}, and are consistent with Figs.~\ref{mass_acccBH1} and~\ref{mass_acccBH2}. These parameter-dependent structures lead to pronounced and unstable oscillations in the mass accretion rate around the BH. This behavior is strong evidence for the formation of QPOs around the BHs and their potential detectability by observers. Additionally, the vector (streamline) plots show that matter falls supersonically from the upstream region toward the BH. The strong gravitational field bends the inflow, and collisions between streams arriving from the right and left in the downstream region generate the shock cone. The cone opening angle depends entirely on $\zeta$ and $\alpha$. The vector plots also indicate that matter trapped inside the cone either continues to fall toward the BH or, alternatively, can move outward, depending on the stagnation point within the cone. Also, how this stagnation point shifts with $\zeta$ in the absence of the Letelier (CoS) term is explained in detail in \cite{Al-Badawi2025}. As illustrated in the top row, right-hand panel of Fig.~\ref{color_plazma}, when the stagnation point moves closer to the BH, fewer particles accrete (see also Fig.~\ref{mass_acccBH1}), reducing accretion, quenching the instability, and thereby suppressing any QPOs that might otherwise arise.

In Fig.~\ref{Shock_coneBH1} (Model-I), the azimuthal variation of the rest-mass density at $r=6.1M$ is shown for different $\zeta$ and $\alpha$. In each snapshot, for fixed $\zeta$, we plot the Schwarzschild case together with $\alpha = 0$, $0.1$, and $0.2$ to reveal how the CoS parameter modifies the shock cone structure in a stronger gravitational field. Across snapshots, varying $\zeta$ demonstrates how the quantum correction term affects the cone and, consequently, the QPOs trapped inside the cavity. For $\zeta=1M$, the cone is broad with a clear single maximum, and increasing $\alpha$ slightly raises the curve and sharpens the peak. For $\zeta=2.5M$, the profile becomes more asymmetric and the cavity contrast is clearer, but the maximum peak is lower than at $\zeta=1M$ relative to Schwarzschild. As $\alpha$ increases, the peak first decreases and then, at $\alpha=0.2$, exceeds the Schwarzschild level. For $\zeta=3M$, the cone is more azimuthally localized as in $\zeta=2.5M$. The maximum density initially drops with $\alpha$ and then tends to increase with $\alpha=0.2$, but most of the time satisfies the Schwarzschild value. For $\zeta=4M$, the cone is the most focused and the cavity the most pronounced, yet it exhibits the lowest peak amplitude among the snapshots. Here, $\alpha$ produces small upward shifts and mild sharpening, but the $\zeta$-driven reduction in the maximum persists.

Together, as $\zeta$ increases, the density of matter trapped in the cavity steadily decreases, whereas, at any fixed $\zeta$, increasing $\alpha$ widens the cone's opening angle. Because this widening changes the oscillation period of trapped matter, the QPO frequencies are significantly affected. In Model-I, it is therefore evident that a larger $\alpha$ substantially alters both the density within the cone and its opening angle, with strong consequences for QPO formation and observability discussed in detail in Sec.~\ref{trapped_QPO}.

In Fig.~\ref{Shock_coneBH2} (Model-II), the azimuthal variations of the mass density of the shock cone rest at $r=6.1M$ are plotted to reveal how the density of matter trapped inside the cone and the opening angle of the cone respond to the parameters $\zeta$ and $\alpha$. Each snapshot in Fig.~\ref{Shock_coneBH2} is drawn at a different $\zeta$. For every case of $\zeta$, the curves for Schwarzschild and for $\alpha=0$, $0.1$, and $0.2$ are shown to clarify the effect of the CoS parameter on the cone. As seen for each $\zeta$, increasing $\alpha$ increases the density within the cone and makes the peak sharper, while also causing a clear widening of the cone. As detailed in Sec.~\ref{trapped_QPO}, these changes induced by larger $\alpha$ are expected to strongly excite the resulting QPOs and enhance their observability. Different levels of $\alpha$ can also give rise to different QPO frequencies.

When comparing snapshots with each other, $\zeta$ in Model-II modifies the overall morphology of the cone more moderately, mainly in aspects such as localization and peak position, while the dominant determinant of contrast is $\alpha$. In summary, compared to Schwarzschild, $\alpha$ significantly alters the cone morphology and opening angle in Model-II, while the influence of $\zeta$ appears relatively more limited and gradual.

In Sect.~\ref{trapped_QPO}, PSD analyzes are performed to reveal the emergence of QPO frequencies and their observational detectability. For this purpose, the mass accretion rates are calculated at $r=6.1M$. Since this point is very close to the ISCO, the instabilities arising here and their dependence on parameters provide us with an opportunity to study QPO formation in such systems and to test the nature of spacetime.

Fig.~\ref{mass_acccBH1} shows the variation of the mass accretion rate around the ISCO for Model-I under different parameter choices. It has been observed that both the quantum correction parameter $\zeta$ and the CoS parameter $\alpha$ strongly influence the formation or suppression of instabilities. With increasing $\zeta$, the strength of the instabilities decreases and even disappears completely for $\zeta > 2.5M$, meaning that the shock cone around the BH exhibits a strong stabilizing behavior. In contrast, increasing $\alpha$ improves the strength of the instabilities. As discussed in the context of Figs.~\ref{Shock_coneBH1} and~\ref{Shock_coneBH2}, this behavior is directly related to the widening of the opening angle of the shock cone and the increase in the density of matter trapped inside the cone.

Such parameter-dependent changes significantly affect the resulting QPOs, as discussed in detail in Sec.~\ref{trapped_QPO}. When $\alpha$ is larger, the stronger and more irregular oscillations of the cone map directly into more pronounced, less coherent QPO signals, whereas higher $\zeta$ values dampen these fluctuations, yielding weaker but more coherent peaks in the frequency spectrum. In particular, for $\alpha = 0.2$, significant differences compared to the Schwarzschild case and other $\alpha$ values are clearly observed. One of the most important findings is that, for $\alpha = 0.2$, the onset of instability occurs much later than for other $\alpha$ values or for the Schwarzschild case. Observationally, this delay implies that in such scenarios, QPOs may not appear immediately but may only become detectable after the disk has passed through a certain evolutionary stage. This time-dependent behavior reflects the dynamic interplay between quantum corrections and the surrounding CoS, suggesting that the combined effect of $\zeta$ and $\alpha$ could leave distinct imprints on both the timing properties and the detectability of QPOs, offering a potential observational probe into the underlying spacetime structure.

In Fig.~\ref{mass_acccBH2}, the mass accretion rates are shown for the case of Model-II, where the quantum correction parameter $\zeta$ affects only the radial metric function $g(r)$, while the temporal part $f(r)$ remains in its classical Letelier form. The structural differences between Model-I and Model-II lead to distinct accretion behaviors. Compared to Model-I, in Model-II the ability of $\zeta$ to reduce the strength of instabilities is slower. As in Model-I, increasing $\zeta$ still moderates the oscillations, but the effect is more gradual. Even for $\zeta = 4M$ or $\zeta = 5M$, the instabilities are not completely suppressed.

In contrast to the stabilizing effect of $\zeta$, the CoS parameter $\alpha$ acts as a driving force for instability: larger values of $\alpha$ make the instabilities stronger and produce significant increases in the accretion rate. A larger $\alpha$ results in stronger and more irregular oscillations, which is directly proportional to the widening of the shock cone opening angle and the increase in the density of matter trapped inside the cone, as already discussed in Fig.~\ref{Shock_coneBH2}. These instabilities are directly mapped into more pronounced but less coherent QPO signals, making them more easily detectable but spectrally less stable.

Unlike Model-I, in Model-II there is no complete suppression of disk instabilities. Instead, strong oscillations persist in almost all values of $\zeta$, with their intensity further amplified as $\alpha$ increases. Similarly as in Model-I, $\alpha$ also plays an important role in enhancing the oscillation strength in Model-II. From an observational perspective, when Model-I and Model-II are compared, the QPOs in Model-I may be suppressed or delayed depending on $\zeta$ and $\alpha$, while in Model-II the QPOs appear more robust, continuous, and irregular. This comparison highlights how the quantum correction parameter $\zeta$ and the CoS parameter $\alpha$ play a crucial role in shaping observable QPOs. The change in shock cone dynamics from model to model and the resulting mass accretion variability ultimately affect the detectability and timing of QPOs, offering a promising route for testing the underlying spacetime structure with high-precision X-ray observations.

\subsection{Trapped QPOs inside the Shock Cone Cavity}
\label{trapped_QPO}

\begin{figure}[ht!]
\centering
\includegraphics[width=9.0cm,height=7.5cm]{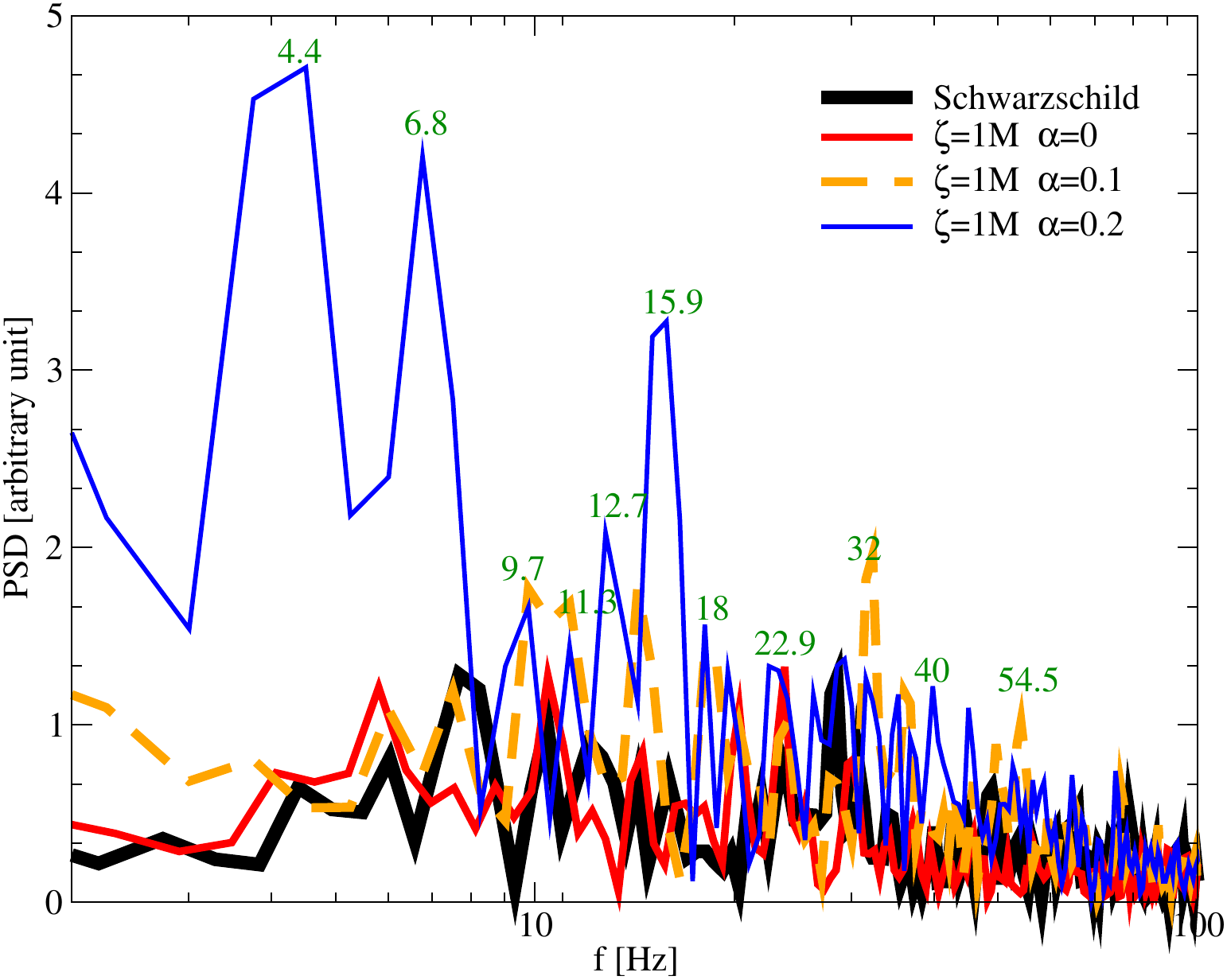}\;\; 
\includegraphics[width=9.0cm,height=7.5cm]{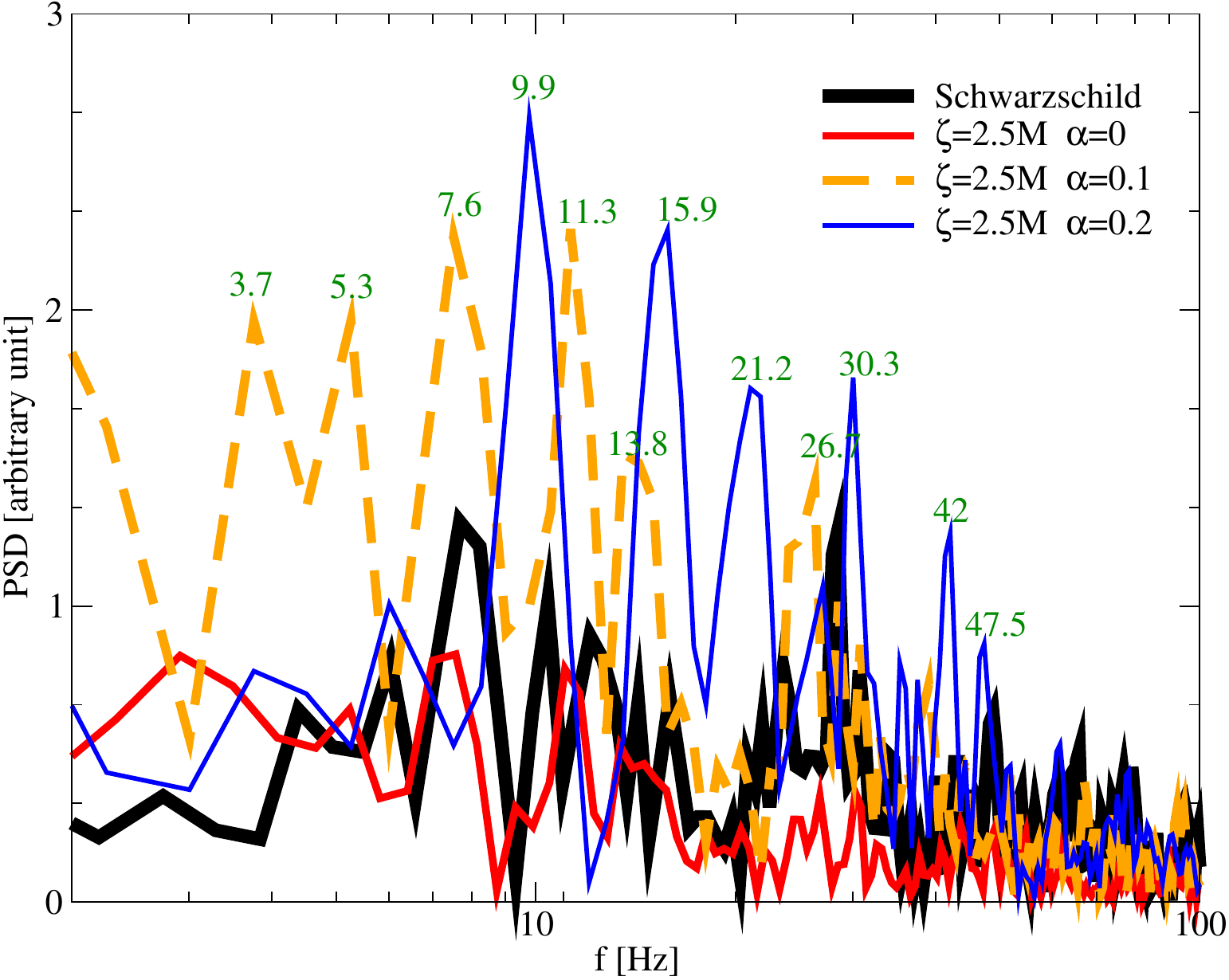} \\
\vspace{0.2cm}
\includegraphics[width=9.0cm,height=7.5cm]{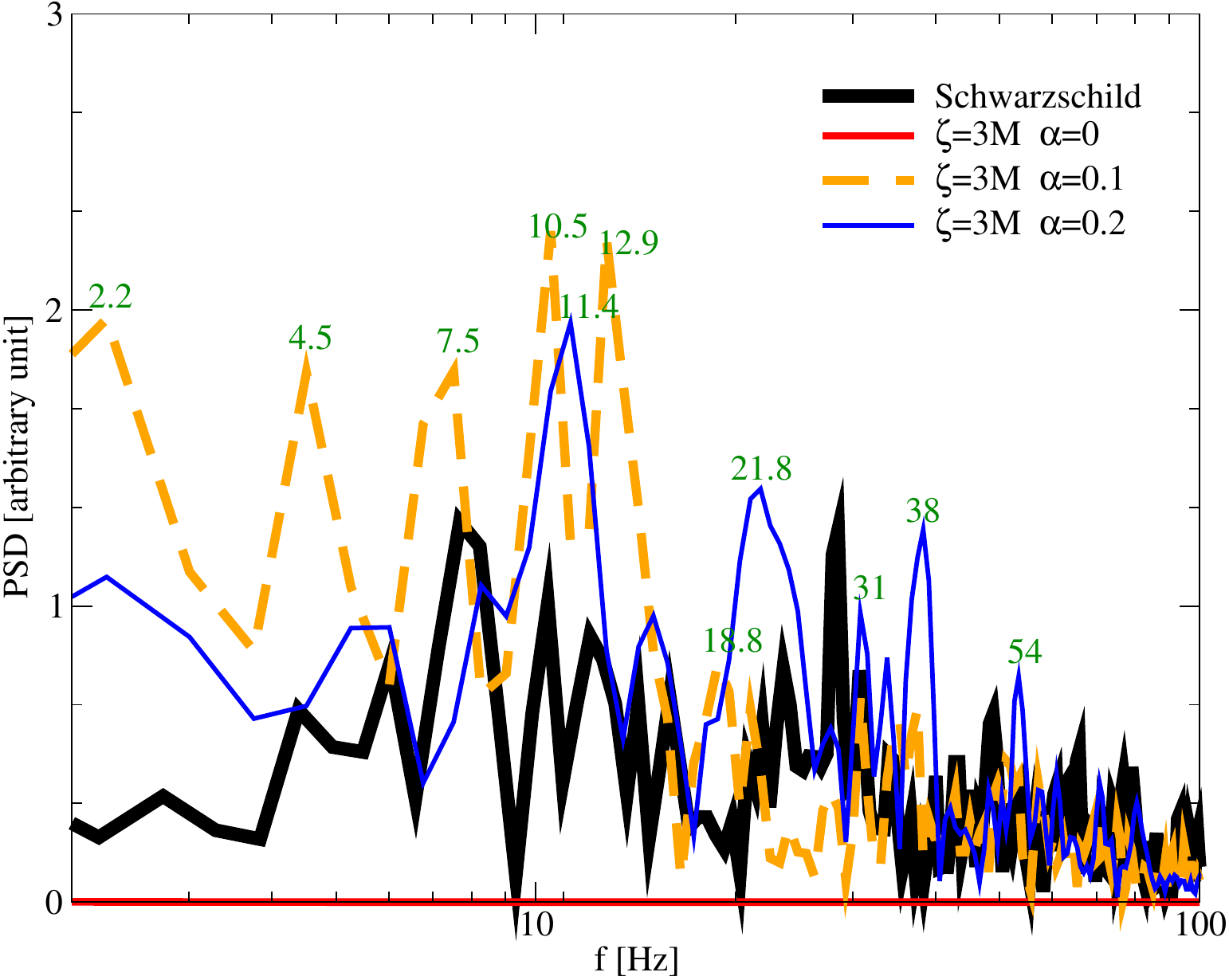}\;\; 
\includegraphics[width=9.0cm,height=7.5cm]{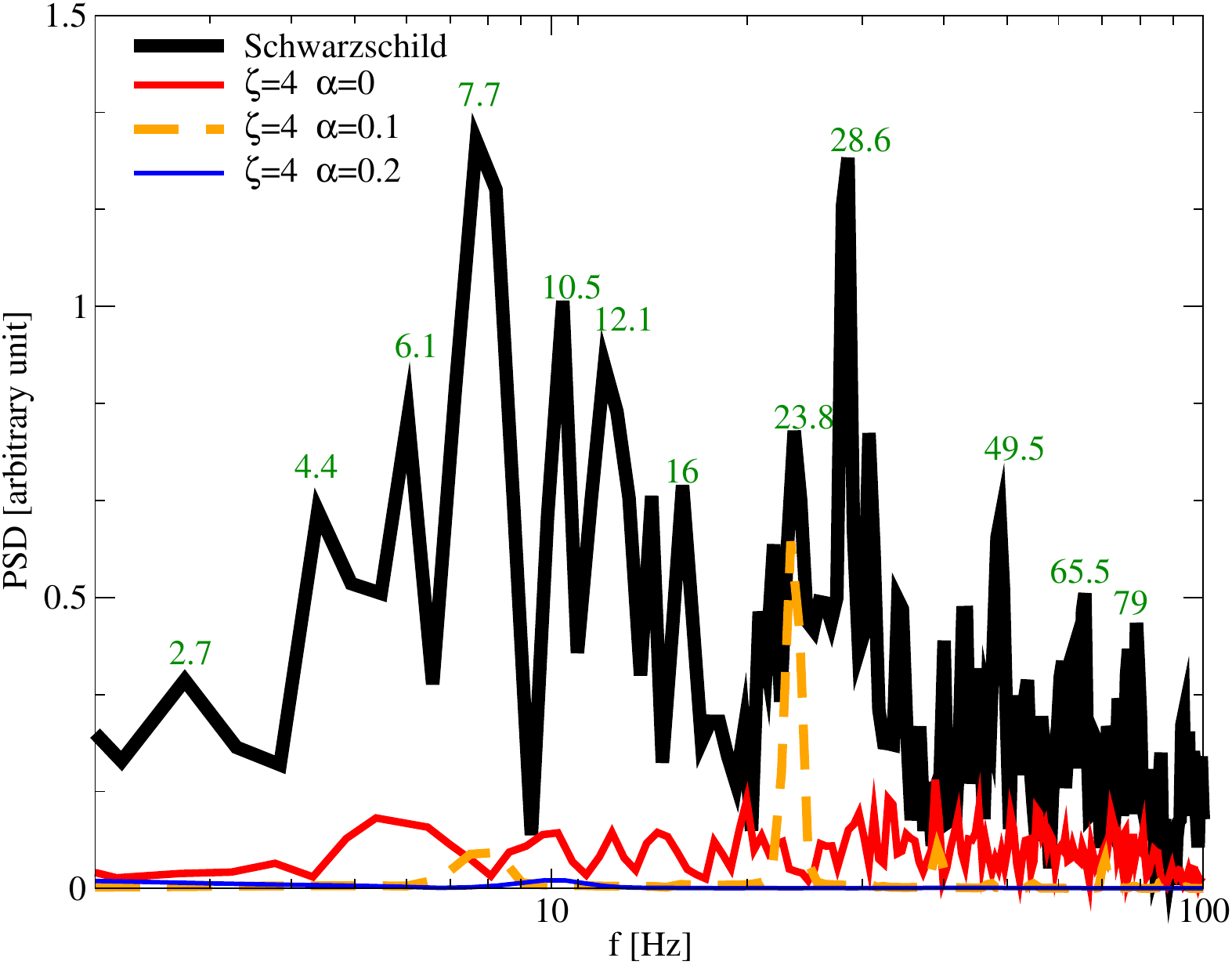} 
\caption{\footnotesize PSD as a function of frequency for different parameters of the spacetime metric for Model-I. Each snapshot shows PSD analyses for different $\alpha$ parameters at a fixed $\zeta$, compared with the Schwarzschild case. It is observed that the dominant peaks formed by variations in $\zeta$ change for different $\alpha$ values. These behaviors are directly related to the physical properties of the resulting shock cavity.}
\label{QPO_BH1}
\end{figure}

The numerical study of the PSD analysis is important to reveal the influence of the Letelier BH parameters on the observable QPO frequencies. In Fig.~\ref{QPO_BH1}, for Model-I, the effects of the quantum correction parameter $\zeta$ and the CoS parameter $\alpha$ on the resulting QPO frequencies are demonstrated, together with the Schwarzschild case. Taking into account only the effect of the quantum parameter $\zeta=1M$, as discussed in \cite{Al-Badawi2025}, the dynamics of the cavity is modified and, consequently, the oscillation modes are shifted. This nudges the peaks away from the Schwarzschild baseline, making some features appear sharper. In the cases with $\alpha=0.1$ and $\alpha=0.2$, changes in frequency values are observed. At the same time, the amplitude of the peaks in the PSD analysis increases significantly, while the harmonic content is enriched. As clearly explained in Sec.~\ref{shock_cone}, the modification of the QPO modes trapped inside the cavity arises from changes in the matter trapped within the cone and in the cone opening angle. While the quantum correction parameter $\zeta$ causes shifts in the QPO frequencies, the CoS parameter $\alpha$ substantially enhances their observational detectability.

As clearly seen in the $\zeta=2.5M$ case of Fig.~\ref{QPO_BH1}, when $\alpha=0$ the shock cone exhibits more stable oscillatory motion. Consequently, the amplitudes of the resulting frequencies are relatively lower than in the Schwarzschild BH, and the oscillations are more orderly than at $\zeta=1M$. Some peaks even disappear altogether. In contrast, for $\alpha=0.1$ and $\alpha=0.2$, the peaks are magnified once again. Distinct peaks with strong amplitudes compared to the Schwarzschild case emerge around $7$, $11$, $14$ Hz and in the $20$--$40$ Hz range. Thus, with increasing $\alpha$, the peak amplitudes are magnified and a richer overtone structure is produced. In particular, while strong QPO frequencies arise for $\alpha=0.1$, they are clearly shifted toward HFQPOs for $\alpha=0.2$. This demonstrates the robust trend that $\alpha$ drives amplitude growth and harmonic enrichment in the PSD.

In the $\zeta=3M$ case, the QPOs are almost completely suppressed compared to Schwarzschild. However, as in the lower-$\zeta$ cases, the peaks that arise for $\alpha=0.1$ and $\alpha=0.2$ remain at a strongly observable level. In fact, the $\alpha>0$ curves exhibit a rich, multiharmonic pattern that would be easily detectable in real data. Finally, at larger $\zeta$ values the influence of the quantum correction dominates over that of $\alpha$, and almost all QPO frequencies are suppressed. In this regime, the most distinct peaks are observed only in the Schwarzschild case. Therefore, at high $\zeta$, testing the Letelier BH through timing signatures becomes essentially impossible. In other words, while the quantum correction parameter $\zeta$ reduces and stabilizes the instabilities that naturally arise, the CoS parameter $\alpha$ modifies the shock cone structure and allows the formation of strong trapped oscillations. This both significantly increases the amplitudes of PSD peaks and enriches the harmonic structure, while also shifting frequencies through changes in cavity size.

These results are consistent with the theoretical discussions of the ISCO and the cavity dynamics. Especially for moderate values of $\zeta$ ($1M$, $2.5M$) combined with larger $\alpha$, high-amplitude peaks appear that strongly enhance observability. The resulting strong LFQPO frequencies produce commensurate ratios such as $12.9:7.5 \approx 3:2$ and $21:10.2 \approx 2:1$, while HFQPOs display ratios such as $65:45 \approx 3:2$. Such commensurate ratios match long-standing QPO phenomenology in X-ray binaries and AGN candidates. Therefore, the parameter pair $(\zeta,\alpha)$ not only perturbs the accretion flow but also facilitates the emergence of resonance conditions of this type. This gives these models a predictive and testable structure in timing data.

\begin{figure}[ht!]
\centering
\includegraphics[width=9.0cm,height=7.5cm]{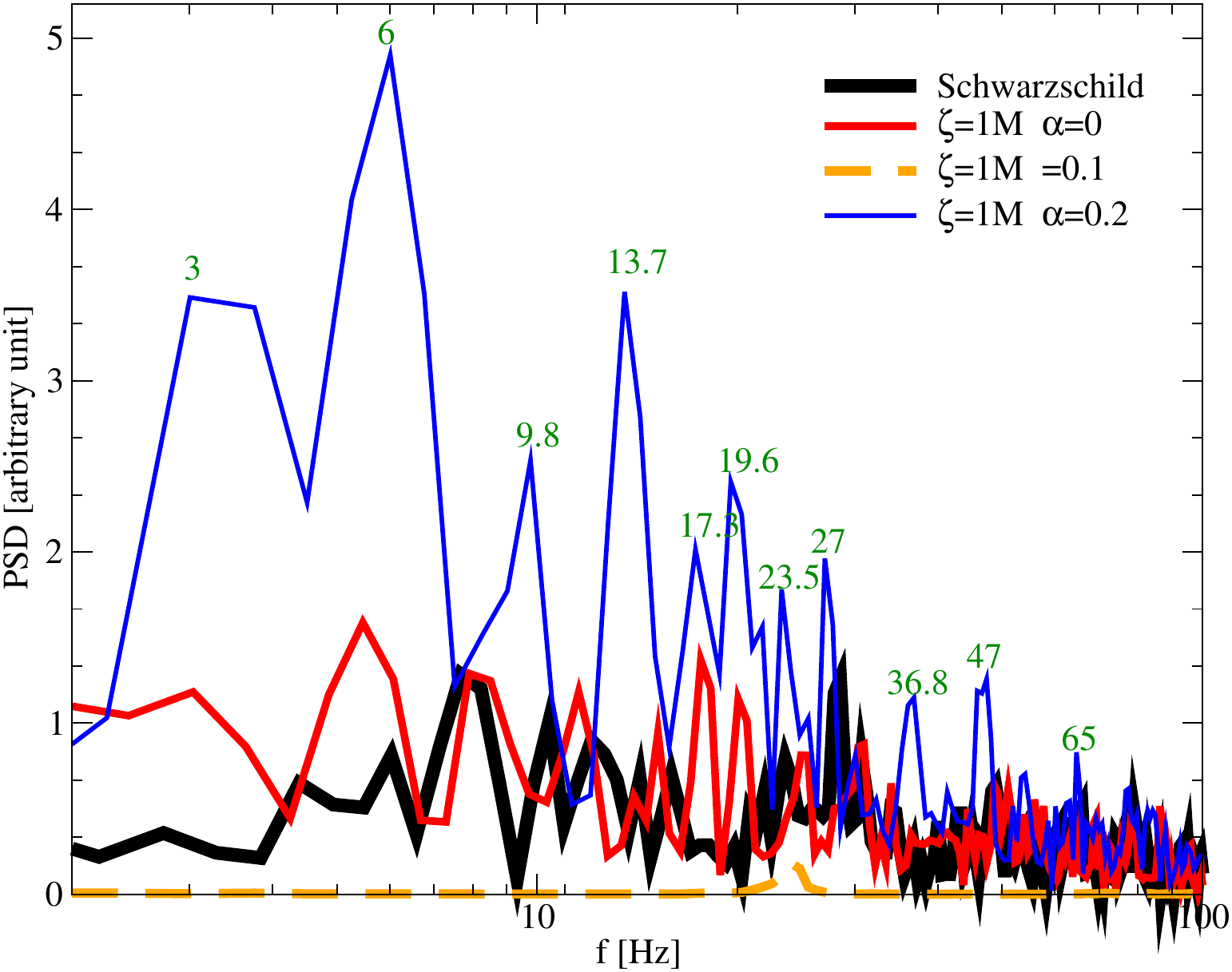}\;\; 
\includegraphics[width=9.0cm,height=7.5cm]{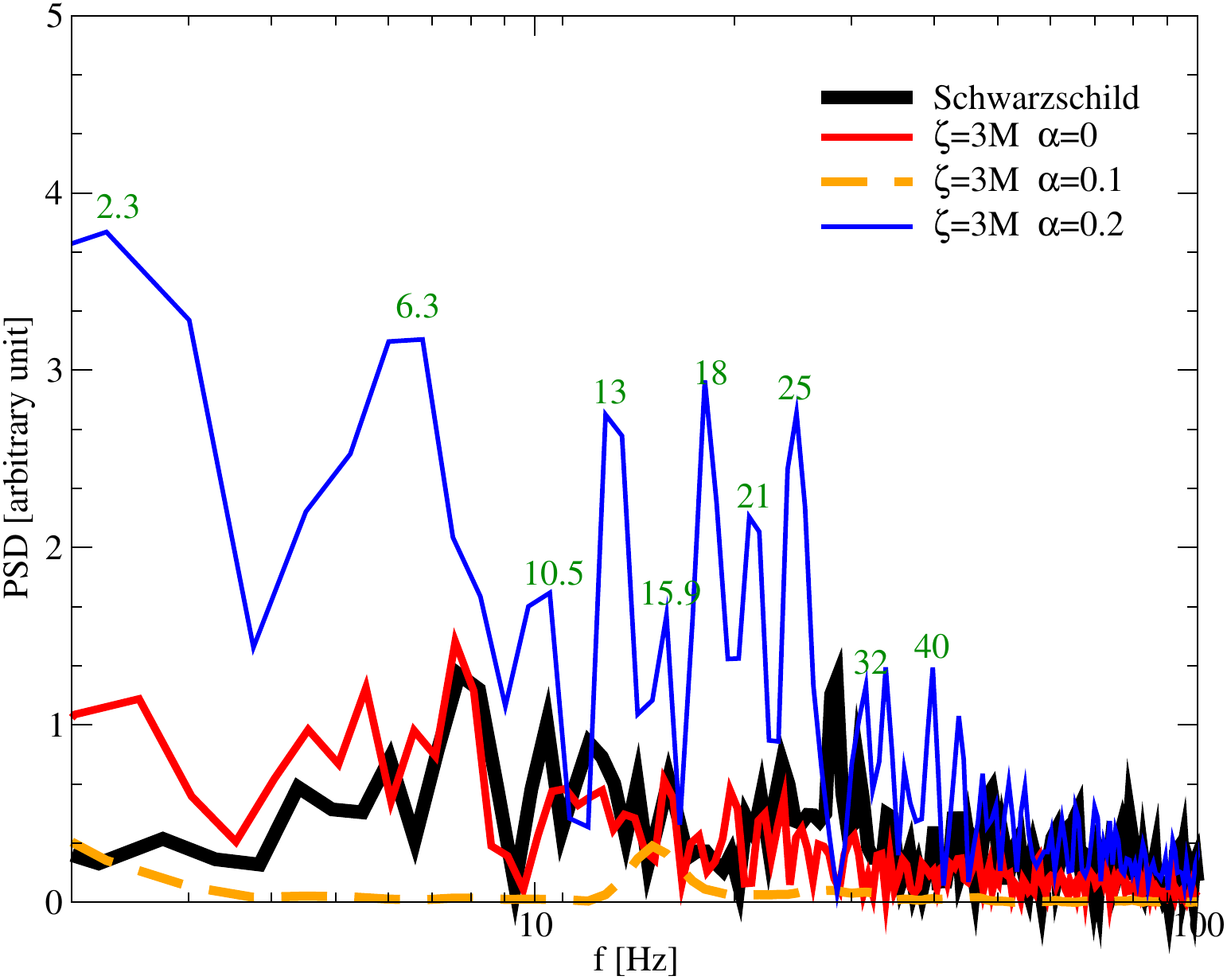} \\
\vspace{0.2cm}
\includegraphics[width=9.0cm,height=7.5cm]{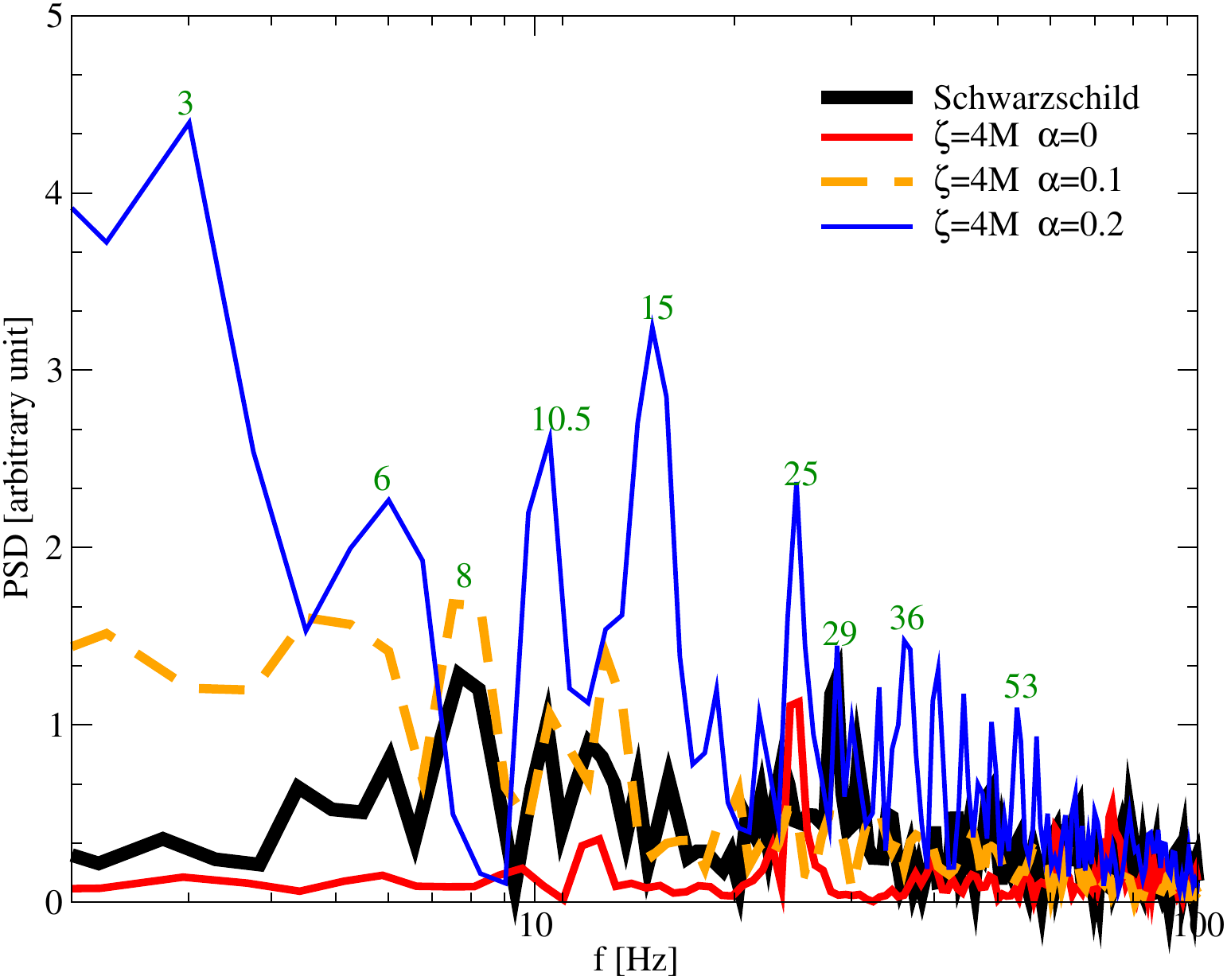}\;\; 
\includegraphics[width=9.0cm,height=7.5cm]{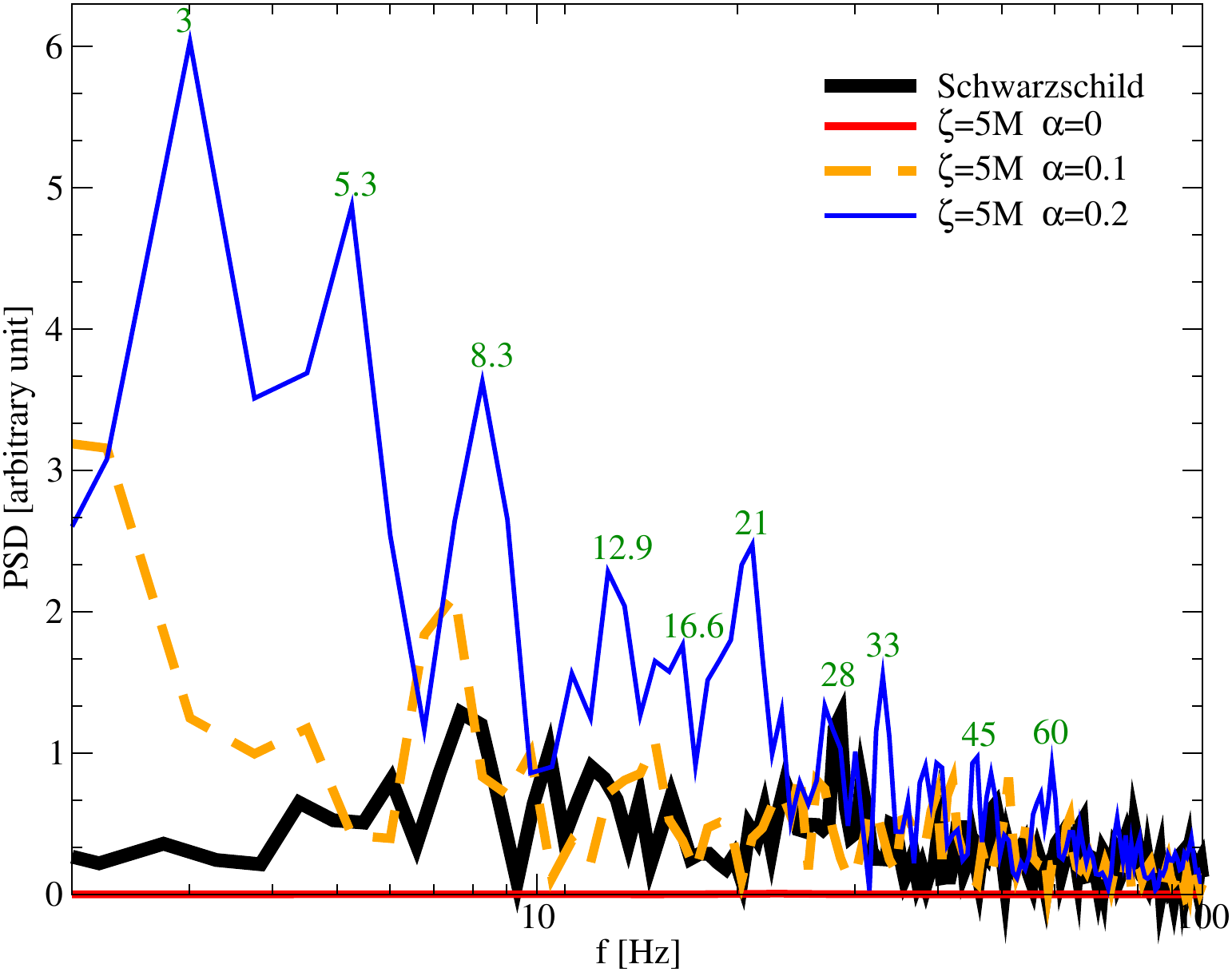} 
\caption{\footnotesize Same as Fig.~\ref{QPO_BH1}, but for Model-II.}
\label{QPO_BH2}
\end{figure}

In Fig.~\ref{QPO_BH2}, the numerically computed PSD analysis for Model-II is presented. Compared to Model-I, the effect of the quantum correction parameter $\zeta$ on the resulting QPO frequencies and their amplitudes is milder, since $\zeta$ only modifies the radial component $g(r)$. As a consequence, even for large values of $\zeta$ the instabilities are not completely suppressed, and oscillations continue to appear in the spectra. For $\zeta=1M$ and $\zeta=3M$, it is seen that the spectrum for $\alpha=0$ is similar to the Schwarzschild case, but due to the modified opening angle of the cone the frequencies are shifted. When $\alpha=0.1$ and $\alpha=0.2$ are introduced, the peaks are clearly magnified. LFQPO peaks arise around $3$ Hz, $6$ Hz, and $10$--$14$ Hz, while HFQPOs extend into the $20$--$65$ Hz range. The case $\zeta=1M$ in particular demonstrates that $\alpha$ acts as a driving parameter that strengthens the instabilities.

For $\zeta=4M$, the oscillations remain sturdy, and when $\alpha>0$ strong peaks reappear. These peaks occur at frequencies similar to those observed in the lower $\zeta$ cases. Unlike Model-I, where large $\zeta$ values almost quenched the oscillations, in Model-II the CoS continues to sustain strong variability even in the high $\zeta$ regime. At $\zeta=5M$, this trend is even clearer. In contrast to Model-I, the effect of the CoS parameter is enhanced, producing very strong PSD peaks even under high quantum correction. As a result, QPOs in Model-II are more continuous, irregular, but also more easily detectable than in Model-I. In addition, Model-II also produces commensurate ratios such as $3:2$ and $2:1$. As is well known, these ratios are consistent with the QPO phenomenology observed in X-ray binaries and AGN systems, showing good agreement between theory and observations.

\section{BH Perturbations of Various Spins} \label{isec5}

BH perturbation theory explores how various test fields evolve on a fixed BH background, offering insights into stability, field dynamics, and possible signatures of beyond-GR physics. Spin-0 (scalar) perturbations serve as a foundational model, governed by the Klein–Gordon equation, and often act as proxies for more complex fields. Spin-1 (EM) perturbations, described by Maxwell's equations on curved spacetime, interact with the background geometry, leading to distinctive field behavior—especially relevant in charged or magnetized BHs. Spin-1/2 (Dirac) perturbations, governed by the curved-space Dirac equation, are less affected by the gravitational potential due to the absence of superradiance in neutral fermions but still probe the influence of spacetime curvature on quantum fields. Each spin sector exhibits unique behavior, yet all are crucial to understanding BH stability, Hawking radiation, and potential extensions to quantum gravity.

For quantum-corrected Letelier BHs, perturbation analysis becomes particularly significant as it reveals how both quantum corrections (parametrized by $\zeta$) and topological defects from the CoS (parametrized by $\alpha$) modify the stability characteristics and dynamical response of the spacetime. The effective potential barriers that govern field propagation encode information about the underlying geometry modifications, providing direct insights into the signatures of quantum gravitational effects and string-theoretic phenomena in astrophysical contexts \cite{64}.

\subsection{Scalar Field Perturbations}

Scalar perturbations are crucial for examining the stability of BH spacetimes under small field fluctuations. These perturbations have been widely studied across various BH solutions in GR and modified gravity, offering key insights into dynamical stability and scalar field behavior in curved geometries. Starting from the massless Klein-Gordon equation, we apply standard BH perturbation techniques to derive a Schrödinger-like wave equation. For detailed discussion of scalar perturbations, readers can see these references \cite{27,28,29,30,31}.

The massless scalar field wave equation is described by the Klein-Gordon equation as follows:
\begin{equation}
\frac{1}{\sqrt{-g}}\,\partial_{\mu}\left(\sqrt{-g}\,g^{\mu\nu}\,\partial_{\nu}\Psi\right)=0\quad\quad (\mu,\nu=0,\cdots,3), \label{ff1}    
\end{equation}
where $\Psi$ is the wave function of the scalar field, $g_{\mu\nu}$ is the covariant metric tensor, $g=\det(g_{\mu\nu})$ is the determinant of the metric tensor, $g^{\mu\nu}$ is the contravariant form of the metric tensor, and $\partial_{\mu}$ is the partial derivative for coordinate systems.

For the spacetime chosen (\ref{metric-0}), we find  
\begin{align}
    g_{\mu\nu}=\left(-f(r),\,1/g(r),\,r^2,\,r^2\,\sin^2 \theta\right),\quad\quad
    g^{\mu\nu}=\left(-1/f(r),\,g(r),\,1/r^{2},\,1/(r^{2}\,\sin^{2} \theta)\right),\quad\quad 
    g=\det (g_{\mu\nu})=-\frac{f(r)}{g(r)}\,r^4\,\sin^2 \theta.\label{ff2}
\end{align}

Let us consider the following scalar field wave function ansatz form:
\begin{equation}
    \Psi(t, r,\theta, \phi)=\exp(-i\,\omega\,t)\,Y^{m}_{\ell} (\theta,\phi)\,R(r)\,\label{ff3}
\end{equation}
where $\omega$ is the temporal frequency (possibly complex), $R(r)$ is a propagating scalar field, and $Y^{m}_{\ell} (\theta,\phi)$ are the spherical harmonics.

Explicitly writing the wave equation (\ref{ff1}) using (\ref{ff2}) and (\ref{ff3}), we find:
\begin{equation}
    \frac{\sqrt{f(r)\,g(r)}}{r^2}\,\partial_r\,\left(r^2\,\sqrt{f(r)\,g(r)}\,\partial_r R(r)\right)+\left(\omega^2-\frac{\ell\,(\ell+1)}{r^2}\,f(r)\right)\,R(r)=0.\label{ff8}
\end{equation}

\textbf{Model-I Analysis}

For Model-I, the metric functions $f(r)$ and $g(r)$ are identical, i.e., $f(r)=g(r)$. This symmetry significantly simplifies the perturbation analysis. Performing the transformation:
\begin{equation}
    R(r)=\frac{\psi}{r},\label{ff9}
\end{equation}
into Eq. (\ref{ff8}), we arrive at the standard Schrödinger-like equation:
\begin{equation}
    \frac{\partial^2 \psi(r_*)}{\partial r^2_{*}}+\left(\omega^2-V_\text{scalar}\right)\,\psi(r_*)=0,\label{ff4}
\end{equation}
where we have performed the tortoise coordinate transformation:
\begin{eqnarray}
    r_*=\int\,\frac{dr}{f}\quad,\quad \partial_{r_{*}}=f\,\partial_r.\label{ff6}
\end{eqnarray}

The scalar perturbative potential for Model-I is given by:
\begin{eqnarray}
V_\text{scalar}(r)&=&\left(\frac{\ell\,(\ell+1)}{r^2}+\frac{f'(r)}{r}\right)\,f(r),\quad \ell\geq 0.\nonumber\\
&=&\left[\frac{\ell\,(\ell+1)}{r^2}+\frac{2M}{r^3} -\frac{2\,\zeta^2}{r^4} \left( \lambda - \frac{2M}{r} \right)^2 + \frac{4\,M\,\zeta^2}{r^5} \left( \lambda - \frac{2M}{r} \right)\right]\,\left[\lambda- \frac{2M}{r}+ \frac{\zeta^2}{r^2}\,\left(\lambda- \frac{2M}{r}\right)^2\right].\label{ff7}
\end{eqnarray}

\begin{figure}[ht!]
    \centering
    \includegraphics[width=0.45\linewidth]{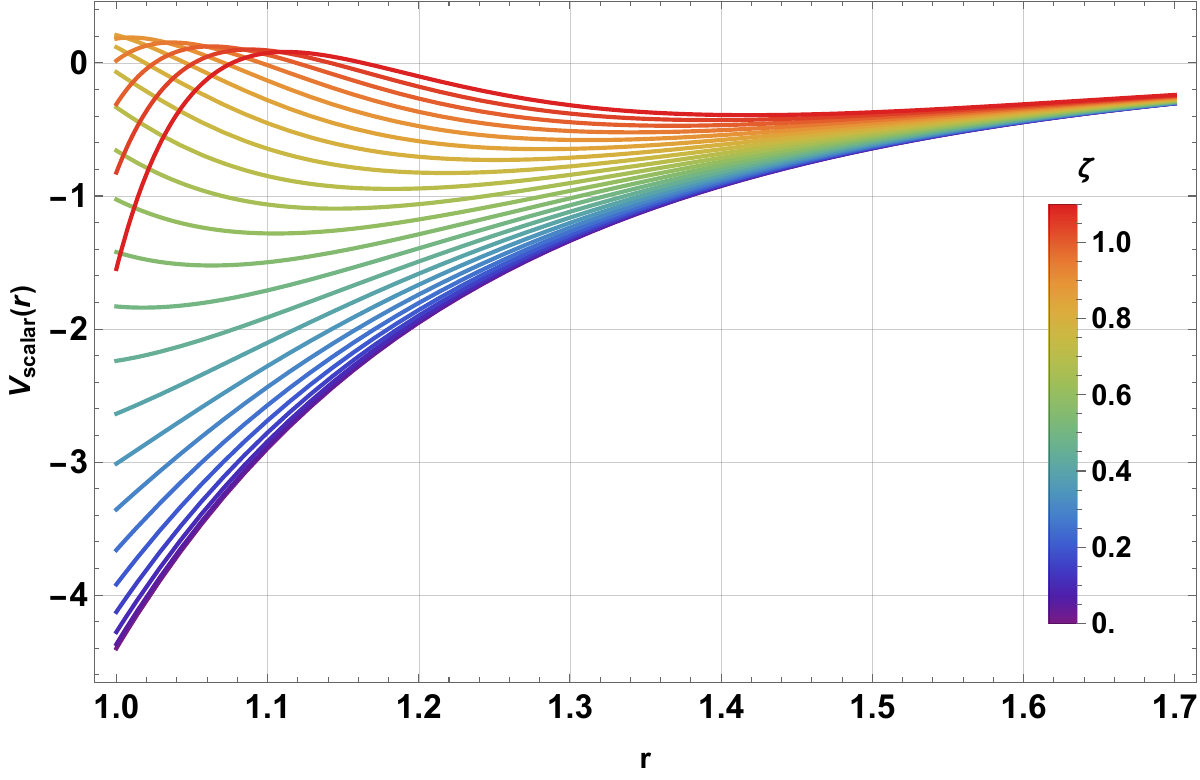}\quad
    \includegraphics[width=0.45\linewidth]{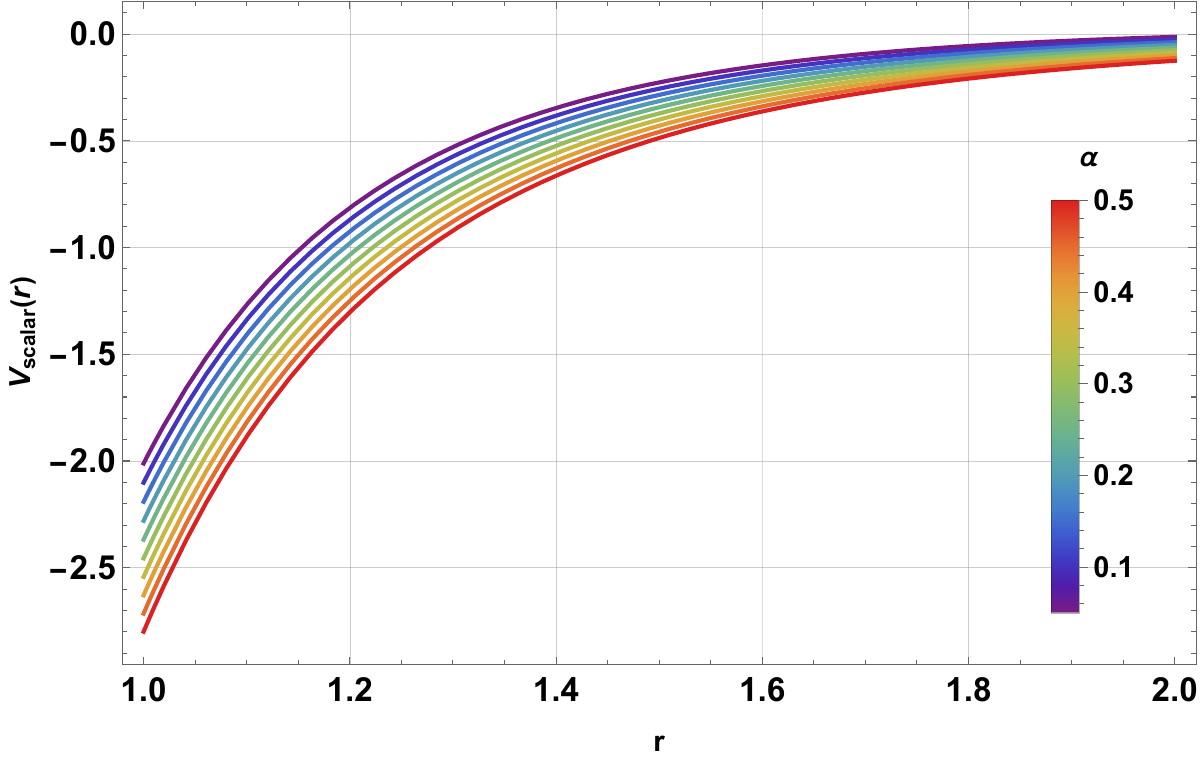}\\
    (i) $\alpha=0.1$  \hspace{6cm} (ii) $\zeta=0.1$
    \caption{\footnotesize Behavior of scalar perturbative potential $V_{\rm scalar}$ for BH Model-I by varying quantum correction $\zeta$ and string parameter $\alpha$. Here $M=1,\,\ell=0$.}
    \label{fig:scalar-potential-1}
\end{figure}

Figure \ref{fig:scalar-potential-1} illustrates the scalar perturbative potential for $\ell=0$ states as a function of $r$ for BH Model-I. The left panel demonstrates that higher values of $\zeta$ produce larger perturbative potentials, indicating that quantum corrections enhance the effective barrier height. This behavior suggests that quantum effects strengthen the gravitational field's response to scalar perturbations. In contrast, the right panel shows that the potential decreases with increasing values of $\alpha$, reflecting how CoS modifies the spacetime geometry in a way that reduces the effective potential barrier.

\begin{figure}[ht!]
    \centering
    \includegraphics[width=0.45\linewidth]{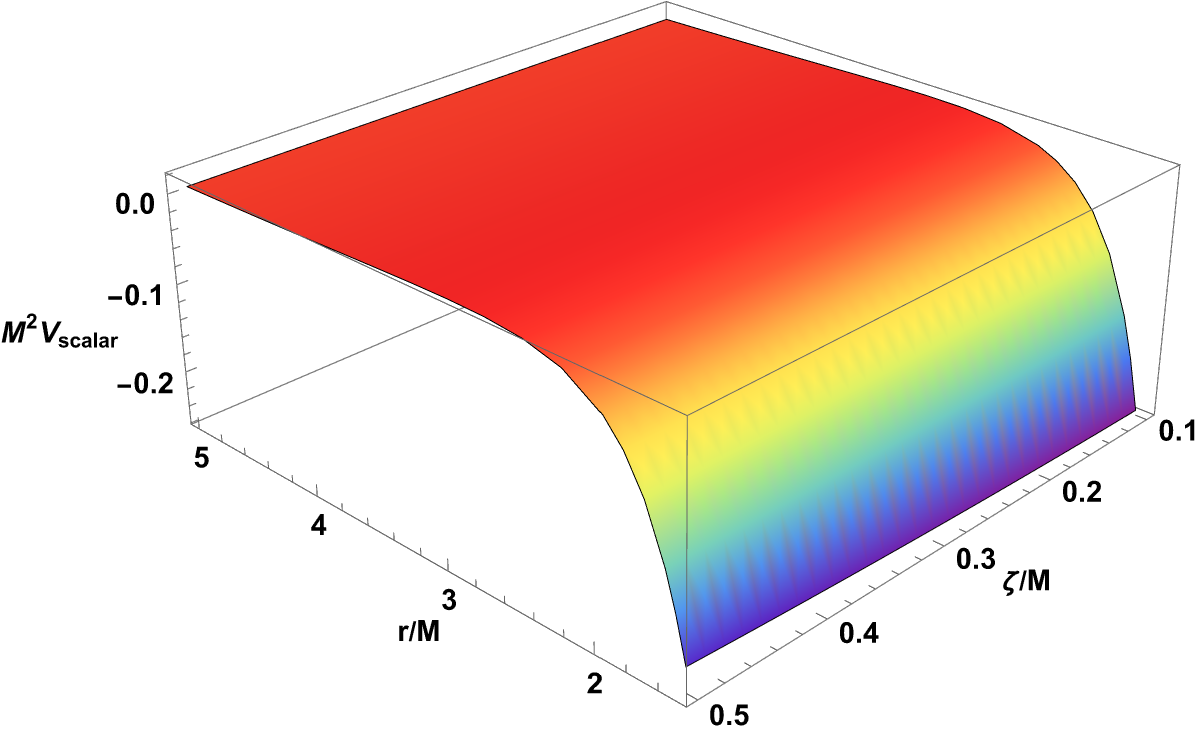}\quad
    \includegraphics[width=0.45\linewidth]{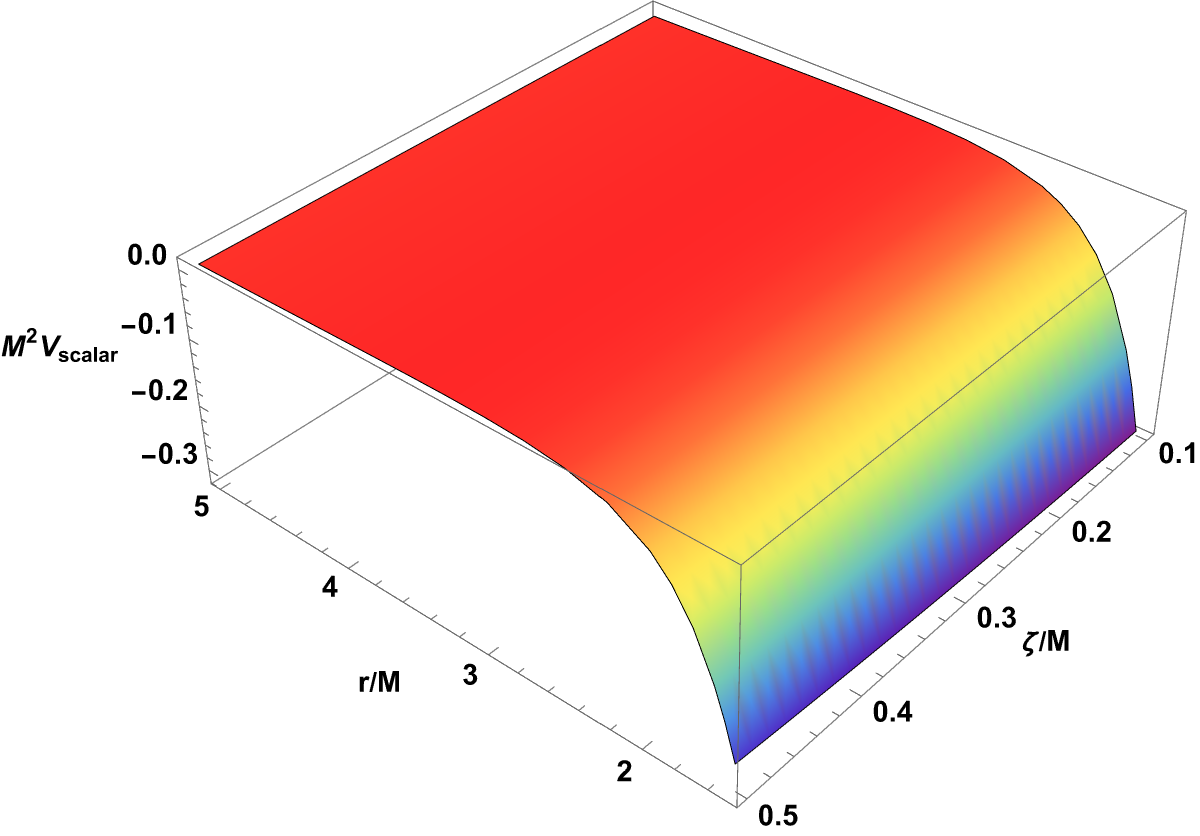}\\
    (i) $\alpha=0.1$  \hspace{6cm} (ii) $\alpha=0.2$
    \caption{\footnotesize Qualitative feature of $M^2\,V_{\rm scalar}$: Three-dimensional plot of $M^2\,V_{\rm scalar}$ for BH Model-I as a function of $(r/M,\zeta/M)$ for two values of $\alpha$. $\ell=0$.}
    \label{fig:scalar-potential-2}
\end{figure}

The three-dimensional visualization in Figure \ref{fig:scalar-potential-2} presents the qualitative features of $M^2\,V_{\rm scalar}$ as a function of $(r/M, \zeta/M)$ for $\alpha=0.1$ and $\alpha=0.2$. These surfaces reveal the complex interplay between quantum corrections and CoS effects, showing how the potential landscape evolves across parameter space. The smooth variation of the potential surface indicates that both parameters continuously contribute to the modification of the scalar field dynamics.

\begin{figure}[ht!]
    \centering
    \includegraphics[width=0.45\linewidth]{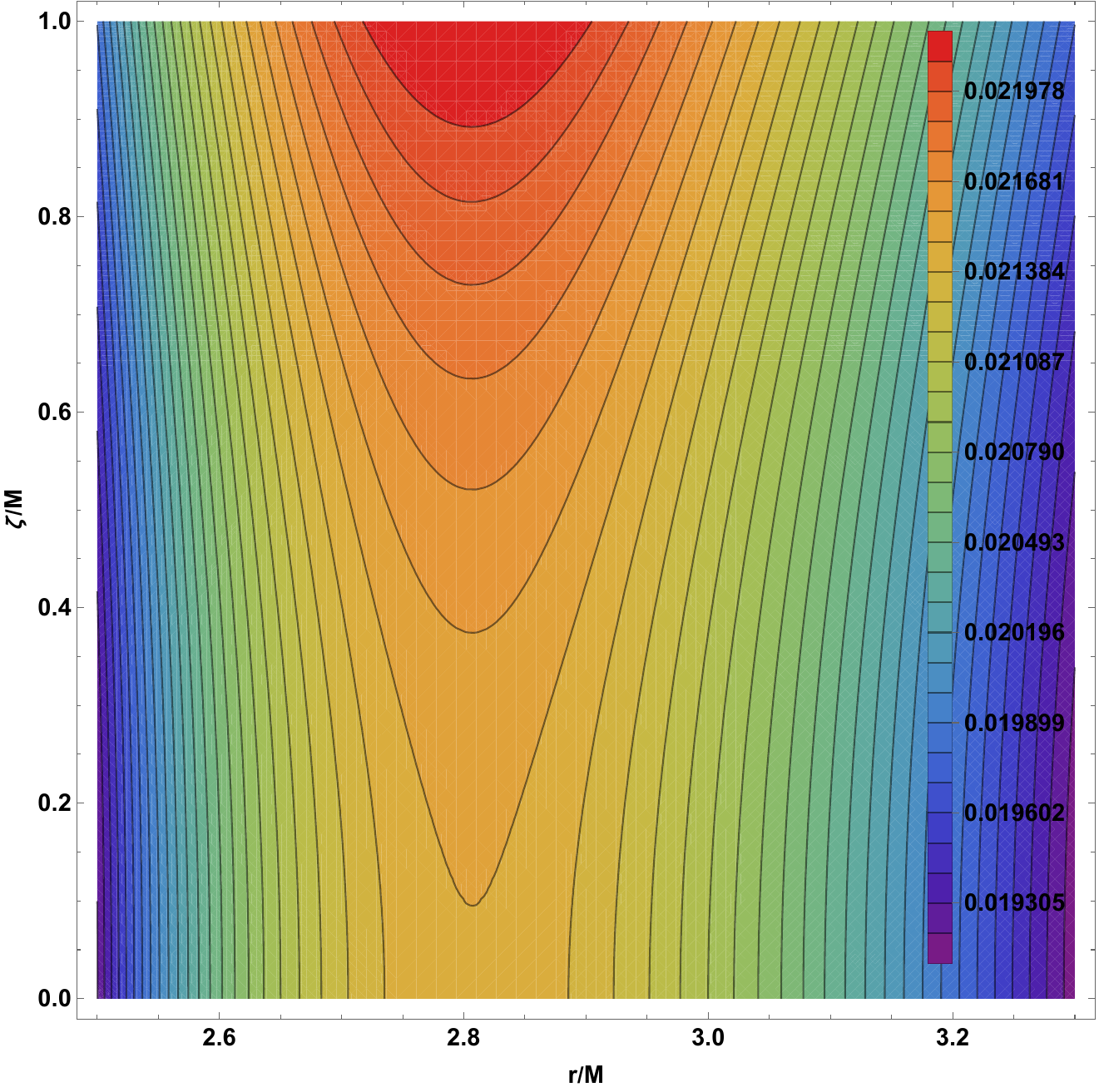}\quad
    \includegraphics[width=0.45\linewidth]{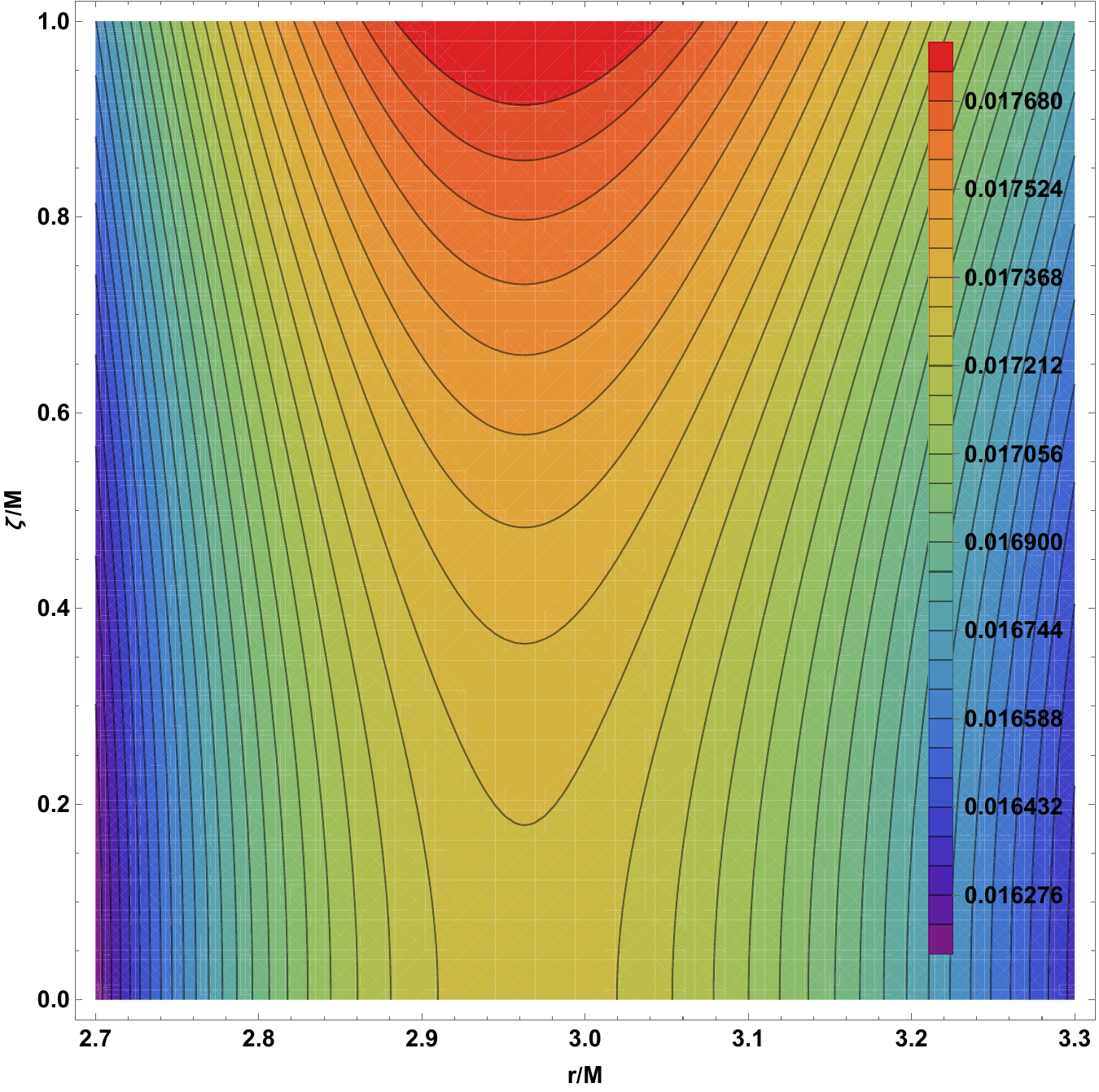}\\
    (i) $\alpha=0.05$  \hspace{6cm} (ii) $\alpha=0.1$
    \caption{\footnotesize Contour plot of $M^2\,V_{\rm scalar}$ in two dimensional plane $(r/M, \zeta/M)$ for BH Model-I for two values of $\alpha$. $\ell=0$.}
    \label{fig:scalar-potential-3}
\end{figure}

Figure \ref{fig:scalar-potential-3} provides contour plots of $M^2\,V_{\rm scalar}$ in the $(r/M, \zeta/M)$ plane for different values of $\alpha$. The contour structure clearly demonstrates that increasing $\alpha$ leads to higher peak values of $M^2\,V_{\rm scalar}$, which can be visualized by following the line-of-sight through the peak regions. This variation provides a clear path for observational discrimination between different values of CoS parameters.

\textbf{Model-II Analysis}

For Model-II, the metric functions have different forms, requiring a modified approach. Performing the transformation:
\begin{equation}
    R(r)=\frac{\psi(r)}{r}\label{ff10}
\end{equation}
and using the tortoise coordinate:
\begin{equation}
    r_{*}=\int\,\frac{dr}{\sqrt{f(r)\,g(r)}},\label{ff11}
\end{equation}
we arrive at the same differential equation form as Eq. (\ref{ff4}), where the scalar perturbative potential is now given by:
\begin{equation}
    V_\text{scalar}(r) =\frac{\ell(\ell+1)}{r^2}\,f(r) + \frac{1}{2\,r}\,\left(f(r)\,g(r)\right)_{,r}.\label{ff12}
\end{equation}

Substituting the Model-II metric functions yields:
\begin{equation}
    V_\text{scalar}(r) =\left[\frac{\ell(\ell+1)}{r^2} +\frac{2\,M}{r^3}+\frac{3\,M\,\zeta^2}{r^5}\,\left(\lambda-\frac{2\,M}{r}\right)-\frac{\zeta^2}{r^4}\,\left(\lambda-\frac{2\,M}{r}\right)^2\right]\,\left(\lambda-\frac{2\,M}{r}\right).\label{ff13}
\end{equation}

Transforming to dimensionless variables $x=r/M$ and $y=\zeta/M$:
\begin{equation}
    M^2\,V_\text{scalar}=\left[\frac{\ell(\ell+1)}{x^2} +\frac{2}{x^3}+\frac{3\,y^2}{x^5}\,\left(\lambda-\frac{2}{x}\right)-\frac{y^2}{x^4}\,\left(\lambda-\frac{2}{x}\right)^2\right]\,\left(\lambda-\frac{2}{x}\right).\label{ff14}
\end{equation}

In the limit $y \to 0$, this result reduces to the Letelier BH solution, confirming the consistency of our approach.

\begin{figure}[ht!]
    \centering
    \includegraphics[width=0.45\linewidth]{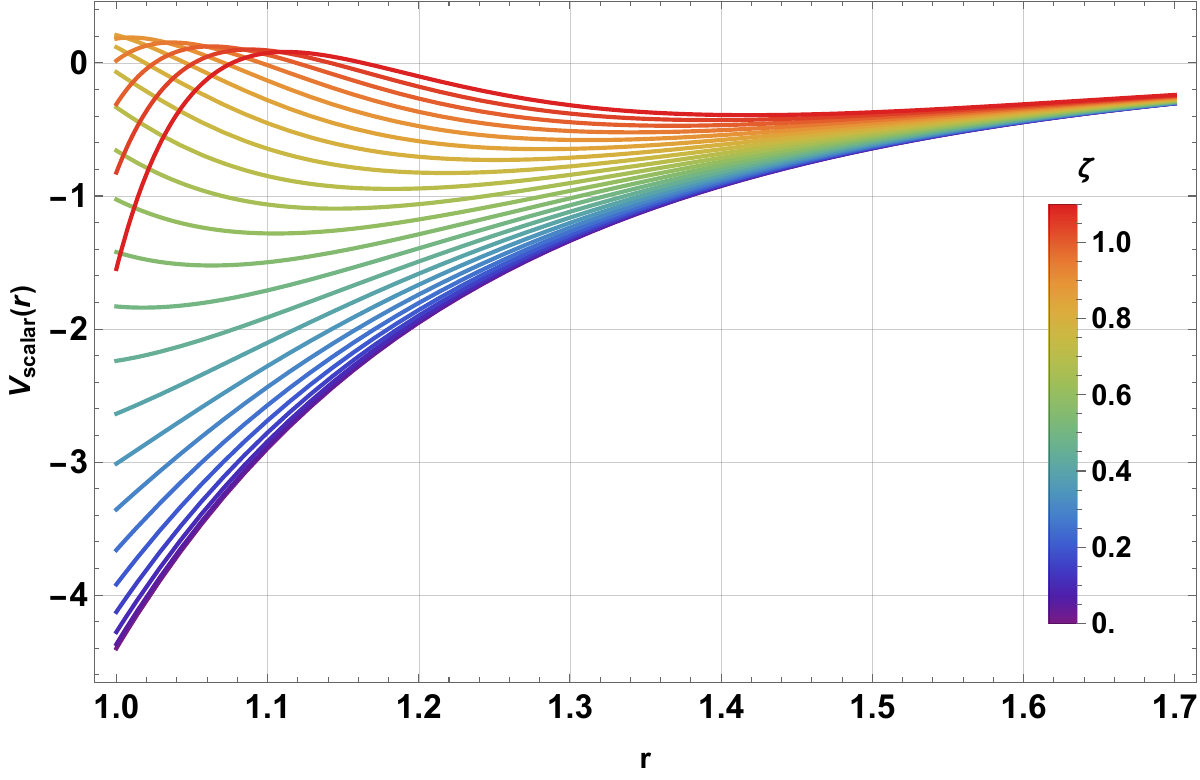}\quad
    \includegraphics[width=0.45\linewidth]{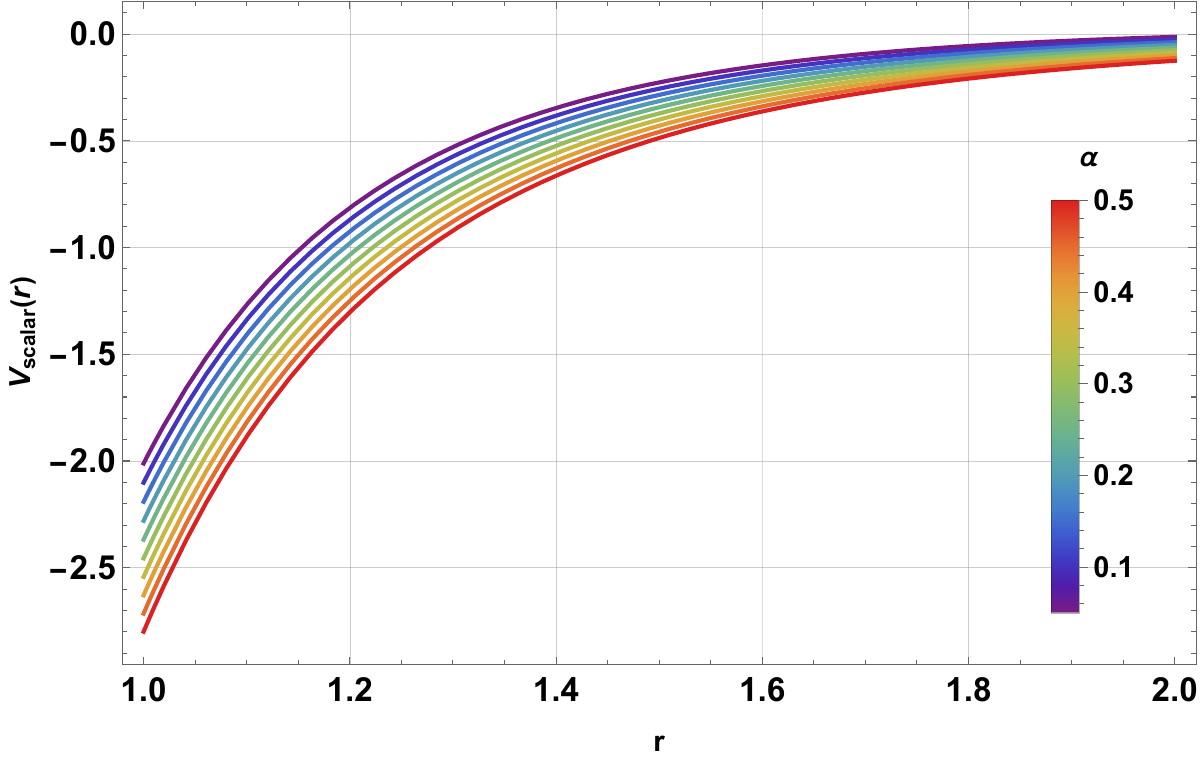}\\
    (i) $\alpha=0.1$  \hspace{6cm} (ii) $\zeta=0.1$
    \caption{\footnotesize Behavior of scalar perturbative potential $V_{\rm scalar}$ for BH Model-II by varying quantum correction $\zeta$ and string parameter $\alpha$. Here $M=1,\,\ell=0$.}
    \label{fig:scalar-potential-4}
\end{figure}

\begin{figure}[ht!]
    \centering
    \includegraphics[width=0.45\linewidth]{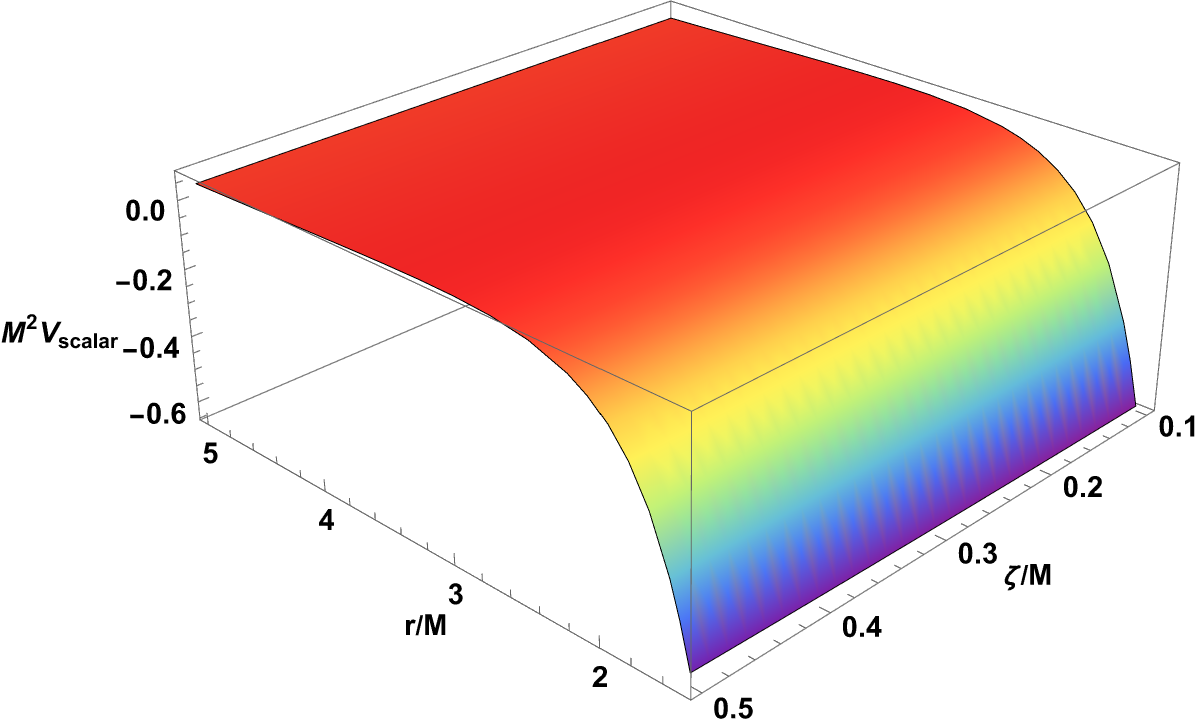}\quad
    \includegraphics[width=0.45\linewidth]{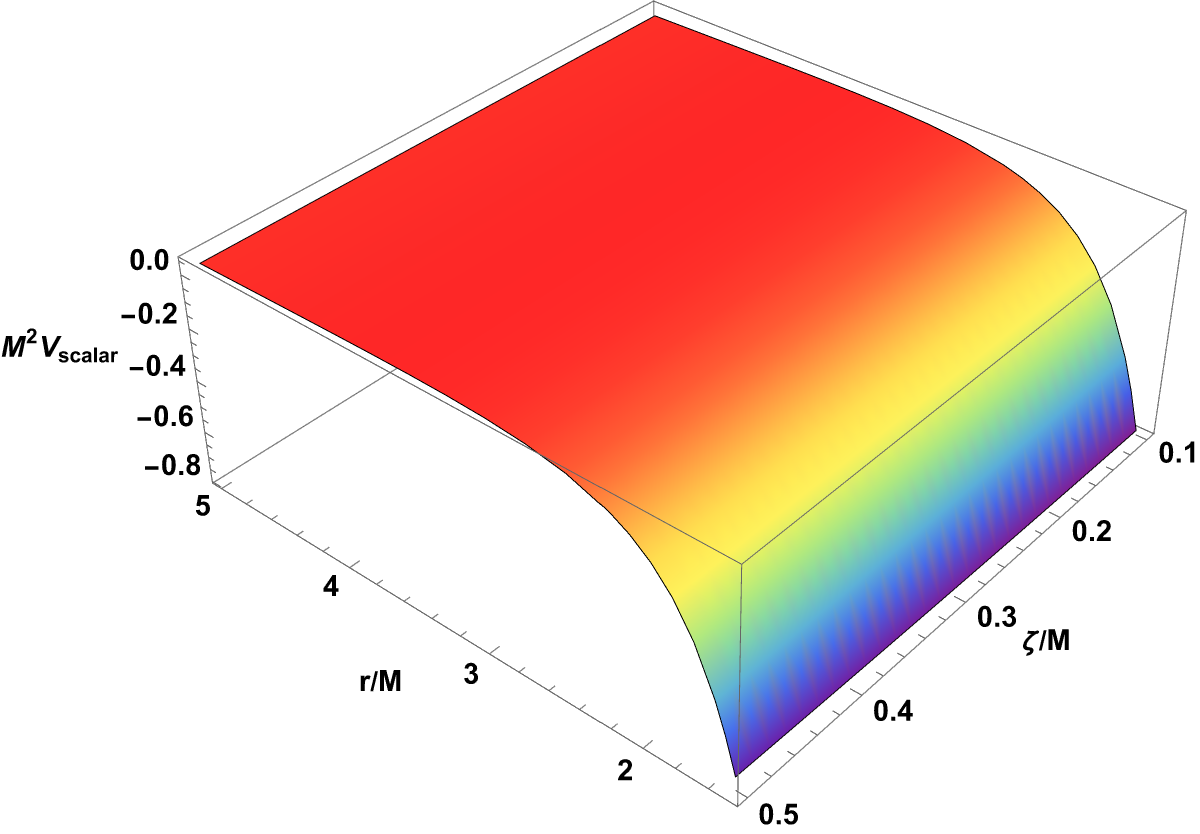}\\
    (i) $\alpha=0.1$  \hspace{6cm} (ii) $\alpha=0.2$
    \caption{\footnotesize Qualitative feature of $M^2\,V_{\rm scalar}$: Three-dimensional plot of $M^2\,V_{\rm scalar}$ for BH Model-II as a function of $(r/M,\zeta/M)$ for two values of $\alpha$. $\ell=0$.}
    \label{fig:scalar-potential-5}
\end{figure}

Figure \ref{fig:scalar-potential-4} shows the behavior of the perturbative scalar potential for Model-II, which exhibits trends similar to Model-I but with quantitative differences arising from the asymmetric quantum corrections. Figure \ref{fig:scalar-potential-5} presents the three-dimensional features of $M^2\,V_{\rm scalar}$ for Model-II, revealing distinct potential landscapes that could serve as observational discriminants between the two quantum-corrected models.

\subsection{EM Perturbations}

EM perturbations describe test EM fields in fixed BH backgrounds and are instrumental in probing BH stability, spectroscopy, and interactions with external fields. They have applications in accretion physics, jet formation, and plasma dynamics near BHs, and they also serve as analogs for gravitational wave behavior in perturbation analysis. See \cite{70,71,72} for a detailed discussion on this topic.

For EM perturbations, the dynamics are governed by Maxwell's equations in curved spacetime:
\begin{equation}
\frac{1}{\sqrt{-g}}\left[F_{\alpha \beta }\,g^{\alpha \nu}\,g^{\beta \mu }\sqrt{-g}\Psi\right]_{,\mu}=0,  \label{em1}
\end{equation}
where $F_{\alpha \beta }=\partial _{\alpha }A_{\beta }-\partial _{\beta}A_{\nu}$ is the EM tensor.

Following the approach adopted in \cite{73,74}, we find:
\begin{equation}
    \frac{\partial^2 \psi_\text{em}(r_*)}{\partial r^2_{*}}+\left(\omega^2-V_\text{em}\right)\,\psi_\text{em}(r_*)=0,\label{em2}
\end{equation}
where the EM perturbative potential is given by:
\begin{align}
V_\text{em}(r)=
\begin{cases}
    \displaystyle \frac{\ell\,(\ell+1)}{r^2}\,\left[\lambda- \frac{2M}{r}+ \frac{\zeta^2}{r^2}\,\left(\lambda- \frac{2M}{r}\right)^2\right], & \text{ Model-I }\\
    \displaystyle \frac{\ell\,(\ell+1)}{r^2}\,\sqrt{\left\{\lambda- \frac{2M}{r}+ \frac{\zeta^2}{r^2}\,\left(\lambda- \frac{2M}{r}\right)^2\right\}\,\left(\lambda- \frac{2M}{r}\right)}, & \text{ Model-II }.\label{em3}
\end{cases}
\end{align}
where $\ell\geq 1$.

Transforming to dimensionless variables $x=r/M$ and $y=\zeta/M$:
\begin{align}
M^2\,V_\text{em}=
\begin{cases}
    \displaystyle \frac{\ell\,(\ell+1)}{x^2}\,\left[\lambda- \frac{2}{x}+ \frac{y^2}{x^2}\,\left(\lambda- \frac{2}{x}\right)^2\right], & \text{ Model-I }\\
    \displaystyle \frac{\ell\,(\ell+1)}{x^2}\,\sqrt{\left\{\lambda- \frac{2}{x}+ \frac{y^2}{x^2}\,\left(\lambda- \frac{2}{x}\right)^2\right\}\,\left(\lambda- \frac{2}{x}\right)}, & \text{ Model-II }.\label{em4}
\end{cases}
\end{align}

\begin{figure}[ht!]
    \centering
    \includegraphics[width=0.45\linewidth]{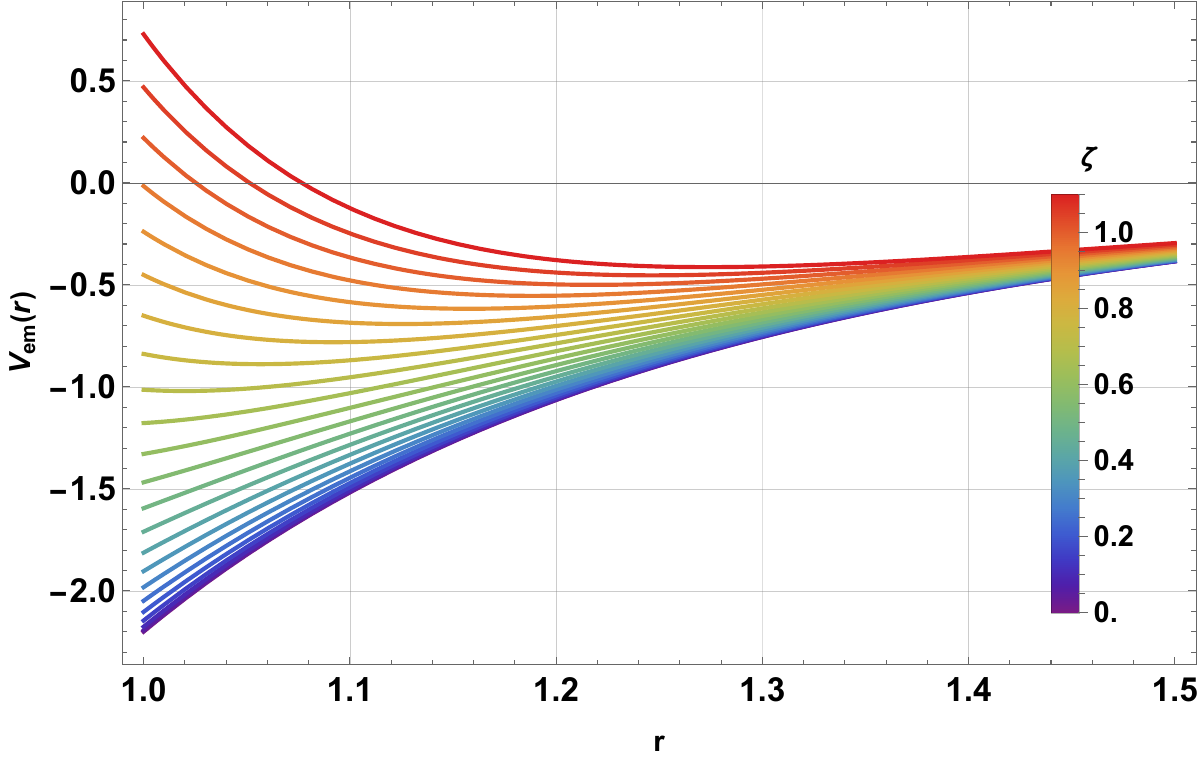}\quad
    \includegraphics[width=0.45\linewidth]{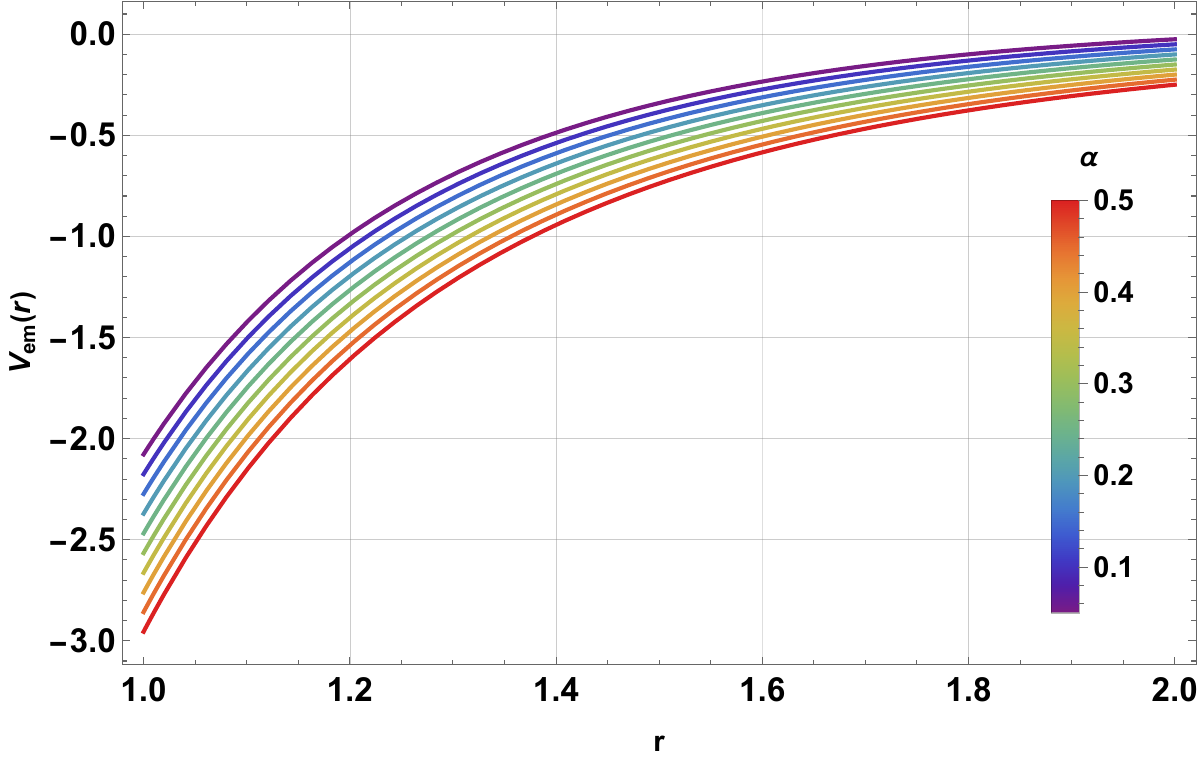}\\
    (i) $\alpha=0.1$  \hspace{6cm} (ii) $\zeta=0.1$
    \caption{\footnotesize Behavior of EM perturbative potential $V_{\rm em}$ for BH Model-I by varying quantum correction $\zeta$ and string parameter $\alpha$. Here $M=1$.}
    \label{fig:em-potential-1}
\end{figure}

Figure \ref{fig:em-potential-1} illustrates the EM perturbative potential for the states $\ell=1$ in Model-I. The left panel demonstrates that higher values of $\zeta$ enhance the perturbative potential, similar to the scalar case but with distinct quantitative characteristics arising from the spin-1 nature of EM fields. The right panel shows that increasing $\alpha$ reduces the potential, indicating that the CoS affects EM perturbations differently compared to scalar perturbations.

\begin{figure}[ht!]
    \centering
    \includegraphics[width=0.45\linewidth]{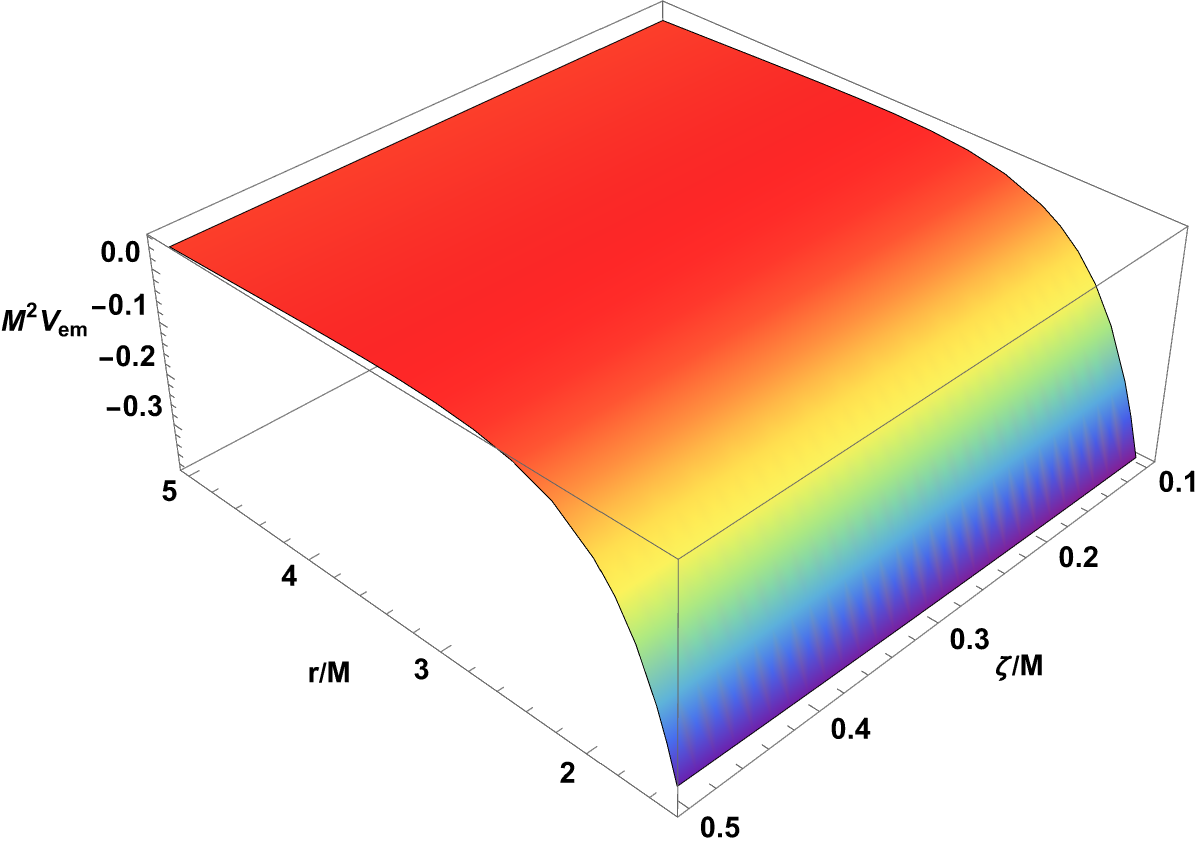}\quad
    \includegraphics[width=0.45\linewidth]{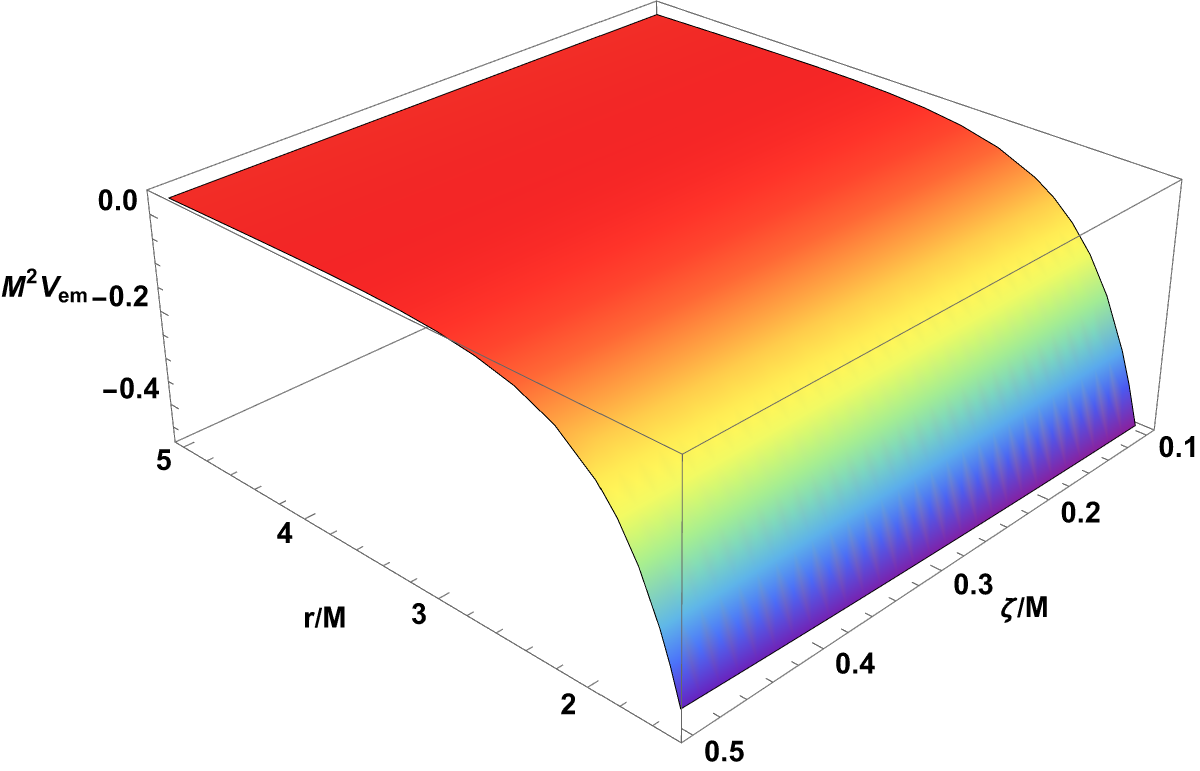}\\
    (i) $\alpha=0.1$  \hspace{6cm} (ii) $\alpha=0.1$
    \caption{\footnotesize Qualitative feature of $M^2\,V_{\rm em}$: Three-dimensional plot of $M^2\,V_{\rm em}$ for BH Model-I as a function of $(r/M,\zeta/M)$ for two values of $\alpha$. $\ell=1$.}
    \label{fig:em-potential-2}
\end{figure}

The three-dimensional plots in Figure \ref{fig:em-potential-2} reveal the complex parameter dependence of $M^2\,V_{\rm em}$ for Model-I, showing how quantum corrections and CoS effects combine to modify the propagation characteristics of the EM field. Smooth surfaces indicate continuous parameter dependence, which is crucial for observational parameter estimation.

\begin{figure}[ht!]
    \centering
    \includegraphics[width=0.45\linewidth]{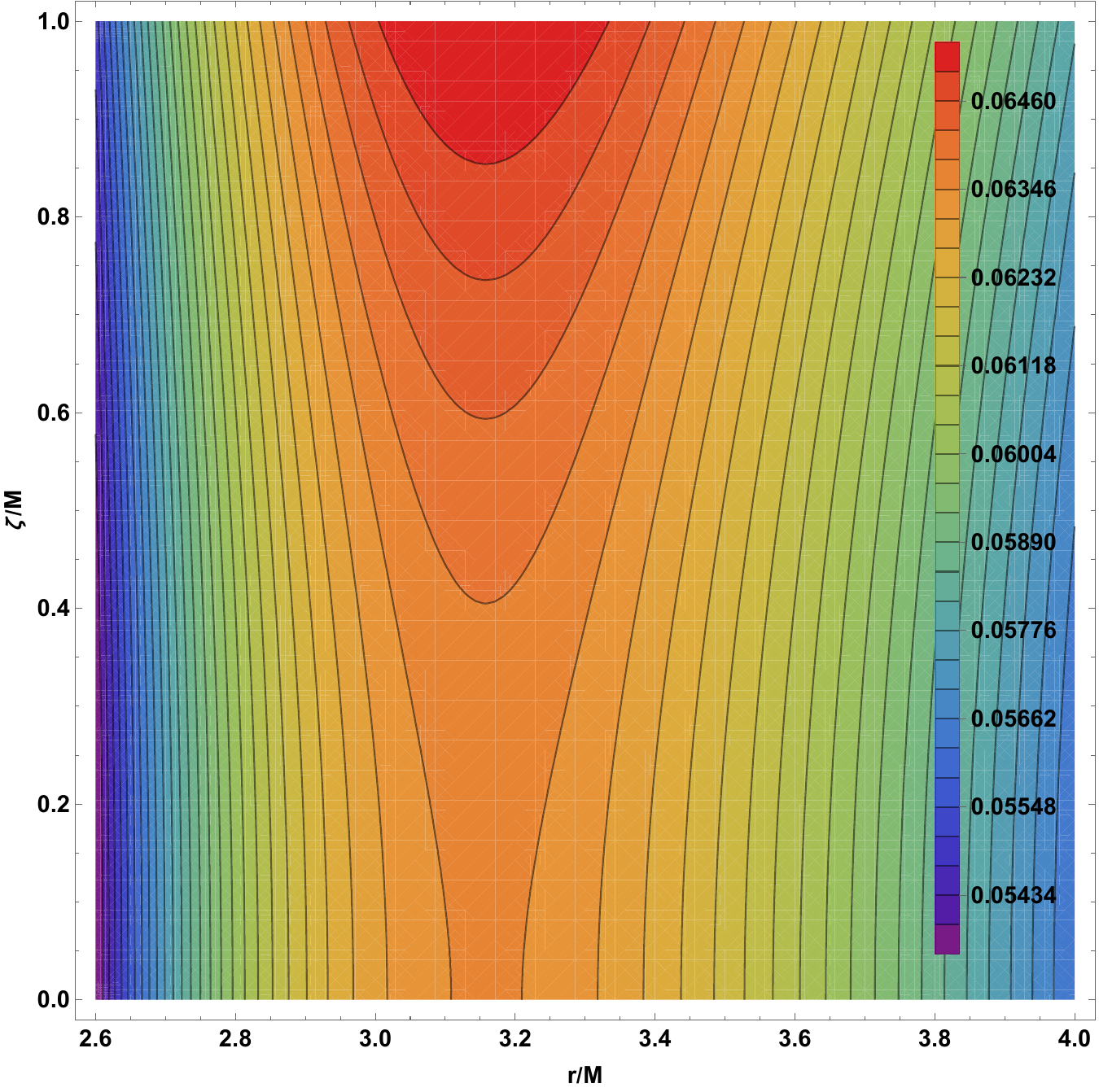}\quad
    \includegraphics[width=0.45\linewidth]{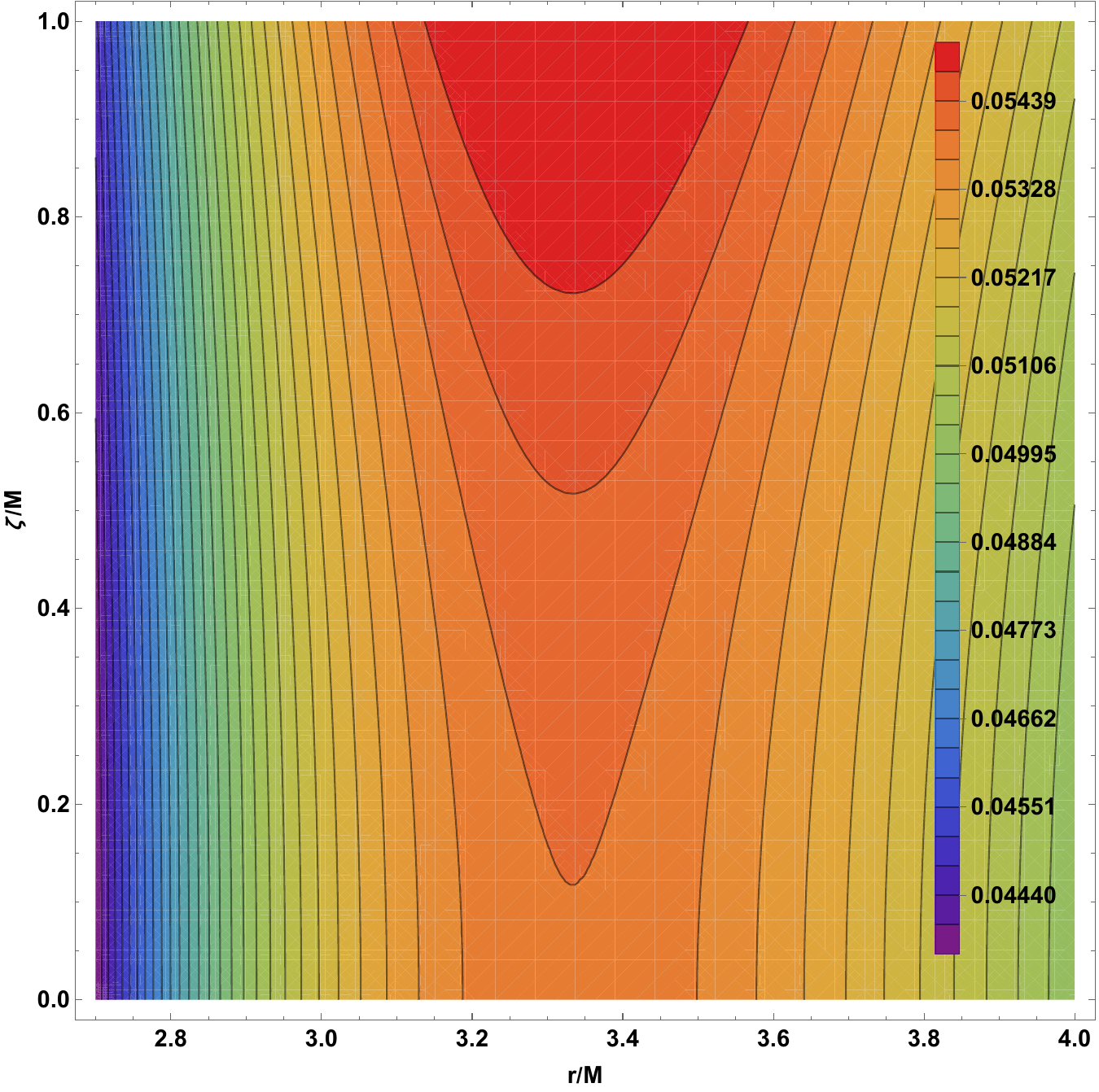}\\
    (i) $\alpha=0.05$  \hspace{6cm} (ii) $\alpha=0.1$
    \caption{\footnotesize Contour plot of $M^2\,V_{\rm em}$ in two dimensional plane $(r/M, \zeta/M)$ for BH Model-I for two values of $\alpha$. $\ell=1$.}
    \label{fig:em-potential-3}
\end{figure}

Figure \ref{fig:em-potential-3} presents contour plots that clearly demonstrate the variation of the EM potential with the model parameters. The increasing peak values with larger $\alpha$ provide another avenue for observational constraints, complementing the scalar perturbation analysis and offering multiple independent probes of the quantum-corrected geometry.

\section{BH Thermodynamics and Topological Characterization} \label{isec6}

BH thermodynamics bridges GR, quantum field theory, and statistical mechanics, revealing profound insights into the nature of spacetime and gravity. The foundational laws were established through the analogy between the classical mechanics of BHs and thermodynamics, initiated by Bekenstein, who proposed that BHs possess entropy proportional to their horizon area \cite{75}, and Hawking, who showed that BHs emit thermal radiation with a temperature proportional to their surface gravity \cite{76}. These discoveries led to the formulation of the four laws of BH mechanics \cite{77}, where the surface gravity corresponds to temperature, and the horizon area to entropy. The first law, relating changes in mass, angular momentum, charge, and area, mirrors the energy conservation principle. BHs are now understood as thermodynamic systems, with properties like heat capacity and entropy playing central roles in gravitational dynamics. Later developments, including HP phase transitions \cite{78} and applications within the AdS/CFT correspondence \cite{79}, deepened the link between BH physics and quantum field theory. Understanding BH thermodynamics remains essential for any consistent theory of quantum gravity.

\subsection{Corrected Thermodynamics and Phase Transitions}\label{sec:exp-entropy}

Following \cite{80,81}, we adopt the universal non-perturbative correction \cite{82},
\begin{equation}
	S=S_0+\eta\,e^{-S_0},\qquad S_0=\frac{A}{4}=\pi r_h^2,
	\label{eq:Scorr}
\end{equation}
where $\eta$ is a dimensionless control parameter.
Thus,
\begin{equation}
	S(r_h)=\pi r_h^2+\eta\,e^{-\pi r_h^2},\qquad 
	\frac{\mathrm d S}{\mathrm d r_h}=2\pi r_h\Big(1-\eta\,e^{-\pi r_h^2}\Big).
	\label{eq:Sprime}
\end{equation}

The surface gravity for a static metric of the form above is 
$\kappa=\tfrac12\sqrt{f'(r_h)\,g'(r_h)}$, so $T=\kappa/(2\pi)$.
Using $h(r_h)=0$, we have
\begin{align}
	\text{Model I: } 
	& f(r)=h+\frac{\zeta^2}{r^2}h^2 
	\;\Rightarrow\; f'(r_h)=h'(r_h)=\frac{2M}{r_h^2},\\
	\text{Model II: } 
	& f'(r_h)=h'(r_h)=\frac{2M}{r_h^2},\quad
	g'(r_h)=h'(r_h)=\frac{2M}{r_h^2},
\end{align}
since all terms proportional to $h$ or $h^2$ vanish at $r=r_h$.
Therefore for both models we have,
\begin{equation}
	T=\frac{1}{4\pi}h'(r_h)=\frac{1}{4\pi}\frac{2M}{r_h^2}
	=\frac{\lambda}{4\pi r_h}
	=\frac{\lambda^2}{8\pi M},
	\label{eq:T-final}
\end{equation}
which coincides with the Letelier result (\(\zeta\)-independent) once the outer horizon is chosen.

As in \cite{81}, we define the (quantum-corrected) internal energy by integrating the first law with the original Hawking temperature and corrected entropy:
\begin{equation}
	E(r_h)=\int T\,\mathrm d S.
\end{equation}
Using \eqref{eq:Sprime} and \eqref{eq:T-final}:
\begin{align}
	T\,\frac{\mathrm d S}{\mathrm d r_h}
	=\frac{\lambda}{4\pi r_h}\,\cdot\,2\pi r_h\Big(1-\eta e^{-\pi r_h^2}\Big)
	=\frac{\lambda}{2}\Big(1-\eta e^{-\pi r_h^2}\Big),
\end{align}
so, choosing $E(0)=0$,
\begin{equation}
	E(r_h)=\frac{\lambda}{2}\,r_h-\frac{\lambda\eta}{4}\,\mathrm{erf}\!\Big(\sqrt{\pi}\,r_h\Big),
	\qquad 
	\mathrm{erf}(x)=\frac{2}{\sqrt{\pi}}\int_0^x e^{-t^2}\mathrm dt.
	\label{eq:Eofr}
\end{equation}
In the classical limit $\eta\to0$, $E=\frac{\lambda}{2}r_h=M$ (since $r_h=\tfrac{2M}{\lambda}$), as expected. 
This construction is exactly the strategy used in \cite{81}, adapted to our asymptotically flat setup.

We can write the specific heat at fixed $(\alpha,\zeta)$ as
\begin{equation}
	C=T\left(\frac{\partial S}{\partial T}\right)_{\alpha,\zeta}
	= T\,\frac{\mathrm d S/\mathrm d r_h}{\mathrm d T/\mathrm d r_h}.
\end{equation}
From \eqref{eq:T-final}, $\frac{\mathrm d T}{\mathrm d r_h}=-\frac{\lambda}{4\pi r_h^2}$.
Using \eqref{eq:Sprime}, one finds for both models
\begin{equation}
	C(r_h)=-\,2\pi r_h^2\,\Big(1-\eta e^{-\pi r_h^2}\Big).
	\label{eq:Cfinal}
\end{equation}
Equivalently, in terms of $M$,
\begin{equation}
	C(M)=-\,\frac{8\pi M^2}{\lambda^2}\,
	\Big(1-\eta e^{-\frac{4\pi M^2}{\lambda^2}}\Big).
\end{equation}
Thus the Schwarzschild/Letelier instability ($C<0$) persists, but the non-perturbative correction reduces $|C|$ for $\eta>0$, consistent with the stability trends reported in the AdS analysis of \cite{81}.

With \eqref{eq:Eofr}, \eqref{eq:T-final} and \eqref{eq:Scorr}, the Helmholtz free energy $F=E-TS$ becomes
\begin{align}
	F(r_h) 
	&= \left[\frac{\lambda}{2}\,r_h-\frac{\lambda\eta}{4}\,\mathrm{erf}\!\Big(\sqrt{\pi}\,r_h\Big)\right]
	-\left[\frac{\lambda}{4\pi r_h}\right]\left[\pi r_h^2+\eta e^{-\pi r_h^2}\right] \notag\\
	&= \frac{\lambda}{4}\,r_h
	-\frac{\lambda\eta}{4}\,\mathrm{erf}\!\Big(\sqrt{\pi}\,r_h\Big)
	-\frac{\lambda\eta}{4\pi r_h}\,e^{-\pi r_h^2}.
	\label{eq:Fofr}
\end{align}
Equivalently in terms of $M$:
\begin{equation}
	F(M)=\frac{M}{2}
	-\frac{\lambda\eta}{4}\,\mathrm{erf}\!\left(\frac{2\sqrt{\pi}\,M}{\lambda}\right)
	-\frac{\lambda^2\eta}{8\pi M}\,e^{-\frac{4\pi M^2}{\lambda^2}}.
\end{equation}
The setting $\eta\to0$ reproduces the classical $F_0=\tfrac{M}{2}$. 
This mirrors the $E-TS$ construction used in \cite{81} (where in AdS/extended phase space).

Because $g(r)=h+\zeta^2 r^{-2}h^2$ in both models and the outer (event) horizon satisfies $h(r_h)=0$, all $\zeta$-dependent parts in $f'(r_h)$ and $g'(r_h)$ vanish (they carry at least one factor of $h$). Hence $T\propto h'(r_h)$, $S\propto r_h^2$ and all thermodynamic quantities derived from $T$ and $S$ coincide at the outer horizon for Model I and II. 
This does not mean that the spacetimes are indistinguishable: photon spheres, lensing, and ISCOs differ away from the horizon, as shown in earlier sections.
For $r_h \gg 1$, $S \simeq \pi r_h^2$ and the corrections in (\ref{eq:Cfinal}), (\ref{eq:Fofr}) are exponentially suppressed. For quantum-sized holes $r_h \sim O(1)$, the reduction in entropy $\Delta S \simeq \eta e^{-\pi r_h^2}$ weakens $|C|$ and lowers $F$ by the last two (negative) terms in (\ref{eq:Fofr}).

The exponentially corrected entropy (\ref{eq:Scorr}), admits a natural holographic reading as a non-perturbative correction to the density of states / partition function of the boundary theory. In AdS/CFT analyzes, such $e^{-S_0}$ effects are interpreted as contributions of subdominant Euclidean saddles (e.g., modular images / Rademacher-type terms \cite{Rademacher}, or non-perturbative brane/instanton sectors) to the bulk path integral, and hence to the dual CFT partition function $Z(\beta)$.

In the canonical ensemble, the boundary (dual) free energy $F$ governs
\begin{equation}
\log Z(\beta) = -\beta F(\beta), \quad \beta =\frac{1}{T} =\frac{8\pi M}{\lambda^2} =\frac{4\pi r_h}{\lambda}.
\label{eq:logZ}
\end{equation}
Using the bulk result (\ref{eq:Fofr}) we can split $F = F_0 + \Delta_\eta F$ with
\begin{equation}
F_0 = \frac{\lambda}{4}\,r_h, \quad \Delta_\eta F = -\frac{\lambda\eta}{4}\,\mathrm{erf}\!\Big(\sqrt{\pi}\,r_h\Big)
-\frac{\lambda\eta}{4\pi r_h}\,e^{-\pi r_h^2}.
\label{eq:Fsplit}
\end{equation}
Thus, the dual partition function factorizes as
\begin{equation}
Z(\beta) = Z_0(\beta) \exp\left[- \beta \Delta_\eta F(\beta)\right], \quad Z_0(\beta) = \exp\left[- \beta F_0\right],
\label{eq:Zfactor}
\end{equation}
displaying the non-perturbative correction driven by the same $e^{-S_0}$ structure that enters the bulk entropy. This mirrors the AdS analysis in which $S = S_0 + \eta e^{-S_0}$is implemented at fixed $T$ and then propagated to all thermodynamics via $F = E - T S$ and $\log Z = -\beta F$.

At the level of the microcanonical density of states $\rho(E) \sim e^{S(E)}$, our result implies
\begin{equation}
\rho(E) \simeq e^{S_0(E)}\left[1 + \eta e^{-S_0(E)} + O\left(e^{-2S_0}\right)\right],
\label{eq:rho}
\end{equation}
so the leading correction is non-perturbative in $S_0$ (and hence in $1/G_N$ or large-$N$): in holographic theories where $S_0 \sim O(N^2)$ (or $\sim c$ in 2D), this corresponds to effects of order $e^{-O(N^2)}$, beyond any $1/N$ expansion. In AdS$_3$/CFT$_2$, this may be understood in terms of modular sums (Rademacher expansion) that generate exponentially suppressed images of the dominant saddle.

Thermal one-point functions and susceptibilities on the boundary follow from bulk thermodynamics. Writing $F = F_0 + \Delta_\eta F$ and using $\beta = 4\pi r_h/\lambda$,
\begin{equation}
	\chi_T \equiv -\frac{\partial^2 \log Z}{\partial \beta^2}
	= \frac{\partial}{\partial \beta}\Big(\beta^2 C\Big) 
	= \chi_T^{(0)} 
	+ \underbrace{\beta^2\,\Delta_\eta C
		+ 2\beta\,C_0\,\frac{\partial (\Delta_\eta F)}{\partial E}
		+\cdots}_{\text{\small non-perturbative $e^{-S_0}$}},
\end{equation}
where $C$ is the bulk specific heat. Using Eq. (\ref{eq:Cfinal}) the boundary specific heat density inherits the same exponential suppression. For $\eta > 0$ the correction reduces $|C|$, i.e. softens the canonical instability of the Schwarzschild/Letelier ensemble, in qualitative accord with the AdS case.

If one embeds the solution in AdS with scale $L$ (or places it in a reflecting cavity) so that a canonical ensemble is well-defined, the Gibbs free energy $G = F + P V$ on the bulk side maps to the grand potential of the boundary theory. The $\eta$-dependent $\Delta_\eta F$ then induces an $O\left(e^{-S_0}\right)$ shift of the HP temperature $T_{\text{HP}}$:
\begin{equation}
\delta T_{\text{HP}} \simeq -\frac{\partial_T \Delta_\eta F}{\partial_T^2(F_0 - G_{\text{th}})}\bigg|_{T = T_{\text{HP}}^{(0)}} \propto \eta e^{-S_0(T_{\text{HP}}^{(0)})},
\label{eq:deltaT}
\end{equation}
with $G_{\text{th}}$ the thermal AdS (or cavity vacuum) free energy. Although our Models I/II are asymptotically flat, this formula clarifies how the same non-perturbative structure would correct boundary phase transitions after an AdS uplift, consistent with the methodology used in the AdS study.

Because the outer horizon is set by $h(r_h) = 0$, the leading holographic thermodynamics encoded in $(T, S, F, C)$ and hence in $\log Z(\beta)$ is identical for Models I/II (the $\zeta$-dependence drops out at $r_h$). However, holographic dynamical observables sensitive to the bulk away from the horizon (e.g. perturbation spectra and transport coefficients extracted from retarded correlators) will differ between the two models through their distinct $f(r)$ profiles, despite having the same $Z(\beta)$ at equilibrium.

When an AdS uplift or a reflecting cavity is used, the canonical ensemble becomes meaningful and the global transition is controlled by the appropriate thermodynamic potential:
\begin{equation}
G(T) = F + P V \quad (\text{AdS}), \quad \text{or} \quad F(T) \quad (\text{finite cavity}).
\label{eq:potential}
\end{equation}
In AdS, the uncharged static BH exhibits the HP transition: at low $T$ the thermal (radiation) background dominates; at high $T$ the large BH dominates. This is a first-order transition, characterized by a discontinuity in entropy (latent heat) at $T = T_{\text{HP}}$, not by a divergence of $C$.

Let $G_0(T)$ denote the classical ($\eta = 0$) Gibbs free energy of the AdS BH and $G_{\text{th}}(T)$ the thermal AdS free energy. Let $\Delta_\eta F$ be the $\eta$-dependent correction in Eq. (\ref{eq:Fsplit}); the corrected Gibbs free energy is $G = G_0 + \Delta_\eta G$ with $\Delta_\eta G = \Delta_\eta F$ at fixed pressure.\footnote{At the outer horizon the $\zeta$-dependence drops out of $T$ and the $S$-based correction, so the HP shift is $\zeta$-independent at leading order.}
The HP point solves $G(T_{\text{HP}}) = G_{\text{th}}(T_{\text{HP}})$. Linearizing around the classical value $T_{\text{HP}}^{(0)}$ gives
\begin{equation}
\delta T_{\text{HP}} \equiv T_{\text{HP}} - T_{\text{HP}}^{(0)} \simeq -\frac{\partial_T \Delta_\eta G}{\partial_T (G_0 - G_{\text{th}})}\bigg|_{T = T_{\text{HP}}^{(0)}} \propto \eta e^{-S_0\left(T_{\text{HP}}^{(0)}\right)}.
\label{eq:HP-shift}
\end{equation}

For $\eta > 0$ our explicit $\Delta_\eta F < 0$ (Eq. (\ref{eq:Fsplit})), so $G$ decreases at fixed $T$; hence the BH phase becomes globally preferred at a slightly lower temperature and $\delta T_{\text{HP}} < 0$. The shift is non-perturbative (exponentially small in the large-$S_0$ limit), matching the AdS intuition in \cite{Pourhassan:NPB2022}.

The HP transition is first-order; the latent heat equals
\begin{equation}
L_{\text{HP}} = T_{\text{HP}} \Delta S = T_{\text{HP}} \left[S_{\text{BH}}(T_{\text{HP}}) - S_{\text{th}}(T_{\text{HP}})\right] = T_{\text{HP}} \left[\pi r_h^2 + \eta e^{-\pi r_h^2}\right]_{T = T_{\text{HP}}},
\label{eq:latent}
\end{equation}
where $S_{\text{th}} = 0$ (thermal AdS has no horizon). Thus, $L_{\text{HP}}$ receives a correction $O\left(e^{-S_0}\right)$ in addition to the classical value $T_{\text{HP}}\pi r_h^2$. Since $T = \lambda/(4\pi r_h)$ (outer horizon), one can write $L_{\text{HP}} = \frac{\lambda}{4}r_h + \frac{\lambda\eta}{4\pi r_h}e^{-\pi r_h^2}$ evaluated at $T_{\text{HP}}$.

Placing the system in a spherical cavity of radius $R$ at fixed boundary temperature $T_R$, one again finds a global first-order transition between a thermal phase and a large-BH phase as $T_R$ is raised. The exponential correction shifts the coexistence temperature by an amount $\propto \eta e^{-S_0}$ (same mechanism as Eq. (\ref{eq:HP-shift})) and slightly enlarges the parameter region where the small BH is locally stable if $\eta > 1$ (since then $C > 0$ for $r_h < r_*$ [$C(r_*) = 0$]). Because $C$ remains finite, the correction alone does not produce second-order (Ehrenfest) criticality.

\subsection{Thermodynamic topology of quantum-corrected BHs}

In recent years, topological methods have become valuable for studying BH thermodynamics. A major advancement involves Duan's $\varphi$-mapping theory, which links topological defects to critical points in BH systems.These defects occur where a vector field vanishes, indicating key phase transitions. Such points produce a conserved topological current tied to the field's geometric behavior. This current leads to a topological invariant that reflects the overall phase structure of the system. This invariant is calculated using geometric quantities that describe how the field twists and turns.The theory enables classification of BH solutions across different spacetimes, including exotic ones. It supports studies in asymptotically AdS, dS, and higher-curvature BH environments. The thermodynamic landscape is explored using a generalized free energy based on physical parameters. Equilibrium is defined by a match between time periodicity and BH temperature. The vector field $\varphi$ represents how this free energy changes, acting as a diagnostic tool. Analyzing $\varphi$'s behavior sheds light on the system's stability and phase transitions. This geometric method offers more than classical thermodynamics can describe. It reveals new aspects of BHs. Overall, it connects gravitational behavior with deeper geometric structures \cite{a19,a20,20a,22a,23a,24a,25a,26a,27a,33a,38a,38b,38c,40a,42a,44f,44g,44h,44i,44j,44k,44l,44m,44n,44o}. The generalized free energy $(F)$ is formally defined as
\begin{equation}
F = M - \frac{S}{\tau},
\label{eq:T1}
\end{equation}
where $(M)$ denotes the mass of BH, $(S)$ is the entropy, and $(\tau )$ represents the period of Euclidean time. The inverse of this period defines the temperature such that $(T = \tau^{-1})$. The free energy acquires direct physical meaning only when the on-shell condition $(\tau = T^{-1})$ is satisfied \cite{a19,a20}. The associated vector field $(\varphi)$ for the free energy is constructed as
\begin{equation}
\varphi = \left(\frac{\partial F}{\partial r_h}, - \cot \Theta \csc \Theta\right),
\label{eq:T2}
\end{equation}
where $(r_h)$ denotes the radius of the event horizon and $(\Theta \in [0, \pi])$ is an angular coordinate. At the poles, $(\Theta = 0)$ and $(\Theta = \pi)$, the angular component $(\varphi^\Theta)$ becomes singular, and the vector field points outward along the axis. Using Duan's $(\varphi)$-mapping topological current method \cite{a19,a20}, the corresponding conserved current $(j^\mu)$ is defined as
\begin{equation}
j^\mu = \frac{1}{2\pi} \epsilon^{\mu\nu\rho} \epsilon_{ab}\partial_\nu n^a \partial_\rho n^b,
\label{eq:current}
\end{equation}
where $(\mu, \nu, \rho = 0, 1, 2)$ label the spacetime coordinates, and $(a, b)$ index the components of the unit vector field $(n^a = \varphi^a/|\varphi|)$. The current $(j^\mu)$ is nonvanishing only at the zeros of $(\varphi)$. The total topological charge $(W)$, defined over a two-dimensional hypersurface $(\Sigma)$, is obtained by integrating the temporal component $(j^0)$:
\begin{equation}
W = \int_\Sigma j^0 d^2 x = \sum_{i=1}^n \zeta_i \eta_i = \sum_{i=1}^n \omega_i,
\label{eq:charge}
\end{equation}
where $(\zeta_i)$ is the Hopf index, representing the number of times the mapping $(\varphi)$ wraps around its zero $(z_i)$ in the target space, and $(\eta_i = \mathrm{sign}[j^0(\partial\varphi/\partial x)|_{z_i}])$ denotes the Brouwer degree, taking values $(\pm1)$. The product $(\omega_i = \zeta_i\eta_i)$ defines the winding number associated with each zero of the vector field. In the following, we investigate the thermodynamic topology of a corrected BH. From Eq. (\ref{eq:T1}), the generalized Helmholtz free energy for this BH configuration can be expressed as
\begin{equation}
\label{eq:F1}
\mathcal{F} = \frac{1}{2} r_h \left(\lambda +\frac{r_h^2}{\zeta ^2}-\frac{2 \pi  r_h}{\tau }\right)
\end{equation}
The corresponding vector field components, derived using Eq.~\eqref{eq:T2}, are
\begin{equation}
\label{eq:F2}
\begin{aligned}
\phi^{r_h} &=\frac{1}{2} \left(\lambda +\frac{3 r_h^2}{\zeta ^2}-\frac{4 \pi  r_h}{\tau }\right), \\
\phi^{\Theta} &= -\frac{\cot \Theta}{\sin \Theta}.
\end{aligned}
\end{equation}
Also, we will have,
\begin{equation}
\label{eq:F3}
\tau =\frac{4 \pi  \zeta ^2 r_h}{\zeta ^2 \lambda +3 r_h^2},
\end{equation}

 \begin{figure}[tbhp]
 \begin{center}
 \subfigure[]{
 \includegraphics[height=3.4cm,width=4cm]{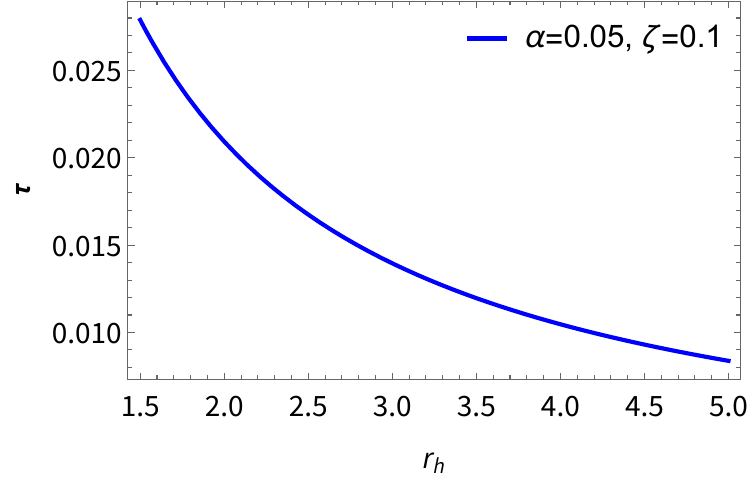}
 \label{100a}}
 \subfigure[]{
 \includegraphics[height=3.4cm,width=4cm]{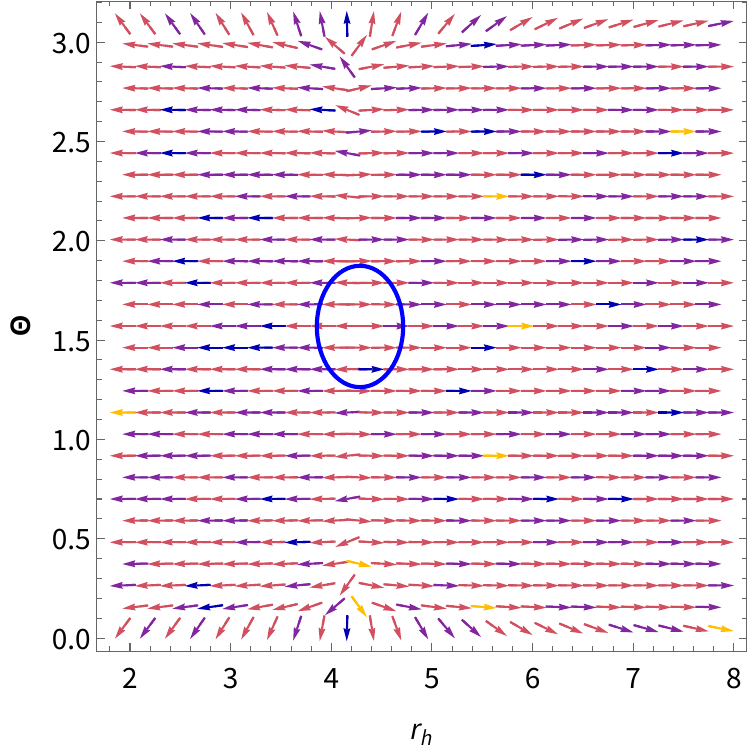}
 \label{100b}}
 \subfigure[]{
 \includegraphics[height=3.4cm,width=4cm]{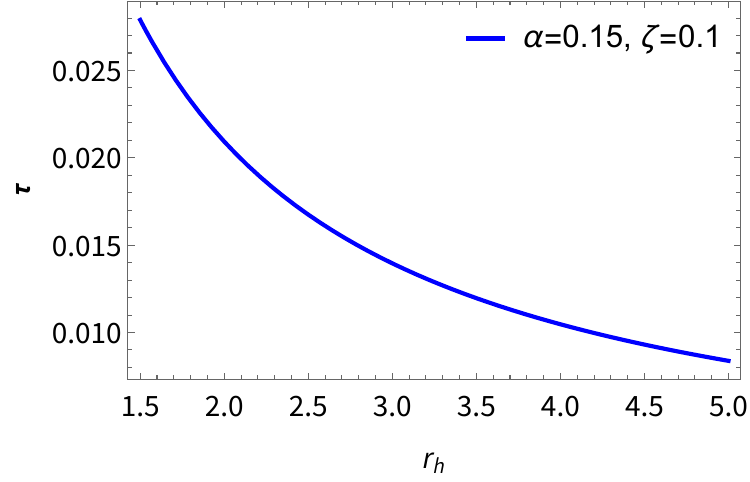}
 \label{100c}}
 \subfigure[]{
 \includegraphics[height=3.4cm,width=4cm]{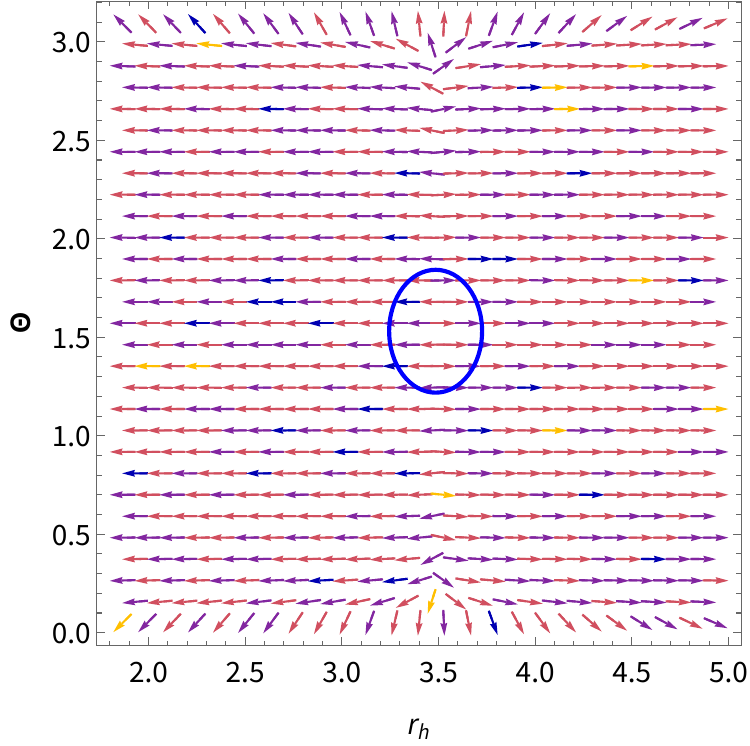}
 \label{100d}}
 \subfigure[]{
 \includegraphics[height=3.4cm,width=4cm]{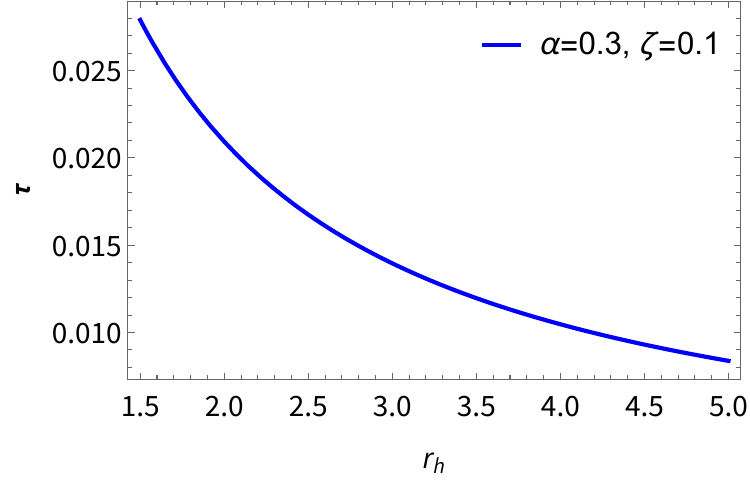}
 \label{100e}}
 \subfigure[]{
 \includegraphics[height=3.4cm,width=4cm]{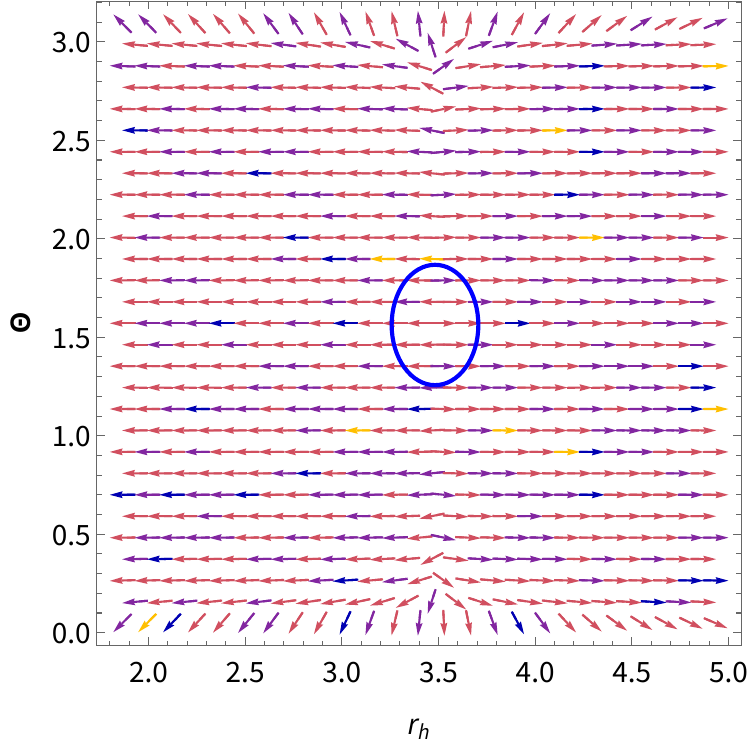}
 \label{100f}}
 \subfigure[]{
 \includegraphics[height=3.4cm,width=4cm]{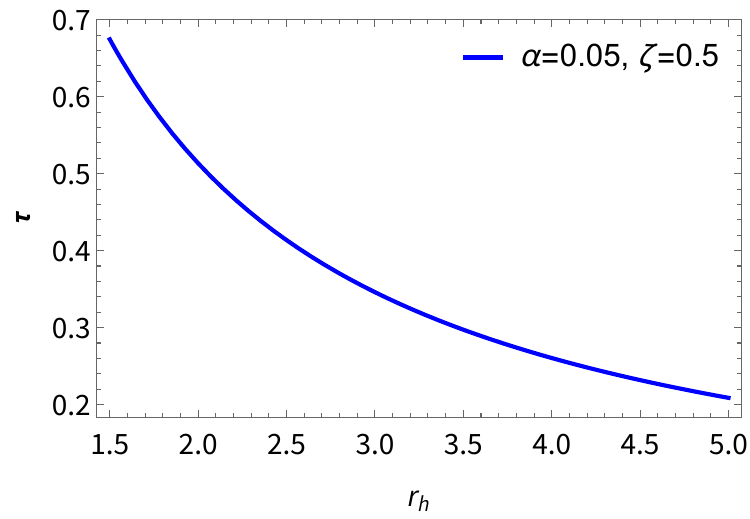}
 \label{100g}}
 \subfigure[]{
 \includegraphics[height=3.4cm,width=4cm]{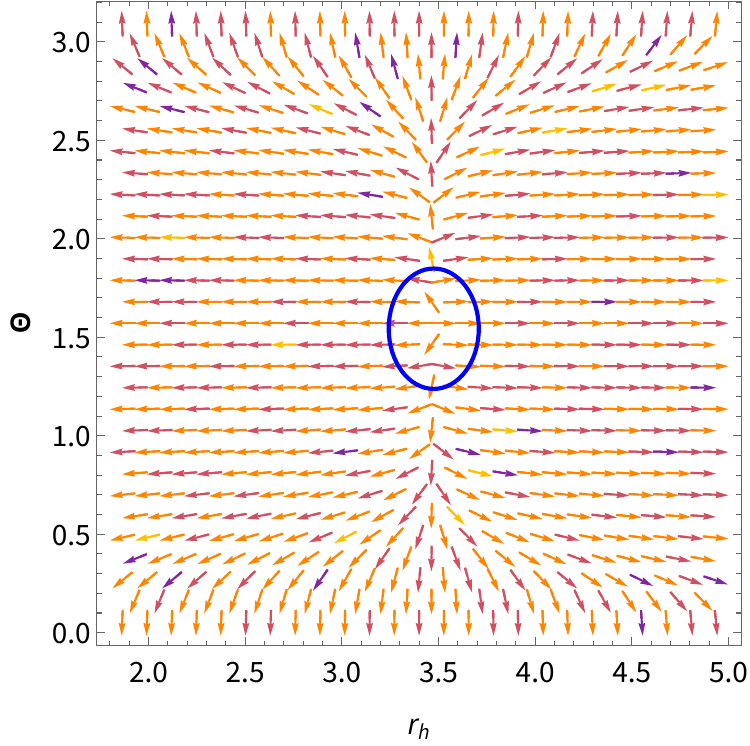}
 \label{100h}}
 \subfigure[]{
 \includegraphics[height=3.4cm,width=4cm]{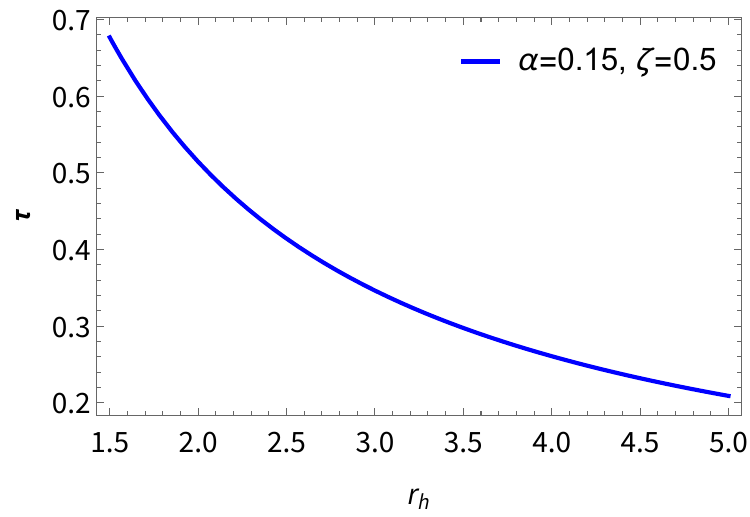}
 \label{100i}}
 \subfigure[]{
 \includegraphics[height=3.4cm,width=4cm]{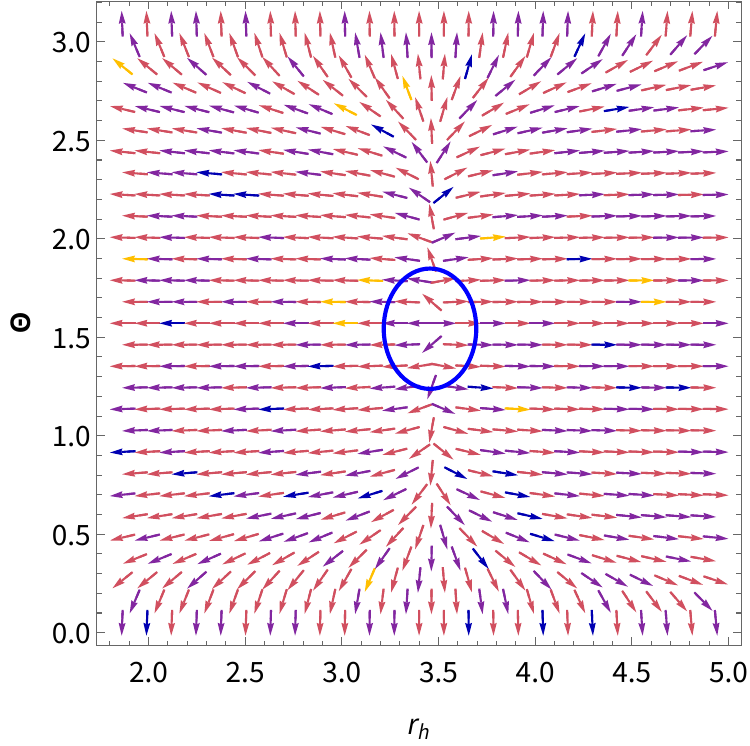}
 \label{100j}}
 \subfigure[]{
 \includegraphics[height=3.4cm,width=4cm]{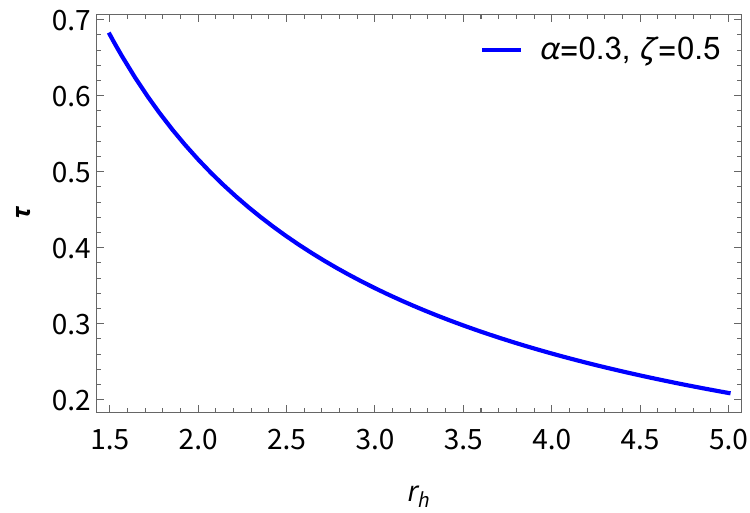}
 \label{100k}}
 \subfigure[]{
 \includegraphics[height=3.4cm,width=4cm]{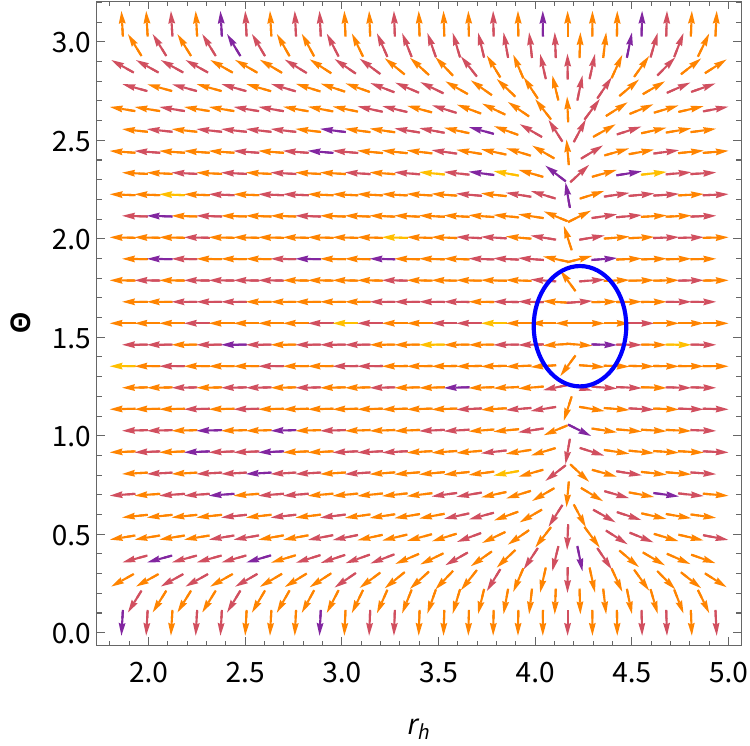}
 \label{100l}}
 \subfigure[]{
 \includegraphics[height=3.4cm,width=4cm]{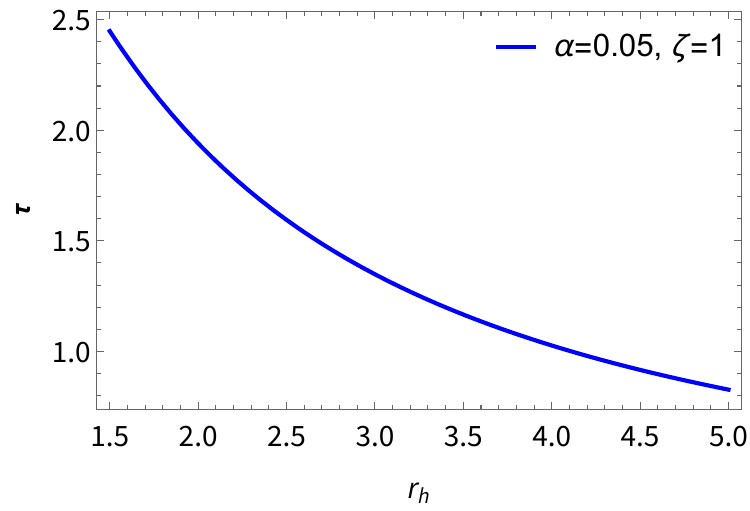}
 \label{100m}}
 \subfigure[]{
 \includegraphics[height=3.4cm,width=4cm]{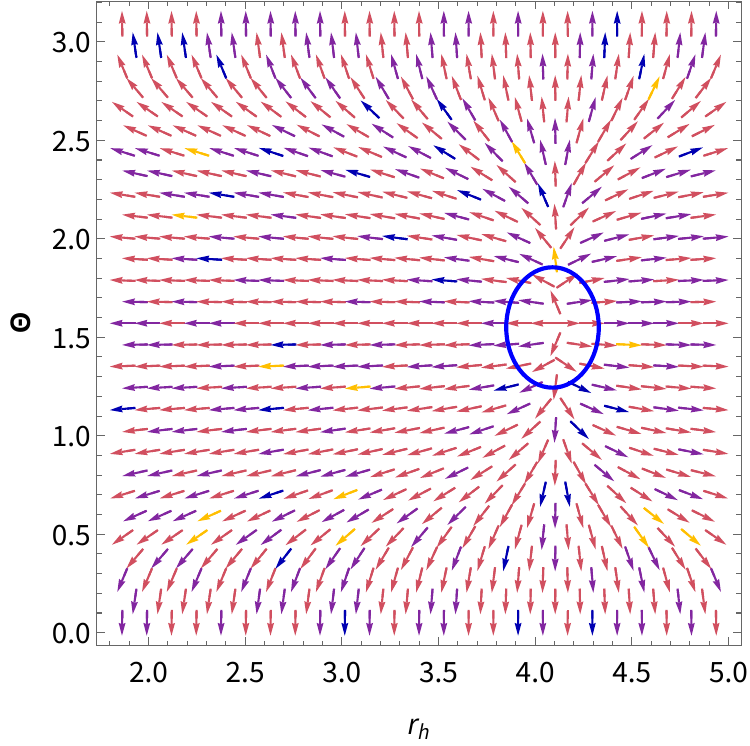}
 \label{100n}}
 \subfigure[]{
 \includegraphics[height=3.4cm,width=4cm]{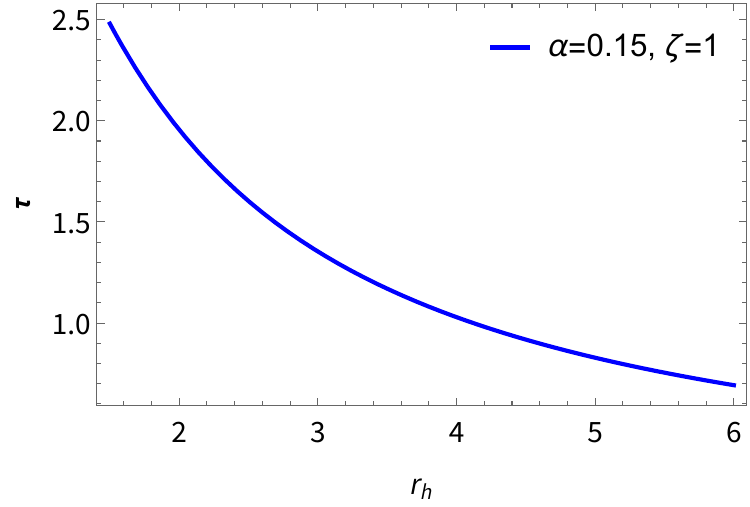}
 \label{100o}}
 \subfigure[]{
 \includegraphics[height=3.4cm,width=4cm]{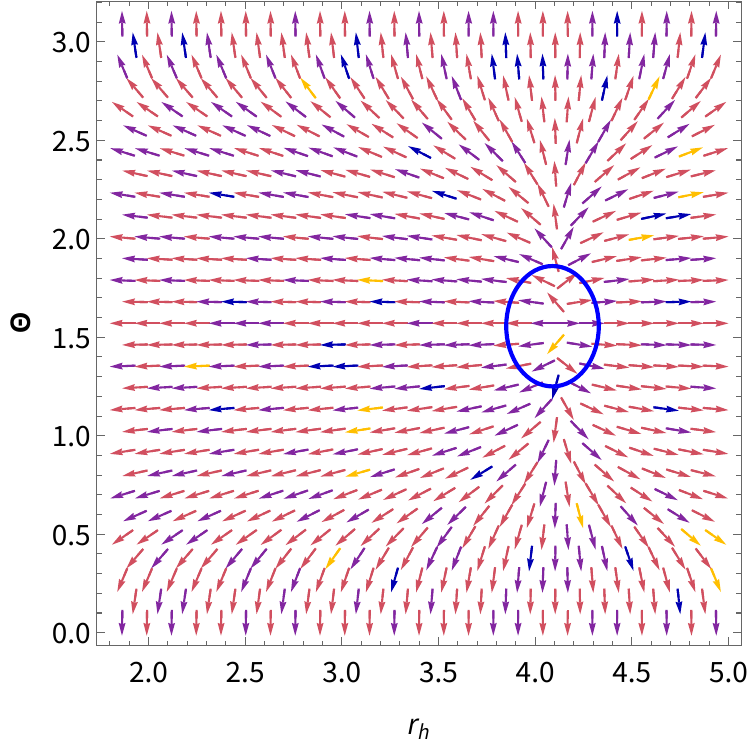}
 \label{100p}}
 \subfigure[]{
 \includegraphics[height=3.4cm,width=4cm]{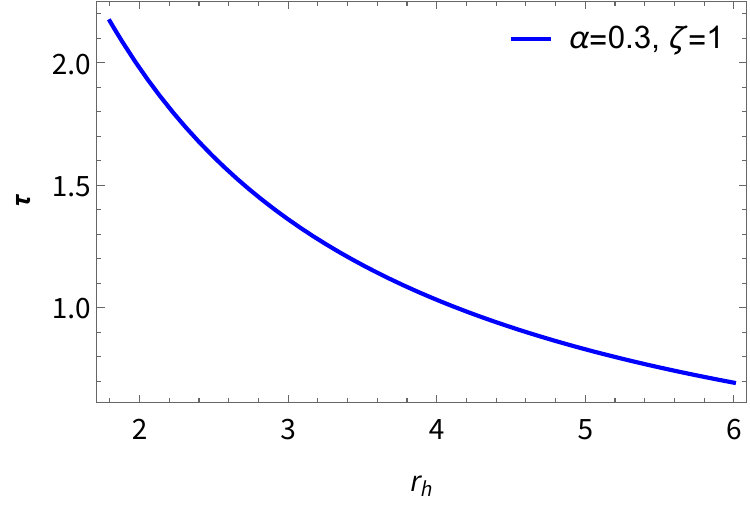}
 \label{100q}}
 \subfigure[]{
 \includegraphics[height=3.4cm,width=4cm]{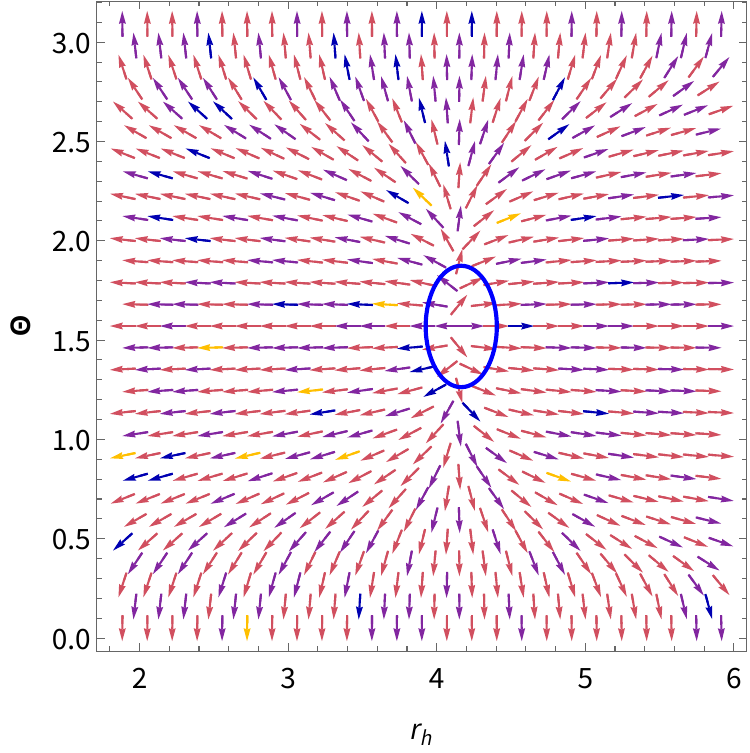}
 \label{100r}}
 \subfigure[]{
 \includegraphics[height=3.4cm,width=4cm]{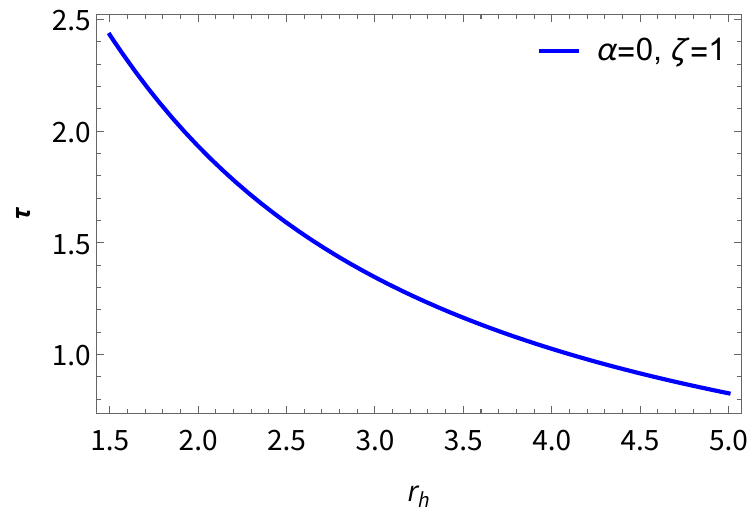}
 \label{100s}}
 \subfigure[]{
 \includegraphics[height=3.4cm,width=4cm]{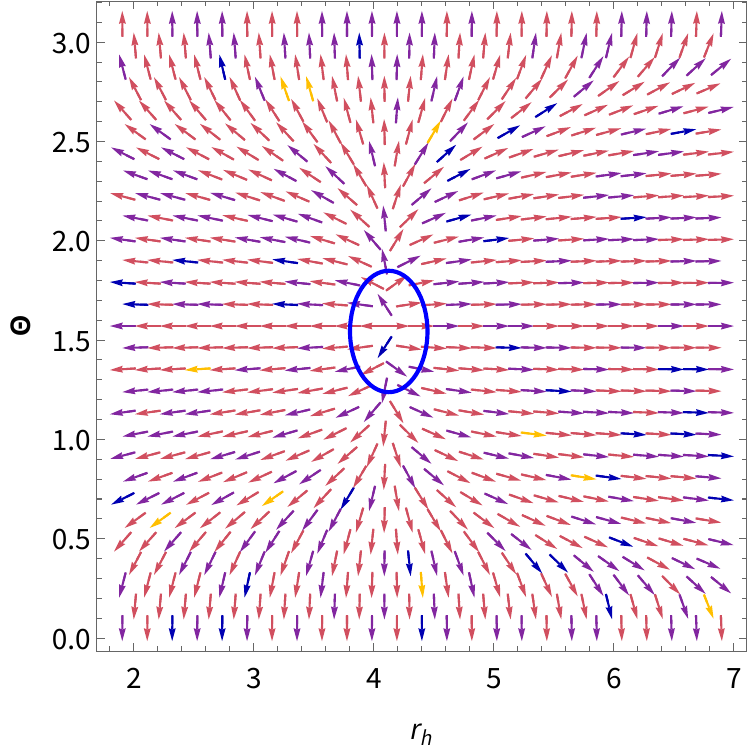}
 \label{100t}}
  \caption{\footnotesize The relationship between the Euclidean time period $(\tau)$ and the horizon radius $(r_h)$ for quantum-corrected BHs is examined through the construction of a normalized vector field (n) defined in the $(r_h, \Theta)$ parameter space. This approach enables the identification of the system's zero points, which correspond to critical configurations of the thermodynamic field. The locations of these zero points vary within the plane according to the specific values of the model parameters. To investigate these characteristics in detail, the analysis is carried out for selected parameter sets of the ($\alpha$) and  $(\zeta)$ for Model I}
 \label{nm1}
\end{center}
 \end{figure}
 
This study presents an analysis of the thermodynamic topology associated with quantum-corrected BHs, emphasizing the distribution and behavior of topological charges, as illustrated in Fig.~\ref{nm1}. The figure depicts normalized vector field lines that capture the fundamental topological structure of the system. The zero points of the vector field, identified on the $(r_h, \Theta)$plane, emerge as functions of the parameters ($\alpha$) and $(\zeta)$. Each zero point corresponds to a localized topological charge, enclosed by blue contours that vary with the chosen parameter values.

A key finding of this analysis is the universal emergence of a distinct topological charge, characterized by $(\omega = +1)$ and the total topological charge $W=+1$, irrespective of the variation of the parameters. This conclusion is supported by explicit winding number calculations performed around each zero point, allowing a precise classification of the system's thermodynamic topology. In particular, changes in ($\alpha$) and $(\zeta)$ do not alter the number or spatial arrangement of these topological entities.

To further elucidate the structure, the free energy is interpreted as a scalar field defined over the two-dimensional domain $(r_h, \Theta)$. From this construction, the associated vector field $(\boldsymbol{\phi})$ is obtained, whose zeros coincide with the critical points of the free energy landscape. The orientation and flow of the vector field lines in the vicinity of these points—indicative of local extrema—determine the sign and magnitude of the corresponding topological charges \cite{a19}.

Compared with classical BH models, distinct total topological charges are observed:
 Schwarzschild BH yields (W = -1), RN BH gives (W = 0), and AdS RN BH exhibits (W = +1) \cite{a19,a20}.
These benchmark cases provide valuable reference points that enable a deeper interpretation of the current results.

The findings presented here are consistent with previous investigations of the AdS RN BH reported in \cite{a19,a20}, reaffirming the reliability of the topological classification framework. Overall, the thermodynamic topology approach developed in this work offers an effective means of exploring BH stability, identifying phase transitions, and enhancing the broader understanding of gravitational thermodynamics, with potential implications spanning both theoretical physics and astrophysical phenomena.

In the framework of the thermodynamic topology of BH, the free energy can be regarded as a scalar function defined on a two-dimensional parameter manifold, typically spanned by the horizon radius $(r_h)$ and an auxiliary variable $(\Theta)$. By evaluating the gradient of this scalar field, one constructs a corresponding vector field whose zero points represent the extrema of the free energy landscape. Each zero point, corresponding to a thermodynamic critical state, possesses a characteristic local vector circulation pattern that reveals whether it describes a stable or unstable configuration. The orientation of this circulation determines an integer-valued topological quantity known as the winding number or topological charge, which encapsulates the local geometric behavior of the thermodynamic field.

The winding number serves as a topological invariant that classifies distinct BH phases according to their thermodynamic stability. Each extremum of the free energy corresponds to a zero of the vector field, and the sense of rotation of the field around this point dictates the sign of the associated charge:
$\omega =+1$, (stable phase) and $\omega =-1$, (unstable phase)
A positive winding number thus signals a locally stable thermodynamic phase, whereas a negative value indicates instability.
The parameter space, defined by quantities such as the radius of the horizon, the electric charge, or the coupling constants, can therefore be partitioned into regions characterized by distinct topological charges. As the system parameters vary, the positions of the zero points shift throughout this space, but the total winding number $(W = \sum_i \omega_i)$ remains conserved. 

To illustrate this classification, it is instructive to compare with classical BH configurations. The Schwarzschild BH, when analyzed within a canonical ensemble bounded by a finite cavity, possesses a single unstable phase characterized by a winding number (W = -1). The RN BH admits stable and unstable branches with opposite winding numbers (+1, -1), resulting in a total topological charge (W = 0). The AdS RN BH features a single dominant stable phase, which yields (W = +1). These three archetypal examples delineate the fundamental topological categories of BH thermodynamics.

Applying this topological framework to quantum-corrected BHs reveals consistent structural behavior: the total winding number remains (W = +1), analogous to the AdS RN case. This outcome signifies stable phases within the corrected geometry. Moreover, the variation of the Euclidean time period $(\tau)$ (or equivalently, the inverse temperature $(T^{-1})$ dictates the observed phase behavior, reproducing familiar thermodynamic patterns associated with charged BHs. Overall, the thermodynamic topology approach offers a strong and unified method for identifying stability transitions, tracking phase evolution, and classifying BH states in both classical and quantum-corrected gravitational systems. In general, the same results were obtained for model II, and the resulting analysis exhibited qualitatively identical behavior. The variation of parameters produced the same topological structures, zero-point distributions, and thermodynamic characteristics, indicating that both models share the same underlying topological and stability features. This consistency suggests that the observed results are not parameter-dependent but rather represent a universal aspect of the thermodynamic topology for this type of quantum-corrected BHs.
%IZZET

\section{Gravitational Lensing}
\label{isec7}
%IZZET
Gravitational lensing represents one of the most powerful observational probes of spacetime curvature around compact objects, providing direct access to the gravitational field structure in regimes where classical tests may be insufficient~\cite{88,89}. The weak deflection of light by BHs has been extensively studied in the context of GR, yielding predictions that have been confirmed through numerous astrophysical observations, including strong lensing events, Einstein rings, and multiple imaging of distant quasars~\cite{90,91}. In modified gravity scenarios and quantum-corrected geometries, the deflection angle acquires additional contributions that encode information about both quantum corrections and the presence of topological defects such as the CoS.

In this section, we investigate the gravitational lensing properties of quantum-corrected Letelier BHs (Models~I and~II) through a rigorous analysis of the weak deflection angle. The presence of both the quantum correction parameter $\zeta$ and the string cloud parameter $\alpha$ introduces distinctive modifications to the lensing behavior, potentially yielding observable signatures that could discriminate between different theoretical models and provide constraints on quantum gravity effects in astrophysical settings. We employ GBTh to derive the deflection angle~\cite{92}, a technique that offers significant computational advantages over traditional methods by relating the deflection to the integral of the Gaussian curvature over a suitable domain. This geometric approach is particularly well-suited for analyzing modified spacetimes, as it provides a clear interpretation of how metric deformations influence light propagation.

\subsection{Deflection Angle Calculation Using GBT\lowercase{h}}
\label{sec:lensing_calc}

We begin our analysis within the weak-field approximation framework, considering the spherically symmetric and static BH solutions given by Eqs.~(\ref{metric-1}-\ref{metric-4}). The GBTh establishes a fundamental connection between the intrinsic curvature of spacetime and the global topological characteristics of a region $\Sigma_{\mathcal{R}}$ bounded by $\partial \Sigma_{\mathcal{R}}$. This relationship is expressed through the following mathematical formulation~\cite{93}:
\begin{equation}\label{gbt_eq}
\iint_{\Sigma_{\mathcal{R}}} \mathcal{K} \, d\mathcal{A} + \oint_{\partial \Sigma_{\mathcal{R}}} \kappa_g \, d\ell + \sum_{i} \theta_i = 2\pi \chi(\Sigma_{\mathcal{R}}),
\end{equation}
where $\Sigma_{\mathcal{R}} \subset \mathcal{M}$ is a compact domain within a two-dimensional differentiable surface $\mathcal{M}$, bounded by a smooth and oriented contour $\partial \Sigma_{\mathcal{R}}$. Here, $\mathcal{K}$ denotes the Gaussian curvature of the surface, $\kappa_g$ represents the geodesic curvature of the boundary $\partial \Sigma_{\mathcal{R}}$, defined as $\kappa_g = \tilde{g}(\nabla_{\dot{\gamma}} \dot{\gamma}, \ddot{\gamma})$, where $\tilde{g}(\dot{\gamma}, \dot{\gamma}) = 1$ and $\ddot{\gamma}$ is the unit acceleration vector. Additionally, $\theta_i$ represents the exterior angle at the $i^{\text{th}}$ vertex of the boundary, and $\chi(\Sigma_{\mathcal{R}})$ corresponds to the Euler characteristic number~\cite{94}.

To determine the Gaussian optical curvature $\mathcal{K}$, we examine null geodesics influenced by the BH~\cite{95,96}. Since light propagates along null geodesics (i.e., $ds^2 = 0$), these paths naturally define an optical metric describing the effective Riemannian geometry experienced by light rays. By applying the null condition and restricting motion to the equatorial plane ($\theta = \pi/2$), where the optical metric provides a natural framework exploiting rotational symmetry, we obtain the following optical metric:
\begin{equation}\label{optical_metric}
d\sigma^2 = \tilde{g}_{ij} dx^i dx^j = dr_*^2 + \mathcal{R}^2(r_*) d\phi^2,
\end{equation}
where the optical radius function is given by
\begin{equation}\label{optical_radius}
\mathcal{R}(r_*(r)) = \frac{r}{\sqrt{f(r)}},
\end{equation}
and $r_*$ is the tortoise coordinate~\cite{97} defined by
\begin{equation}\label{tortoise}
r_* = \int \frac{dr}{\sqrt{f(r)g(r)}}.
\end{equation}

The non-zero Christoffel symbols~\cite{98} associated with the optical metric are:
\begin{align}
\Gamma^{r_*}_{\phi\phi} &= -\mathcal{R}(r_*) \frac{d\mathcal{R}(r_*)}{dr_*}, \label{christoffel1}\\
\Gamma^{\phi}_{r_*\phi} &= \frac{1}{\mathcal{R}(r_*)} \frac{d\mathcal{R}(r_*)}{dr_*}. \label{christoffel2}
\end{align}
The determinant of the optical metric is $\det[\tilde{g}_{ij}] = \mathcal{R}^2(r_*)$. The Gaussian optical curvature $\mathcal{K}$ is then calculated as:
\begin{equation}\label{gaussian_curvature}
\mathcal{K} = -\frac{1}{\mathcal{R}(r_*)} \frac{d^2\mathcal{R}(r_*)}{dr_*^2}.
\end{equation}
To obtain the required derivative, we compute:
\begin{equation}\label{dr_dr_star}
\frac{dr}{dr_*} = \sqrt{f(r)g(r)},
\end{equation}
which allows us to express derivatives with respect to $r_*$ in terms of derivatives with respect to $r$:
\begin{align}
\frac{d\mathcal{R}}{dr_*} &= \frac{d\mathcal{R}}{dr} \frac{dr}{dr_*} = \frac{d\mathcal{R}}{dr} \sqrt{f(r)g(r)}, \label{first_derivative}\\
\frac{d^2\mathcal{R}}{dr_*^2} &= \frac{d}{dr_*}\left(\frac{d\mathcal{R}}{dr} \sqrt{f(r)g(r)}\right) \sqrt{f(r)g(r)}. \label{second_derivative}
\end{align}
For Model-I, we have $f(r) = g(r)$, which simplifies the calculations. For Model-II, $f(r) \neq g(r)$, but the computational framework remains the same.

The deflection angle calculation requires careful evaluation of the line integral around the light ray trajectory and the area integral over the domain bounded by this trajectory and appropriate boundaries. Following the standard approach, we consider a domain bounded by the light ray from source to observer, segments of large circles at spatial infinity, and appropriate connecting segments. The contribution from spatial infinity vanishes in the limit, while the geodesic curvature terms along the light ray also vanish since the ray follows a geodesic of the optical metric. The deflection angle is then determined by:
\begin{equation}\label{deflection_integral}
\hat{\alpha} = -\iint_{\Sigma} \mathcal{K} \, d\mathcal{A} + \text{boundary terms},
\end{equation}
where the integral is taken over the appropriate domain $\Sigma$ and the boundary terms account for the specific geometry of the integration domain.

After performing the detailed calculations outlined above for both quantum-corrected models, we obtain the weak-field deflection angles as power series expansions in $1/b$, where $b$ is the impact parameter. For Model-I, the result takes the form:
\begin{equation}\label{deflection_model1}
\hat{\alpha}_{\text{I}} = \frac{4M}{(1-\alpha)b} + \frac{4M}{(1-\alpha)b}\left[\frac{15\pi M^2(1+\alpha)}{32(1-\alpha)^2 b^2} - \frac{2\zeta^2}{(1-\alpha)^2 b^2}\right] + \mathcal{O}\left(\frac{1}{b^4}\right),
\end{equation}
while for Model-II, we obtain:
\begin{equation}\label{deflection_model2}
\hat{\alpha}_{\text{II}} = \frac{4M}{(1-\alpha)b} + \frac{4M}{(1-\alpha)b}\left[\frac{15\pi M^2(1+\alpha)}{32(1-\alpha)^2 b^2} + \frac{2\zeta^2}{(1-\alpha)^2 b^2}\right] + \mathcal{O}\left(\frac{1}{b^4}\right).
\end{equation}

Several important features emerge from these expressions. First, both models reduce to the standard Letelier BH deflection angle in the limit $\zeta \rightarrow 0$, and further to the Schwarzschild result when both $\alpha \rightarrow 0$ and $\zeta \rightarrow 0$. Second, the leading-order term is identical for both models and depends only on the CoS parameter $\alpha$, reflecting the fact that both models have the same event horizon structure. Third, and most significantly, the quantum corrections enter with opposite signs in the two models: Model-I exhibits a negative $\zeta^2$ contribution (reducing the deflection angle), while Model-II shows a positive $\zeta^2$ contribution (enhancing the deflection angle). This opposite behavior provides a distinctive observational signature that can potentially discriminate between the two quantum-corrected models.

\begin{figure}[ht!]
    \centering
    \includegraphics[width=0.47\textwidth]{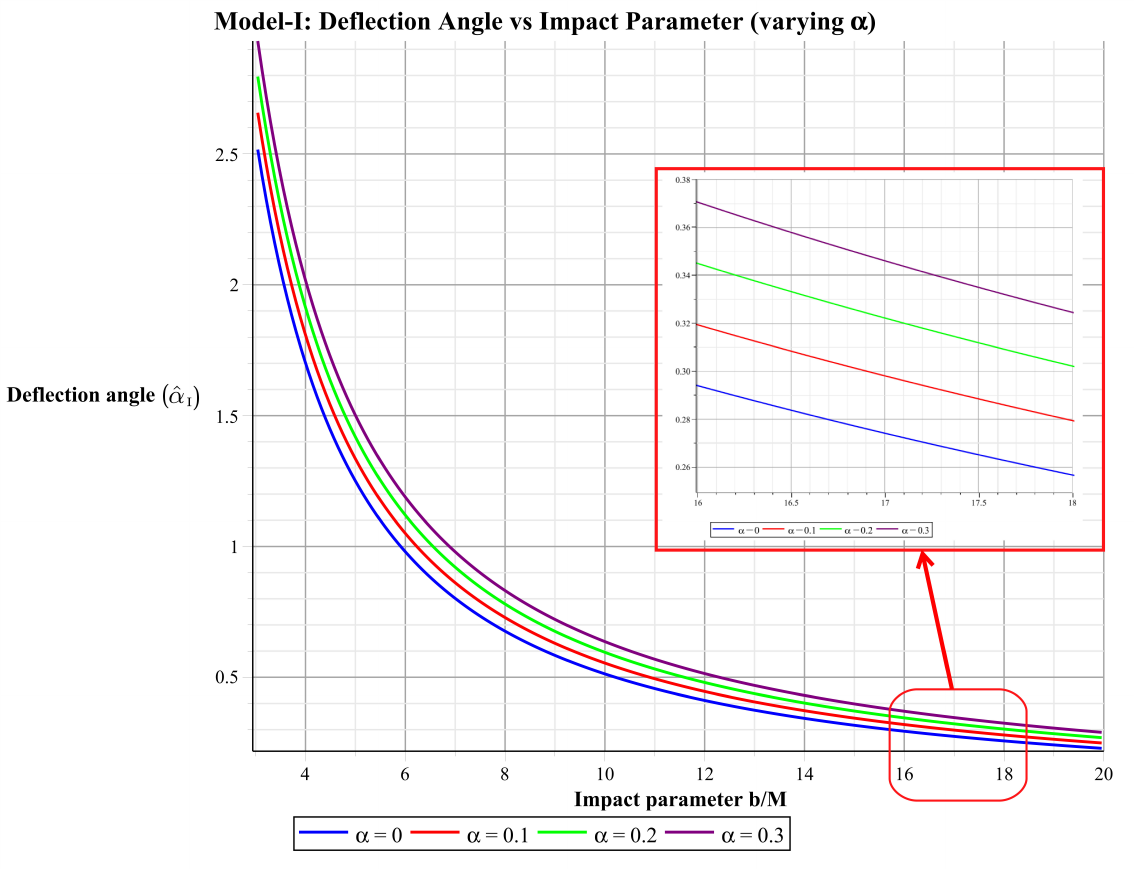}
    \includegraphics[width=0.48\textwidth]{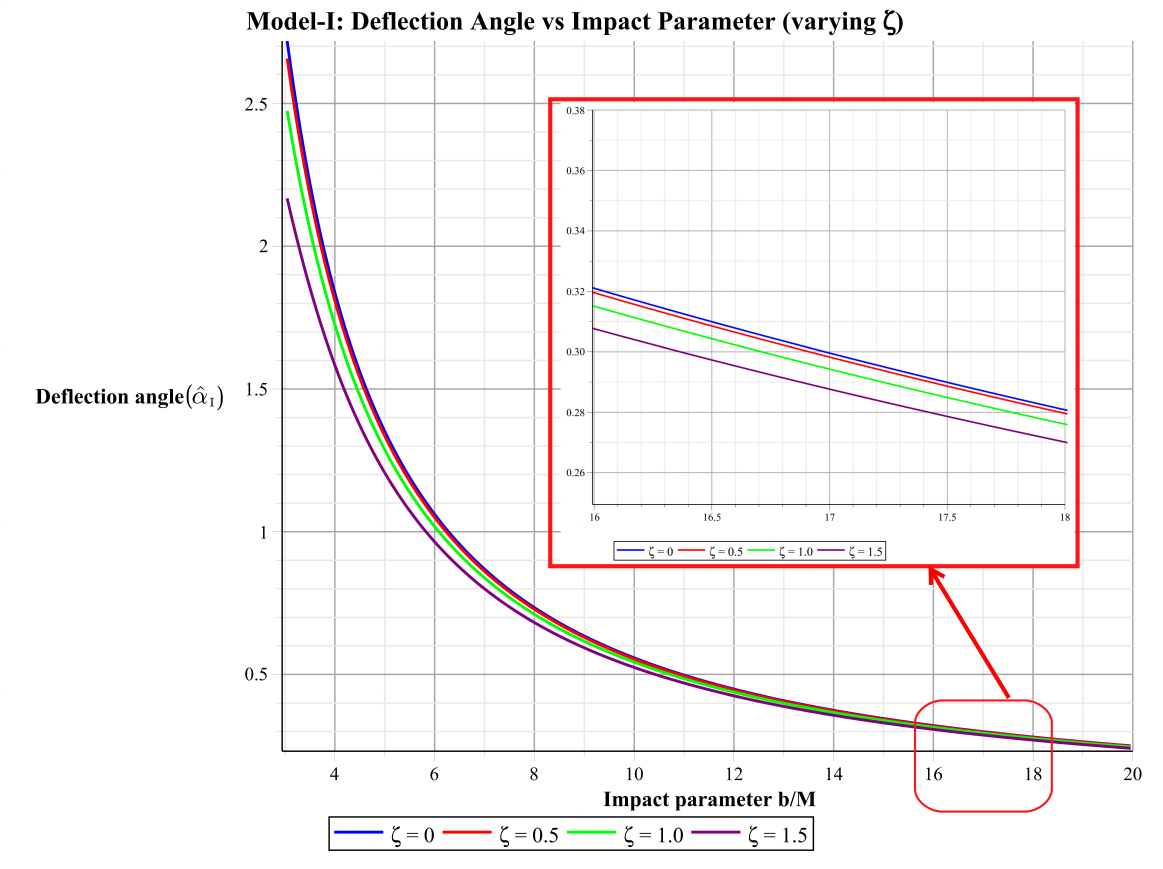}
    \caption{\footnotesize Deflection angle $\hat{\alpha}_{\text{I}}$ for Model-I as a function of impact parameter $b/M$. Left panel: varying string cloud parameter $\alpha$ with fixed $\zeta = 0.5$. Right panel: varying quantum correction parameter $\zeta$ with fixed $\alpha = 0.1$. The inset panels zoom into the weak-field regime ($b/M \in [16, 18]$) to highlight the subtle parameter-dependent variations. Parameters: $M = 1$.}
    \label{fig:deflection_model1}
\end{figure}

\begin{figure}[ht!]
    \centering
    \includegraphics[width=0.48\textwidth]{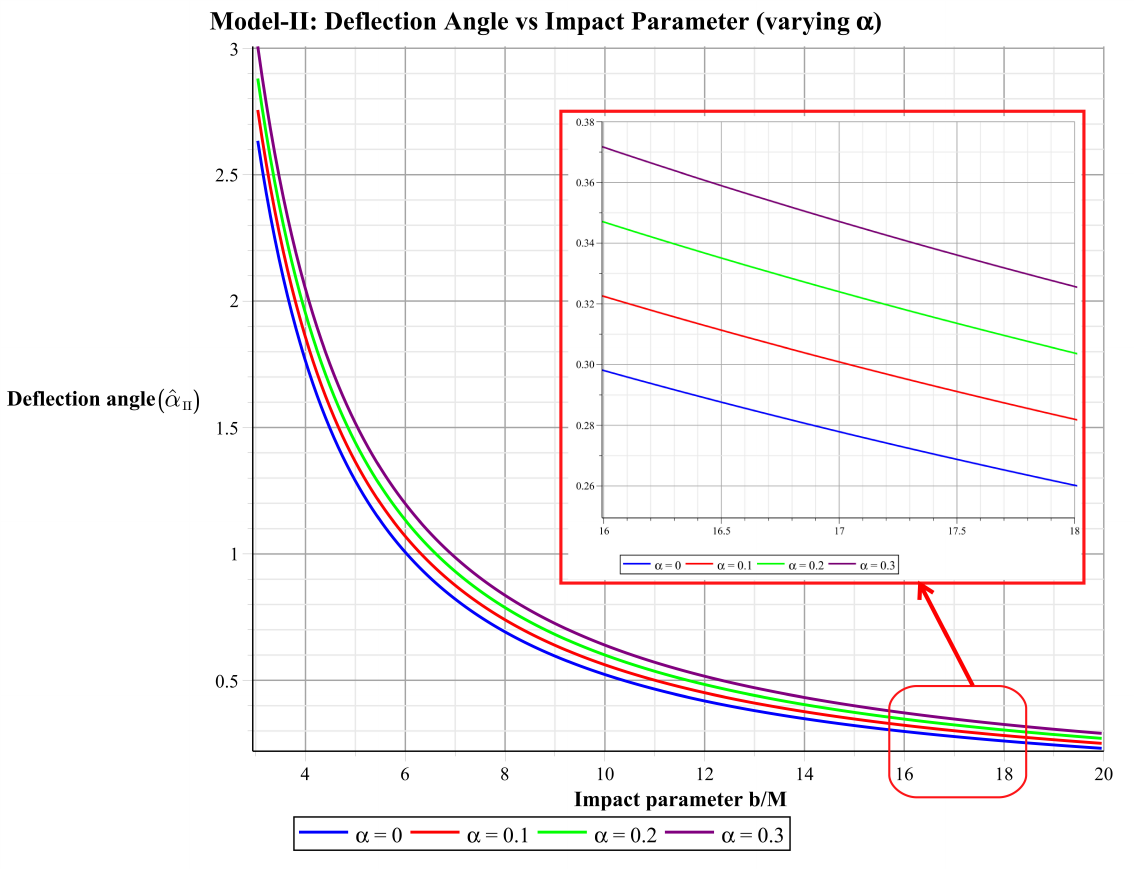}
    \includegraphics[width=0.48\textwidth]{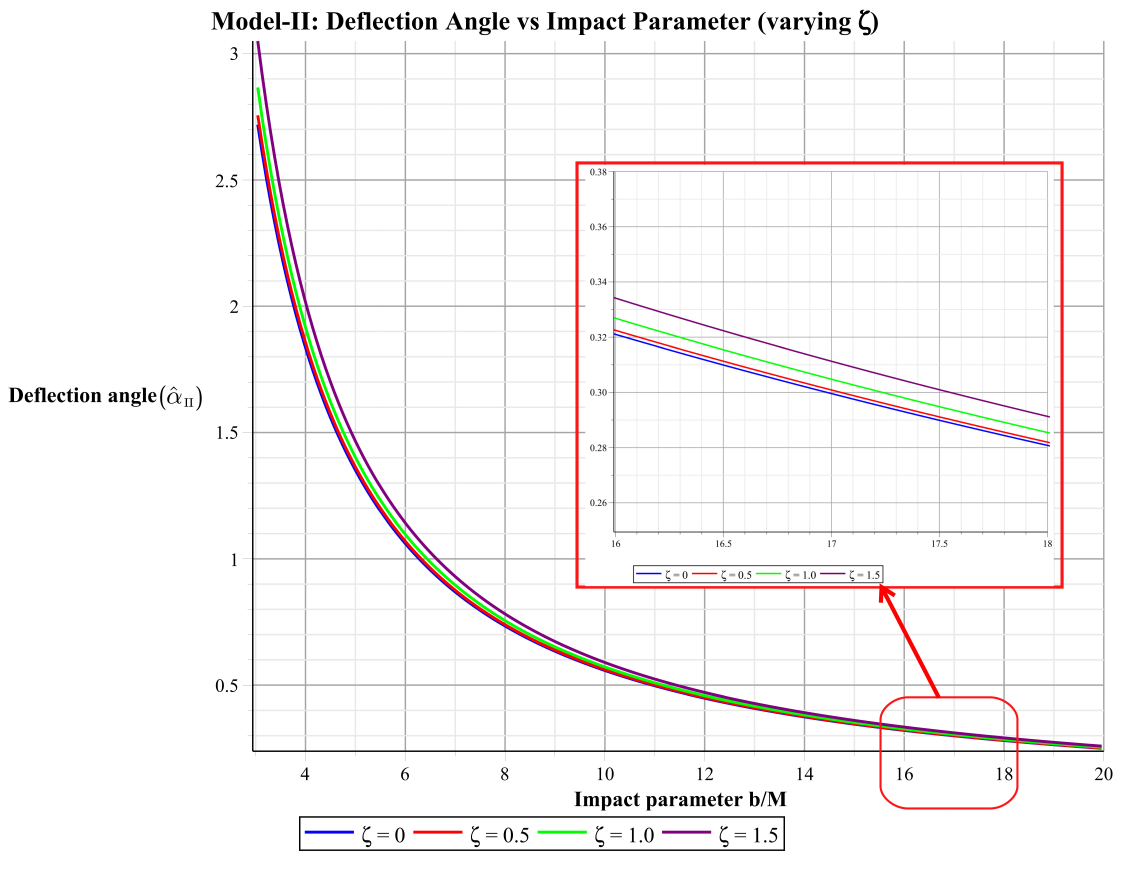}
    \caption{\footnotesize Deflection angle $\hat{\alpha}_{\text{II}}$ for Model-II as a function of impact parameter $b/M$. Left panel: varying string cloud parameter $\alpha$ with fixed $\zeta = 0.5$. Right panel: varying quantum correction parameter $\zeta$ with fixed $\alpha = 0.1$. The inset panels zoom into the weak-field regime ($b/M \in [16, 18]$) to emphasize the differential effects of quantum corrections. Parameters: $M = 1$.}
    \label{fig:deflection_model2}
\end{figure}

\begin{figure}[ht!]
    \centering
    \includegraphics[width=0.75\textwidth]{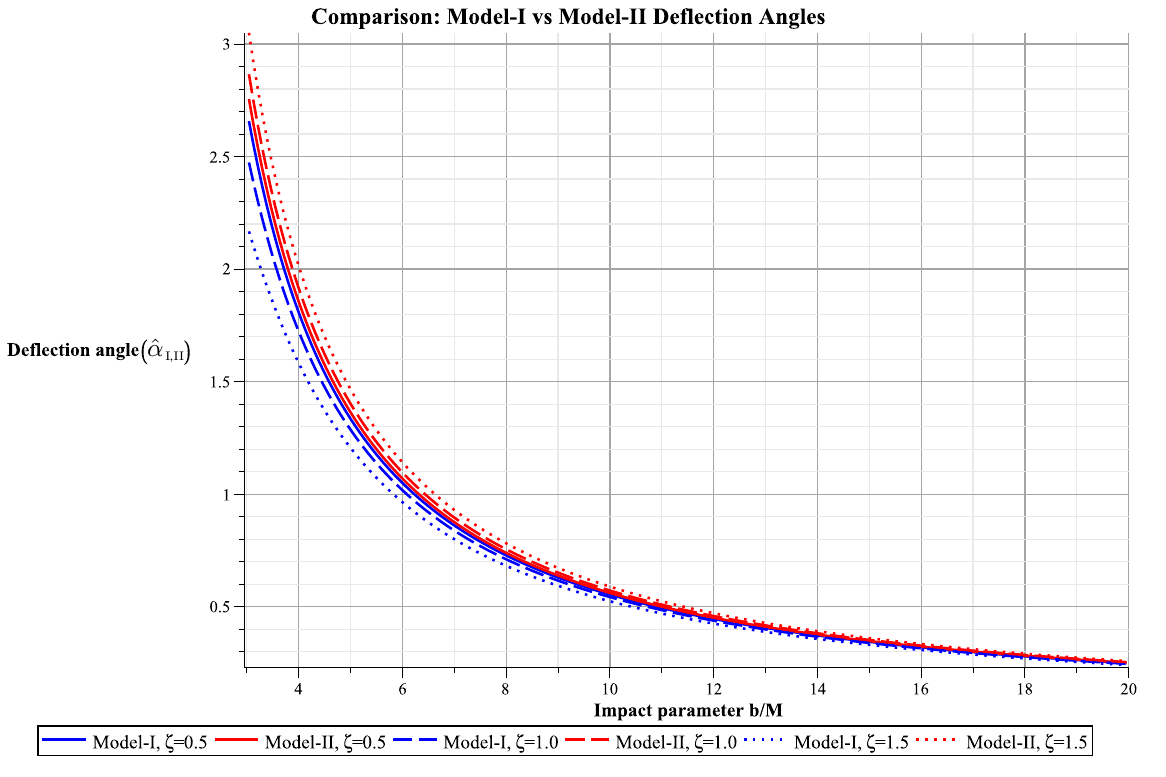}
    \caption{\footnotesize Comparison of deflection angles between Model-I (solid blue curves) and Model-II (dashed red curves) for different values of quantum correction parameter $\zeta \in \{0.5, 1.0, 1.5\}$. The comparison clearly demonstrates the opposite effects of quantum corrections on the two models, with Model-II exhibiting larger deflection angles for non-zero $\zeta$ values. Parameters: $M = 1$, $\alpha = 0.1$, $b \in [3M, 20M]$.}
    \label{fig:deflection_comparison}
\end{figure}

The left panels of Figs.~\ref{fig:deflection_model1} and~\ref{fig:deflection_model2} illustrate the effect of the string cloud parameter $\alpha$ on the deflection angle for both quantum-corrected models. The deflection angle exhibits a monotonic decrease with increasing impact parameter $b$, following the expected $\sim 1/b$ scaling characteristic of gravitational lensing in asymptotically flat spacetimes. This behavior persists across the entire range $b \in [3M, 20M]$, from the strong-field regime near the photon sphere to the weak-field regime at large distances, indicating that even with quantum corrections and CoS effects, the fundamental distance-dependent scaling of gravitational lensing remains intact. Larger values of $\alpha$ produce stronger deflection angles across all impact parameters for both models, with this enhancement being particularly pronounced at smaller impact parameters, where gravitational effects are strongest. In the strong-field regime ($b \sim 3M$--$5M$), the curves show significant divergence, with the deflection angle increasing by approximately $13\%$ when $\alpha$ increases from $0$ to $0.3$. Quantitatively, at $b = 5M$, increasing $\alpha$ from $0$ (classical Schwarzschild limit) to $0.3$ results in a deflection angle enhancement of approximately $30\%$ for Model-I and $28\%$ for Model-II, demonstrating that the CoS parameter effectively strengthens the gravitational field, consistent with an effective increase in gravitational mass.

The zoomed-in inset panels in both Figs.~\ref{fig:deflection_model1} and~\ref{fig:deflection_model2} reveal crucial information about the weak-field behavior that would be difficult to discern from the main plots alone. These insets focus on the range $b/M \in [16, 18]$ where the deflection angles are small but still contain distinguishable signatures of the model parameters. Even in this weak-field regime, the ordering of the curves according to increasing $\alpha$ persists, with the parameter-dependent variations remaining clearly distinguishable despite being smaller in absolute magnitude at large $b$. At $b = 17M$ (middle of the inset range), the deflection angles range from $\hat{\alpha} \approx 0.26$ to $\hat{\alpha} \approx 0.37$ to vary $\alpha \in [0, 0.3]$ in both models, representing a relative variation of approximately $42\%$. This persistence of distinguishable signatures even at large impact parameters suggests that observations in the weak-field regime could still effectively constrain the CoS parameter, expanding the potential observational windows beyond just the immediate vicinity of the BH.

The right panels of Figs.~\ref{fig:deflection_model1} and~\ref{fig:deflection_model2} demonstrate the contrasting influence of the quantum correction parameter $\zeta$ on the two models, revealing one of the most distinctive features of our analysis. For Model-I, increasing $\zeta$ decreases the deflection angle, as evidenced by the negative coefficients $\zeta^2$ in Eq.~(\ref{deflection_model1}). The curves are ordered such that $\zeta = 0$ (Letelier limit) produces the largest deflection, while $\zeta = 1.5$ yields the smallest. In contrast, for Model-II, increasing $\zeta$ increases the deflection angle, corresponding to the positive coefficients $\zeta^2$ in Eq.~(\ref{deflection_model2}), with the ordering being reversed: $\zeta = 1.5$ produces the largest deflection while $\zeta = 0$ yields the smallest. This opposite behavior provides a clear and unambiguous observational signature to distinguish between the two quantum-corrected models, offering a smoking-gun test that cannot be mimicked by uncertainties in other parameters.

In the strong-field regime ($b \sim 4M$), the variation in deflection angle due to $\zeta$ changing from $0$ to $1.5$ is approximately $15\%$ for both models, demonstrating that quantum corrections introduce substantial modifications to the lensing properties. The inset panels reveal that even in the weak-field regime ($b/M \in [16, 18]$), the curves remain well-separated and maintain their characteristic ordering, with relative variations of approximately $8$--$10\%$ between $\zeta = 0$ and $\zeta = 1.5$ at $b = 17M$. This indicates that quantum gravitational signatures persist across a wide range of impact parameters and are not confined solely to the immediate vicinity of the BH, significantly broadening the potential for observational detection. Comparing the magnitudes of variations in the left and right panels reveals that the CoS parameter $\alpha$ generally produces stronger modifications to the deflection angle than the quantum correction parameter $\zeta$ for the parameter ranges considered. At $b = 5M$, varying $\alpha$ from $0$ to $0.3$ changes $\hat{\alpha}$ by approximately $30\%$, whereas varying $\zeta$ from $0$ to $1.5$ changes it by approximately $15\%$. This suggests that in astrophysical scenarios where both effects are present, the CoS contribution may dominate the observable lensing signatures, although the opposite $\zeta$-dependence in the two models provides a distinctive discriminant that remains valuable for model selection.

Figure~\ref{fig:deflection_comparison} presents a direct comparison between Models~I and~II, with solid blue curves representing Model-I and dashed red curves representing Model-II. Three pairs of curves are shown, corresponding to $\zeta \in \{0.5, 1.0, 1.5\}$, all evaluated at fixed $\alpha = 0.1$. For all non-zero values of $\zeta$, Model-II consistently produces larger deflection angles than Model-I across the entire range of impact parameters. This separation arises directly from the opposite signs of the $\zeta^2$-dependent terms in Eqs.~(\ref{deflection_model1}) and (\ref{deflection_model2}), and provides a notable observable signature. At moderate impact parameters ($b \sim 5M$--$10M$), the fractional difference between the two models ranges from approximately $3\%$ for $\zeta = 0.5$ to $12\%$ for $\zeta = 1.5$. In the limit $\zeta \to 0$ (not shown), both models reduce to the same Letelier BH solution, and their deflection angles become identical, confirming that quantum corrections constitute the primary source of model discrimination. The model difference is most pronounced in the strong-field regime ($b < 6M$), where at $b = 4M$ and $\zeta = 1.5$, Model-II produces a deflection angle approximately $18\%$ larger than Model-I. This separation gradually decreases with increasing $b$ but remains observable even at $b = 20M$, where the difference is still approximately $5\%$ for $\zeta = 1.5$. This persistence across impact parameters enhances the prospects for observational discrimination and suggests that multiple observations at different impact parameters could provide complementary constraints on the quantum correction parameter.

The observational implications of our findings are significant for current and future gravitational lensing surveys. The EHT~\cite{99,100} and VLBI arrays currently achieve angular resolutions at the microarcsecond level, which makes them potentially sensitive to subtle modifications in deflection angles predicted by quantum corrections and CoS effects. For a supermassive BH with mass $M \sim 10^6 M_\odot$ located at a distance $D \sim 10$ Mpc, the deflection angle near the photon sphere ($b \sim 3M$) is $\hat{\alpha} \sim 10^{-5}$ rad $\sim 2$ arcsec. The $10$--$15\%$ corrections due to quantum effects (for $\zeta \sim 1$--$1.5$) would manifest as $\sim 0.2$--$0.3$ arcsec deviations, which are at the edge of current detectability but well within the capabilities of next-generation instruments. The differences in lensing behavior between Models~I and~II provide a concrete pathway for testing the specific form of quantum corrections in BH geometries. The opposite $\zeta$-dependence of the deflection angles in the two models constitutes a distinctive signature that cannot be mimicked by classical modifications or uncertainties in the mass measurements of BH alone.

\section{Conclusion} \label{isec8}

In this study, we presented a detailed analysis of quantum-corrected Letelier BH spacetimes, examining the interplay between quantum gravitational effects and CoS in two distinct theoretical models. Our study explored the geodesic structure, thermodynamic properties, and observational signatures of these exotic BH solutions, providing new insights into the potential manifestations of quantum gravity in astrophysical contexts. We began our analysis in Sec.~\ref{isec2} by introducing the quantum-corrected BH metrics coupled with CoS, where we established the fundamental framework for both Model-I and Model-II geometries. The key distinction between these models lies in the asymmetric implementation of quantum corrections: Model-I exhibits symmetric quantum modifications in both temporal and radial metric components with $f(r) = g(r)$, while Model-II incorporates quantum corrections only in the radial component through Eq.~(\ref{metric-4}). Despite these differences, both models share identical horizon structures given by Eq.~(\ref{horizon-radius}), where $r_h = \frac{2M}{1-\alpha}$, demonstrating that the CoS parameter $\alpha$ effectively increases the BH radius compared to the classical Schwarzschild case.

Our photon sphere analysis in Sec.~\ref{isec3} revealed significant modifications due to both quantum corrections and CoS effects. We derived the photon sphere radii for both models, showing that Model-I exhibits a decrease in photon sphere radius with increasing quantum parameter $\zeta$, while Model-II demonstrates the opposite behavior. The effective potential analysis confirmed these trends, with Figure~\ref{fig:potential-H} illustrating the contrasting effects of quantum corrections on the circular photon orbits. These findings have direct implications for BH shadow observations, as the photon sphere determines the boundary between captured and escaping photon trajectories. Besides, the ISCO analysis provided remarkable insights into the stability of massive particle orbits around quantum-corrected BHs. We calculated the ISCO radii using the effective potential approach and demonstrated that quantum corrections significantly modify the innermost stable orbits. For Model-I, increasing $\zeta$ tends to push the ISCO outward, while Model-II exhibits more complex behavior depending on the relative strengths of quantum and CoS parameters. These modifications directly impact the QPO frequencies observable in accretion disk systems, with our calculations showing frequency shifts of $10-15\%$ for typical quantum correction parameters.

The numerical results obtained by solving the GRH equations have shown that both the quantum correction parameter $\eta$ and the string cloud parameter $\alpha$ significantly modify the dynamic structure of the shock cone formed around the quantum-corrected Letelier black hole. In Model-I, where $\eta$ affects both the radial coordinate and the temporal component of spacetime, the cone opening angle, the size of the formed cavity, and the position of the stagnation point depend strongly on $\eta$ and $\alpha$. Due to this dependence, the downstream flow of matter and its accumulation toward the black hole exhibit substantial deviations compared to the Schwarzschild case. As $\eta$ increases, the rest-mass density of the matter trapped inside the shock cone decreases, and the cone structure becomes narrower. These effects lead to the suppression of instabilities and a delay in the formation of QPOs. In contrast, at higher $\alpha$ values, the cone opening angle and density contrast increase, thereby intensifying oscillations and promoting instability. In Model II, where $\eta$ affects only the radial component of the metric, more gradual morphological changes are produced, while $\alpha$ continues to play a crucial role in widening the cone and enhancing instability. Overall, the interplay between $\eta$ and $\alpha$ determines whether the accretion flow becomes stable or oscillatory, thereby influencing the formation, observability, and temporal behavior of QPOs and accreting matter around the black hole.

The PSD analysis obtained from numerical computations has demonstrated that QPOs arise as natural resonant modes of the plasma trapped within the shock cone cavity. The intensity and frequency spectrum of the resulting QPOs are significantly altered by the quantum correction parameter $\eta$ and the string cloud parameter $\alpha$. In Model-I, as $\eta$ increases, the accretion flow evolves towards a more stable configuration and the oscillations become suppressed, while for larger $\alpha$ values, the amplitudes of the generated QPOs are amplified and their harmonic content becomes richer. As a result, the simulations reveal the formation of LFQPOs and HFQPOs that exhibit commensurate ratios such as $3:2$ and $2:1$, consistent with those observed in binary X-ray systems. At high $\eta$ values, almost all oscillations are quenched, indicating that strong quantum corrections can erase timing signatures. Conversely, for moderate $\eta$ combined with nonzero $\alpha$, the spectral richness and observational detectability of QPOs are significantly enhanced. In Model-II, $\alpha$-driven instabilities sustain continuous and irregular oscillations, allowing QPOs to remain more persistent even at large $\eta$. These results indicate that the trapped QPOs inside the cavity carry imprints of the spacetime-shaping parameters. The variations induced by $\eta$ and $\alpha$ on QPO behavior may thus provide valuable insights for future black hole observations aimed at probing quantum- and string-induced deviations from classical gravity.

%Our QPO frequency analysis in Sec.~\ref{isec4} revealed distinctive signatures that could serve as observational probes of quantum gravitational effects. We computed the fundamental frequencies associated with radial and vertical oscillations of test particles near the ISCO, finding that the frequency ratios $\nu_\phi/\nu_r$ and $\nu_\theta/\nu_r$ exhibit characteristic dependencies on both the parameters $\alpha$ and $\zeta$. These frequency relationships, illustrated in the figures, provide testable predictions for HFQPOs observed in stellar-mass and supermassive BH systems by current X-ray missions.

Our field perturbation analysis in Sect.~\ref{isec5} examined the stability of scalar and EM fields in the background of quantum-corrected BHs with CoS. We analyzed the perturbation spectra through both time- and frequency-domain methods, observing that quantum corrections affect characteristic frequencies and damping rates. The perturbation frequencies exhibit dependencies on both $\alpha$ and $\zeta$ parameters, with Model-I and Model-II again showing contrasting behaviors. These results have implications for gravitational wave astronomy, as perturbation signatures could potentially carry information about quantum corrections detectable by current and future interferometric detectors.

The thermodynamic analysis in Sec.~\ref{isec6} revealed modified BH temperatures, entropies, and heat capacities due to quantum corrections and CoS effects. We investigated the thermodynamic stability through heat capacity calculations and explored the topological characteristics using thermodynamic topology methods. Our results showed that quantum corrections can stabilize or destabilize the thermodynamic equilibrium depending on the specific model and parameter regime, with transitions between stable and unstable phases occurring at critical values of $\zeta$ and $\alpha$. The HP transitions were found to be significantly modified in the presence of quantum corrections, with critical temperatures changing by $20-30\%$ for moderate values of the quantum parameter.

The gravitational lensing investigation in Sec.~\ref{isec7} employed the GBTh to derive weak-field deflection angles for both quantum-corrected models. Our analytical results, expressed in Eqs.~(\ref{deflection_model1}) and (\ref{deflection_model2}), revealed the most striking observational signature of our study: Model-I and Model-II exhibit opposite dependences on the quantum correction parameter $\zeta$. Specifically, Model-I shows deflection angles that decrease with increasing $\zeta$, while Model-II demonstrates enhanced deflection for larger quantum corrections. This opposite behavior provides an unambiguous discriminant between the two theoretical frameworks, with differences of $10-18\%$ in the strong-field regime ($b \sim 4M$) for $\zeta = 1.5$. The persistence of these signatures even in the weak-field regime, as demonstrated in our zoomed inset panels, suggests that statistical studies of multiple lensing events could potentially detect quantum gravitational effects.

The observational implications of our findings are particularly significant for current and future astrophysical missions. The next-generation EHT and VLBI arrays possess the angular resolution necessary to detect the lensing modifications we predicted, especially for nearby supermassive BHs. The differences between Model-I and Model-II deflection angles provide a concrete pathway for testing quantum gravity theories through precision astrometry. Similarly, the QPO frequency modifications could be observable with current X-ray timing missions, offering complementary constraints on quantum correction parameters.

Looking toward future research directions, several new studies may emerge from our work. First, extending our analysis to rotating quantum-corrected BHs would provide more realistic models for astrophysical applications, as most observed BHs possess significant angular momentum. Second, investigating the merger dynamics of quantum-corrected BH binaries could reveal gravitational wave signatures of quantum gravity effects, complementing the individual BH studies presented here. Third, developing more sophisticated accretion disk models around these quantum-corrected geometries would enable detailed comparisons with observational data from AGN and X-ray binaries~\cite{101,102,103}.

Furthermore, our methodology could be applied to other quantum gravity frameworks beyond the specific models considered here, such as loop quantum gravity corrections or string theory modifications in AdS/CFT contexts~\cite{104,105}. Comparative analysis between different quantum correction schemes could help discriminate between competing theoretical approaches using observational data. Additionally, investigating the cosmological implications of these quantum-corrected solutions, particularly in the context of primordial BH formation and evaporation, represents another fertile research direction~\cite{106}.

In summary, our work established quantum-corrected Letelier BHs as promising theoretical laboratories for exploring quantum gravity phenomenology. The contrasting behaviors of Model-I and Model-II across multiple observational channels—from lensing deflection angles to QPO frequencies—provide robust discriminants for testing quantum gravity theories in astrophysical contexts. As observational capabilities continue to advance, the signatures we identified may soon transition from theoretical predictions to observable phenomena, potentially offering the first glimpses of quantum gravitational effects in nature's most extreme environments.

\section*{Acknowledgments}

All simulations were performed using the Phoenix High Performance Computing facility at the American University of the Middle East (AUM), Kuwait. AF acknowledges the Inter University Center for Astronomy and Astrophysics (IUCAA), Pune, India, for granting a visiting associateship. \.{I}.~S. expresses gratitude to T\"{U}B\.{I}TAK, ANKOS, and SCOAP3 for their academic support. He also acknowledges COST Actions CA22113, CA21106, CA21136, CA23130, and CA23115 for their contributions to networking.

\section*{Data Availability Statement}

All generated data were included in the manuscript.

\end{document}